\documentclass[fleqn,usenatbib]{mnras}

\usepackage{newtxtext,newtxmath}
\usepackage[T1]{fontenc}

\DeclareRobustCommand{\VAN}[3]{#2}
\let\VANthebibliography\thebibliography
\def\thebibliography{\DeclareRobustCommand{\VAN}[3]{##3}\VANthebibliography}

\usepackage{graphicx}	% Including figure files
\usepackage{amsmath}	% Advanced maths commands
\newcommand{\kms}{\,km\,s$^{-1}$\,}	%km per s
\usepackage{array}
\newcolumntype{!}{>{\global\let\currentrowstyle\relax}}
\newcolumntype{^}{>{\currentrowstyle}}

\title[]{Milliarcsecond structure and variability of methanol maser emission in three high-mass protostars}
\author[A. Aberfelds et al.]{A. Aberfelds,$^{1}$\href{https://orcid.org/0000-0002-6727-2858}{\includegraphics[scale=0.5]{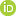}}\thanks{E-mail: artis.aberfelds@venta.lv}
A. Bartkiewicz,$^{2}$\href{https://orcid.org/0000-0002-6466-117X}{\includegraphics[scale=0.5]{orcid.png}}
M. Szymczak,$^{2}$\href{https://orcid.org/0000-0002-1482-8189}{\includegraphics[scale=0.5]{orcid.png}}
J. Šteinbergs,$^{1}$\href{https://orcid.org/0000-0001-9265-3655}{\includegraphics[scale=0.5]{orcid.png}}
G. Surcis,$^{3}$\href{https://orcid.org/0000-0003-2775-442X}{\includegraphics[scale=0.5]{orcid.png}}
A. Kobak,$^{2}$\href{https://orcid.org/0000-0002-1206-9887}{\includegraphics[scale=0.5]{orcid.png}}
\newauthor
M. Durjasz$^{2}$\href{https://orcid.org/0000-0001-7952-0305}{\includegraphics[scale=0.5]{orcid.png}}
and I. Shmeld$^{1}$\href{https://orcid.org/0000-0002-9671-5267}{\includegraphics[scale=0.5]{orcid.png}}
\\
$^{1}$Engineering Research Institute "Ventspils International Radio Astronomy Center", Ventspils University of Applied Sciences, Inzenieru Str. 101, Ventspils, LV-3601, Latvia\\
$^{2}$Institute of Astronomy, Faculty of Physics, Astronomy and Informatics, Nicolaus Copernicus University, Grudziadzka 5, 87-100 Torun, Poland\\
$^{3}$INAF -- Osservatorio Astronomico di Cagliari, Via della Scienza 5, I-09047 Selargius, Italy
}

\date{Accepted XXX. Received YYY; in original form ZZZ}

\pubyear{2023}

\begin{document}
\label{firstpage}
\pagerange{\pageref{firstpage}--\pageref{lastpage}}
\maketitle

\begin{abstract}
{The variability study of 6.7\,GHz methanol masers has become a useful way to improve our understanding of the physical conditions in high-mass star-forming regions.}
{Based on the single-dish monitoring using the Irbene telescopes, we selected three sources with close sky positions.}
{We imaged them using the European Very Long Baseline Interferometer Network and searched available data on VLBI archives to follow detailed changes in their structures and single maser spot variability.}
{All three targets show a few groups of maser cloudlets of a typical size of 3.5\,mas and the majority of them show linear or arched structures with velocity gradients of order 0.22\kms\,mas$^{-1}$. The cloudlets and overall source morphologies are remarkably stable on time scales of 7-15\,yr supporting a scenario of variability due to changes in the maser pumping rate.}
\end{abstract}


%%%%%%%%%%%%%%%%%%%%%%%%%%%%%%%%%%%%%%%%%%%%%%%%%%

%%%%%%%%%%%%%%%%% BODY OF PAPER %%%%%%%%%%%%%%%%%%

\section{Introduction}
The formation of high-mass stars remains an important topic in modern astrophysics, as it is still not yet clear how a star obtains its final mass. It is believed that an embedded protostar can acquire the mass required to evolve into a massive star by either a global collapse or competitive accretion (e.g. \citealt{Zinnecker2007}, for review). High-mass star-forming regions (HMSFRs) are challenging for observational study due to their large distance, obscuration by dust clouds and fast evolution. Our ability to make significant progress in this field has been reinforced by the observations of methanol masers, especially the 6.7 GHz transition, whose spectral features are very bright and compact. The 6.7\,GHz maser emission is primarily a unique signpost of HMSF sites (e.g. \citealt{menten1991}) and powerful tools to determine the trigonometric parallaxes (e.g. \citealt{rygl2010}; \citealt{reid2019}). High angular resolution observations allow us to measure the size and orientation of accretion disks, their kinematics and structure (e.g. \citealt{sanna2010a,sanna2010b,sanna2017}; \citealt{moscadelli2011}; \citealt{sugiyama2014}). Long-term monitoring of the 6.7\,GHz methanol masers has turned out to be an unexpectedly powerful tool for identifying accretion outbursts (\citealt{fujisawa2015}; \citealt{Szymczak2018}; \citealt{macleod2018}; \citealt{burns2020}).

A few detailed studies have been performed with the very-long baseline interferometry (VLBI) technique. Imaging maser emission at a milliarcsecond (mas) scale shows a variety of structures.
The maser emission is frequently located along lines or arcs, sometimes with velocity gradients indicating the presence of edge-on discs (\citealt{norris1993}; \citealt{minier2000}; \citealt{dodson2004}). Elliptical or ring-like morphologies observed in several sources (\citealt{bartkiewicz2005, bartkiewicz2009, bartkiewicz2014, bartkiewicz2016}; \citealt{fujisawa2014a}) can arise from inclined or face-on rotating discs. \cite{bartkiewicz2016} reported that almost half had a complex structure in a sample of 63 objects. A similar result was achieved by \cite{fujisawa2014a}.

Multi-epoch observations allow us to estimate the proper motions of single maser clouds and provide valuable information. 
For example, in G16.59$-$0.05 the bipolar distribution of 6.7~GHz maser line-of-sight velocities was associated with a rotating disk or toroid around a central mass of about 35~\(\textup{M}_\odot\) \citep{sanna2010a}. The velocity field of methanol masers can be explained in terms of a composition of slow (4\kms) motion of radial expansion and rotation about an axis approximately parallel to the jet in the vicinity of massive young stars (e.g., \citealt{sanna2010b}, \citealt{moscadelli2011}).  

Long-term single-dish maser studies have found two different kinds of the variability of the 6.7~GHz methanol maser line: flaring and gradual (including periodic) changes of flux densities (e.g., \citealt{goedhart2004}; \citealt{Szymczak2018}). Known variability periods range from 24 to 600 days and are seen in some or all features. Additionally phase-lags between individual features are also noticed (\citealt{goedhart2004}; \citealt{goedhart2014b}; \citealt{fujisawa2014b} and \citealt{g107_period}). What causes the variability of 6.7~GHz methanol masers is still a topical field of research. 
\cite{caswell1995} suggested that the maser variations are due to changes in the gain path length caused by large-scale motions. A strong correlation of maser and infrared flux densities was observed in the periodic 6.7\,GHz sources (e.g. \citealt{olech2020, olech2022}; \citealt{kobak2023}) indicates that the pumping rate plays a dominant role. Extraordinarily outbursts of the methanol masers in S255IR$-$NIRS3 and G358.93$-$0.03 are powered by accretion bursts (\citealt{caratti2017}; \citealt{stecklum2021}).

A combination of single-dish monitoring and detailed morphology analyses employing the VLBI technique mounts a convincing argument that most of the observed variability is caused by changes in maser pumping rates, furthermore implying the importance of infrared variability (\citealt{szymczak2014}; \citealt{olech2019} and \citealt{durjasz2019} ).

The Irbene programme of 6.7\ GHz methanol maser monitoring, which started in 2016, includes 41 targets from the Torun methanol maser catalogue \citep{Szymczak2012}. The sources were selected based on their declination above $-$10\degr ~and their maser flux density ($>$5\,Jy). The main aim was to study the variability of 6.7\,GHz methanol masers with particular attention to bursting episodes. Observation intervals were targeted every five days for all sources and daily for some exciting sources during their most active periods \citep{aberfelds2018}. This paper presents results from the European VLBI Network\footnote{The European VLBI Network is a joint facility of independent European, African, Asian, and North American radio astronomy institutes. Scientific results from data presented in this publication are derived from the following EVN project code: EA063} in order to: i) map the milliarcsecond structure of the emission of three sources with different variability behaviour, ii) identify maser groups responsible for the flux density changes seen in the single-dish spectra, and iii) put constraints on variability mechanisms going on around these high-mass young stars.
Preliminary result for the source G78.122+3.633 was reported in \cite{aberfelds2021}. 

\section{Targets}
Three 6.7\,GHz methanol maser sources from the Irbene monitoring list were selected for VLBI observations as showing noticeable variability in 2018-2019. 
They were characterised by steady rise/decline of some features, moderate flares and fluctuations on timescales of a few months and years. The objects for VLBI studies are: G78.122+3.633, G90.92+1.49 and G94.602$-$1.796 (hereafter G78, G90 and G94 abbreviations are used accordingly). Below, we present some of their properties relevant in the context of the paper.

{\bf G78} (also known as IRAS 20126$+$4104) is a well studied early B0.5 spectral class star with Keplerian disk around $\sim7 \textup{M}_\odot$ protostar \citep{cesaroni1997}. \citet{moscadelli2011} reported water maser jet structure based on 22~GHz VLBA data and two groups of methanol masers based on the 6.7~GHz EVN observations. Two methanol maser groups appeared at very similar and overlapping LSR velocity ranges, from $-$4.5 to 8.5~km~s$^{-1}$. The blue-shifted one is likely associated with a rotating, narrow ring related to the circumstellar disk, while the second is tracing the disk material marginally entrained by the jet. The trigonometric parallax were measured using the water masers by \citet{moscadelli2011} and \citet{Nagayama2015} providing a mean value of 0.645$\pm$0.030~mas, implying 1.64$^{+0.30}_{-0.12}$\,kpc) \citep{reid2019}. JVLA images show the 6.7~GHz methanol emission in the LSR velocity range of $-$8.4 to $-$4.7~km~s$^{-1}$, distributed in over $\sim0\farcs36\times0\farcs25$ area without any regularities \citep{Hu2016}. 

{\bf G90}. The methanol maser in this source was discovered by \citet{Szymczak_2000}. In the BeSSeL\footnote{http://bessel.vlbi-astrometry.org} survey the parallax was estimated to be 0.171$\pm$0.031\,mas based on the methanol 6.7~GHz line measurements implying the distance of 5.85$^{+1.30}_{-0.90}$~kpc \citep{reid2019}.  
Images from the JVLA show emission distributed over $\sim0\farcs3\times0\farcs6$ area at the LSR velocity of $-$71.5 to $-$68.3\kms with a velocity gradient from S-W to N-E \citep{Hu2016}. They also found a 0.63\,Jy continuum emission about 0\farcs15 S-E of the maser, whereas BeSSeL detected only the brightest part of arched maser distribution of $\sim$130\,mas in size.

{\bf G94} (also known as V645 Cyg or AFGL2789) is an young stellar object (YSO) containing O7 spectral type star \citep{cohen1977} with variable gas outflows \citep{Clarke_2006}. A mean trigonometric parallax of 0.221$\pm$0.013~mas estimated based on the water and methanol masers (\citealt{Oh2010}; \citealt{Choi_2014}; \citealt{Sakai2019}) indicates a distance of 4.5$^{+0.3}_{-0.2}$~kpc \citep{reid2019}. \cite{Slysh2002} observed this source using EVN with 5 antennas in 1998 and 2000 at the 6.7~GHz maser transition. Their images show four groups of masers with an indication of a rotating disc, as it was noted by the authors. JVLA images show emission in LSR velocity range from $-$44.1 to $-$40.4~km~s$^{-1}$, distributed over $\sim0\farcs10\times0\farcs22$ area with a velocity gradient from N to S of the major emission \citep{Hu2016}. They also found a weak radio-continuum emission with an integrated flux of 0.42~Jy.

\section{Observations}\label{sec:obser}
We conducted single-epoch observations  of the methanol maser transition $5_1$-$6_0$ $A^+$ (rest frequency of 6668.51920\,MHz) using the EVN towards three targets: G78, G90 and G94. Observations were carried out on 31 October 2019, and the basic observing parameters are summarized in Table\,\ref{details}. The eleven EVN antennas that took part in these observations, which lasted for 10\,h, are Jodrell, Effelsberg, Medicina, Onsala, Torun, Westerbork, Yebes, Sardinia, Irbene and Tianma. The phase$-$referencing technique was used with the cycle time of 105~s$+$225\,s between a phase calibrator and a target. In total, the on-source time on the targets was $\sim$139, 105 and 98~min, respectively. 3C345 was used as a fringe finder and bandpass calibrator.
To increase the signal-to-noise ratio for the phase-reference sources, two correlator passes were used; one using all eight BBCs per polarization, with a bandwidth of 4 MHz each (corresponding to $\sim$200~km~s$^{-1}$), and 128 correlator channels per BBC; the other using only the single BBC per polarization containing the maser line and 2048 correlator channels. The latter had a spectral resolution of 1.95~kHz ($\sim$0.088~km~s$^{-1}$). Both data sets were processed at the Joint Institute for VLBI ERIC (JIVE) with the SFXC correlator \citep{sfxc_2016}. 
The Astronomical Image Processing System (AIPS) was used for data calibration and reduction. The standard procedures for line observations were used and Effelsberg antenna was set as a reference. The synthesised beam of final images was typically 4~mas~$\times$~3~mas as listed in Table~\ref{details}. We created 1024~px$\times$1024~px image cubes  with a pixel size of 1~mas, so we searched the region of 1''$\times$1'' for the emission. To measure the spot parameters, we used the AIPS task JMFIT which implements a 2D Gaussian fitting procedure. The spectra were extracted from the image cubes using the AIPS task ISPEC. 

Moreover, the targets were observed from March 2017 to September 2022 as part of the maser monitoring project with the Irbene 32-m and 16-m radio telescopes in North$-$West Latvia. The vast majority ($\sim$96\%) of presented data were obtained with the 16-m instrument. The observations were done at irregular time intervals of one day to one month with an average of 2.3, 6.5 and 5.2\,d for G78, G90 and G94, respectively. The basic parameters of the instruments at 6.7\,GHz transition are listed in Table\,\ref{tab:single-dish}. The data were reduced with MDPS package (\citealt{steinbergs2021}).

\begin{table*}
\caption{Parameters of EVN observations. The coordinates are given for the brightest spot.}
\centering
\begin{tabular}{l c c c c c} 
\hline
Name & RA & Dec & Phase calibrator & Synthesized beam &  RMS noise\\
         & (h : m : s)&($\degr$ : $\arcmin$ : $\arcsec$)&    & (mas~$\times$~mas; $\degr$)  & (mJy\,beam$^{-1}$)    \\
\hline
G78.122$+$3.633 & 20:14:26.04441  & 41:13:32.6295  &  J2007+4029     & 4.4 $\times$ 2.9; $-$67 & 3.8 \\
G90.925$+$1.486 & 21:09:12.97472  & 50:01:03.6578  &  J2114+4953     & 4.0 $\times$ 2.8; $-$43 & 3.4 \\
G94.602$-$1.796 & 21:39:58.25505  & 50:14:20.9982  &  J2137+5101     & 3.7 $\times$ 2.8; $-$40 & 4.6 \\
\hline
\end{tabular}\\
\label{details}
\end{table*}

\begin{table}
\caption{Observing parameters with the Irbene 32-m and 16-m radio telescopes.}
\centering
\begin{tabular}{l c c c} 
\hline
Parameter & 32-m & 16-m & Unit\\
\hline
System temperature          & 28 -- 32 & 32 -- 36 & K \\
Half power beam-width       & 6  & 12 & arcmin \\
Number of spectral channels & 4096 & 4096 & \\
Correlator bandwidth        & 1.5625 & 1.5625 & MHz \\
Velocity range              & 35     & 35 & km~s$^{-1}$ \\
Correlator resolution       & 0.017  & 0.017 & km~s$^{-1}$ \\
RMS noise                   & 0.9    & 1.9   & Jy \\
Calibration accuracy        & $-$ & $\sim$20 & \% \\
\hline
\end{tabular}\\
\label{tab:single-dish}
\end{table}

\begin{figure*}
\centering
\includegraphics[width=0.8\paperwidth, height=0.32\textheight]{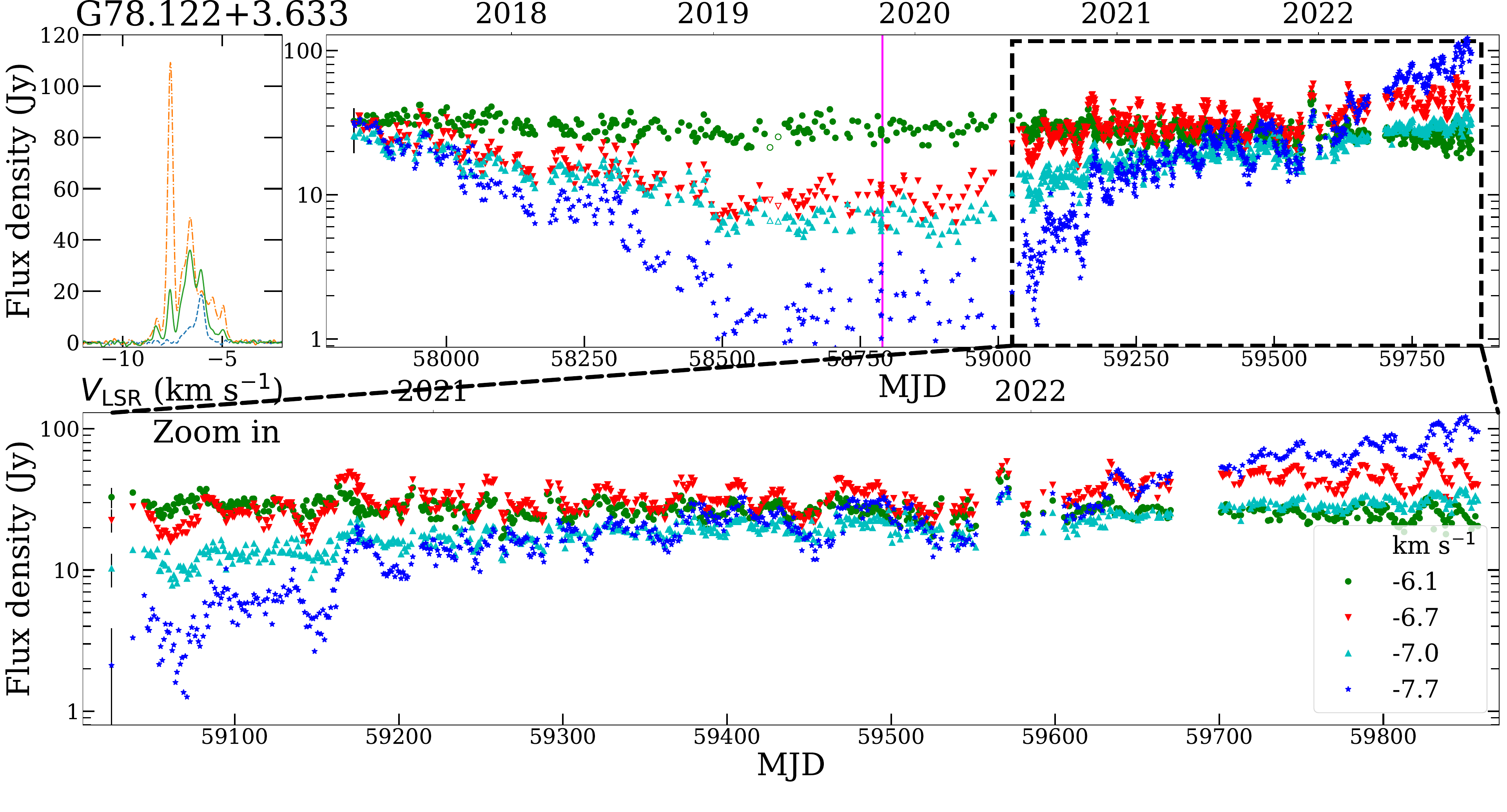}
\includegraphics[width=0.8\paperwidth, height=0.18\textheight]{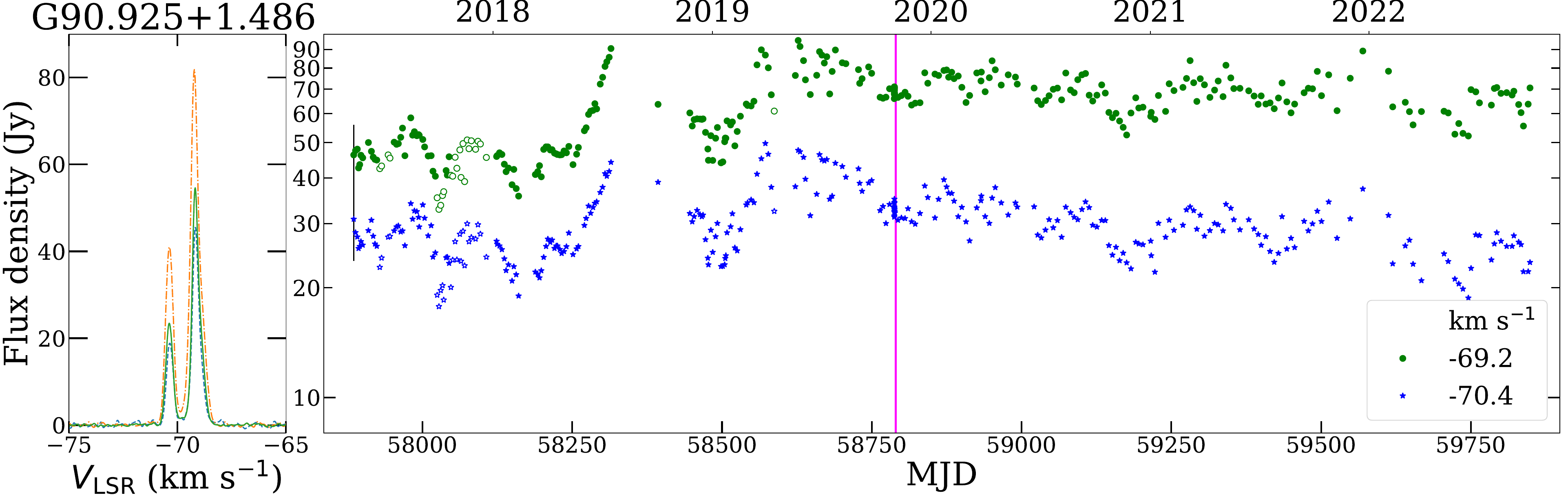}
\includegraphics[width=0.8\paperwidth, height=0.18\textheight]{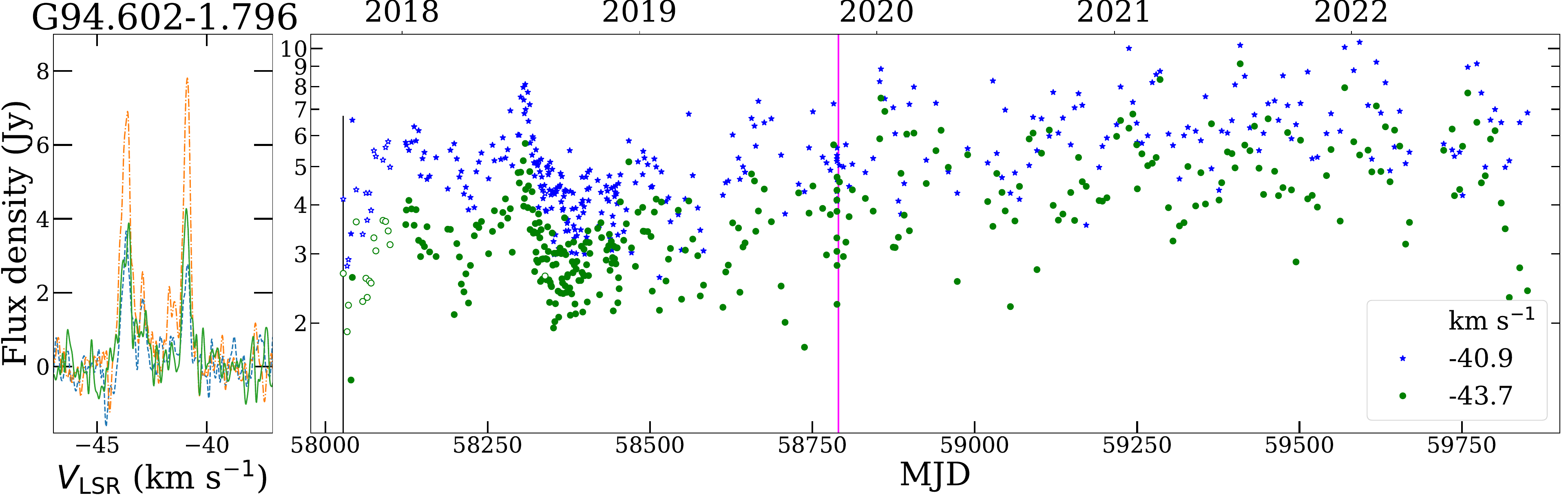}
\caption{Spectra and light curves of the 6.7\,GHz maser emission in the studied objects. Left-hand panels: maximum (orange), minimum (blue) and average (green) spectra. Right-hand panels: light curves of main features. The filled and empty symbols represent the data from the 16-m and 32-m telescopes, respectively. Typical measurement uncertainty is shown by the bar for the first data point. The vertical magenta line denotes the date of EVN observations. ashed rectangle for G78 time series denotes zoomed-in interval  showing light curve during the high-cadence observations, plotted below. }
\label{three_target_var}  
\end{figure*}

\begin{table}
\caption{Variability properties of selected features.}
\centering
\begin{tabular}{l c c c c c} 
\hline
$V_\mathrm{p}$ (km\,s$^{-1}$)  & $S_\mathrm{p}$ (Jy) & $VI$ & $FI$ & $\chi_\mathrm{r}^{2}$ & $\chi_{99.9\%}^{2}$\\
\hline
\multicolumn{4}{l}{G78.122$+$3.633 (Time-span = 2023\,d, N=880)}  & &1.15  \\
-4.9 & 10.4 & 0.76 & 0.71 & 9.77 \\
-5.4 & 7.0  & 0.81 & 0.68 & 8.25 \\
-6.1 & 27.7 & 0.24 & 0.20 & 0.61 \\
-6.7 & 29.1 & 0.74 & 0.67 & 11.66 \\
-7.0 & 18.4 & 0.62 & 0.64 & 4.87 \\
-7.7 & 24.1 & 0.97 & 2.48 & 65.51 \\
-8.3 & 9.7  & 0.82 & 0.74 & 8.76 \\
\multicolumn{4}{l}{G90.925$+$1.486 (Time-span = 1823\,d, N=305)}  & & 1.27 \\
-69.2 & 61.49 & 0.36 & 0.44 & 3.99 \\
-70.4 & 29.79 & 0.37 & 0.39 & 1.6 \\
\multicolumn{4}{l}{G94.602$-$1.796 (Time-span = 1963\,d, N=351)} &  &1.25 \\
-40.9 & 5.30 & 0.14 & 0.45 & 0.68 \\
-43.7 & 3.74 & 0.43 & 0.47 & 0.59 \\
\hline
\end{tabular}\\
\label{tab:monitoring}
\end{table}

\section{Results} 
\subsection{Variability}
Figure\,\ref{three_target_var} shows the 6.7\,GHz maser spectra and the light curves of individual spectral features; in the case of similar intensity and variability patterns, only one light curve is shown.  Table\,\ref{tab:monitoring} summarizes the results of the statistical variability analysis of light curves based on the average flux density of three spectral channels centred at the peak velocity of each feature. There are the variability index, $VI$, which is a degree of the amplitude variations, fluctuation index, $FI$, measuring the spread about the mean flux density and $\chi^{2}$-test giving a probability for the presence of variability. More details on the definitions of these parameters are given in Appendix\,\ref{appendix-A}.

All the features in G78, except for the $-$6.1\,\kms feature, are moderately or highly variable. The feature at $-$7.7\,\kms has the highest variability indices (Table\,\ref{tab:monitoring}). It initially had a flux density of 32\,Jy but decayed to $\sim$1.5\,Jy after 1.6\,yr and remained at a noise level over $\sim$1.5\,yr, then increased to 105\,Jy at the end of our monitoring. The other variable features essentially show a similar pattern of variability. The $-$6.1\,\kms\, feature decreased by about 25\,per\,cent on a timescale of 5.5\,yr, which is marginally above the measurement accuracy. The high cadence observations after MJD 59045 revealed, for the variable features, the occurrence of flux fluctuations by a factor of 0.7$-$3.5 on a timescale range of 9$-$40\,d imposed on the overall growth of flux on a timescale of 2.2\,yr. We conclude that the fluctuations are real because their amplitudes are well above the measurement uncertainty. The Lomb-Scargle periodogram analysis \citep{scargle1982} showed no statistically significant periodicity. These relatively rapid fluctuations are simultaneous within 3$-$4\,d.

The two main features in G90 are moderately variable (Table\,\ref{tab:monitoring}). A likely flare occurred around MJD 58250 when the flux density increased by a factor of two during 58\,d, but the flare profile could not be determined due to a gap in observations. One can only notice that the flux density has returned to a pre-flare level after 7.5\,months. Since MJD $\sim$58570\, the emission shows an overall decrease by a factor of 1.5 during $\sim$3\,yr. 
There are similar changes in the intensity of the two features on timescales 3$-$11\,months, but their amplitudes are within or below uncertainty in the absolute flux density calibration. 

The emission of G94 is faint and not significantly variable within the noise.

\subsection{Maser structure}

\begin{figure*}
\centering
\begin{minipage}{0.31\textwidth}
\includegraphics[scale=0.48]{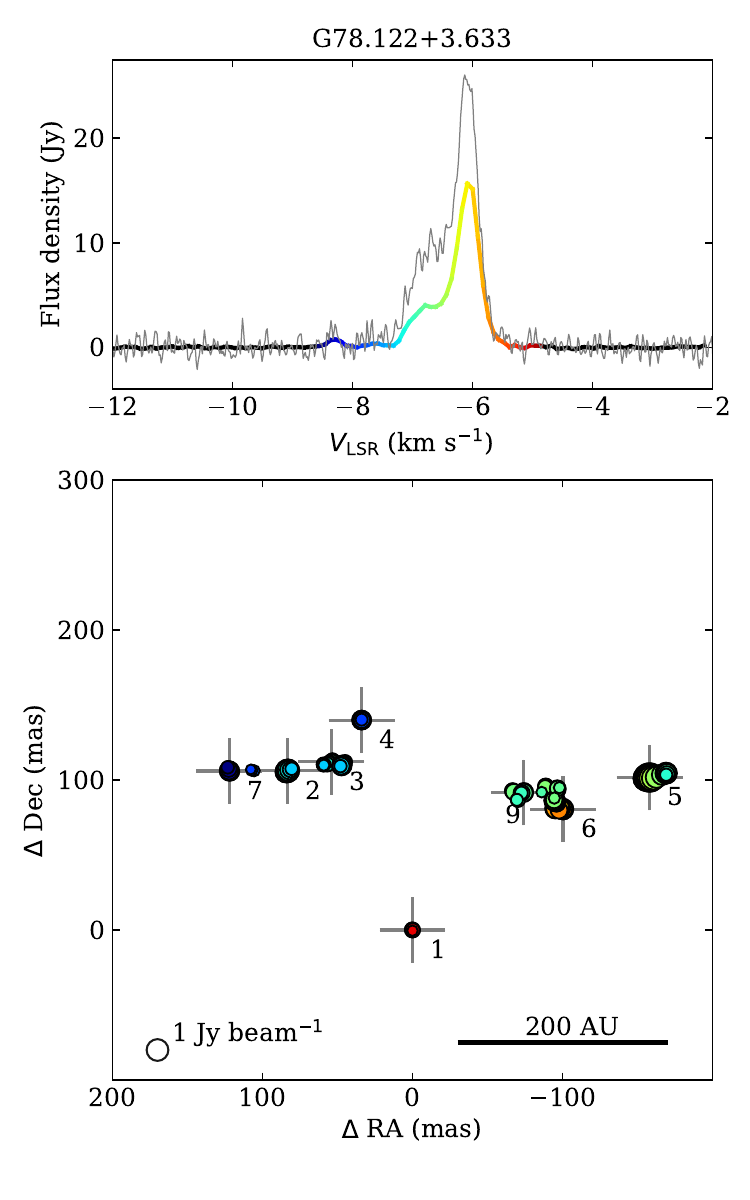}  
\end{minipage}\hfill
\begin{minipage}{0.31\textwidth}
\includegraphics[scale=0.48]{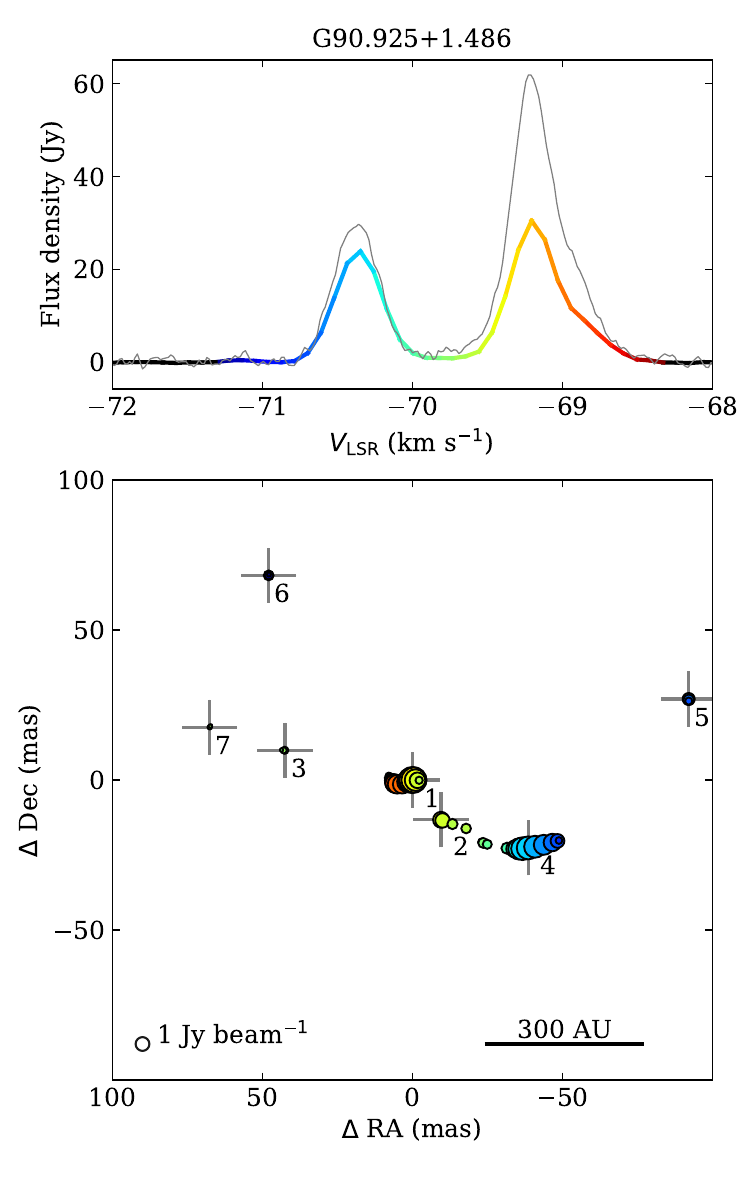} 
\end{minipage}\hfill
\begin{minipage}{0.31\textwidth}
\includegraphics[scale=0.48]{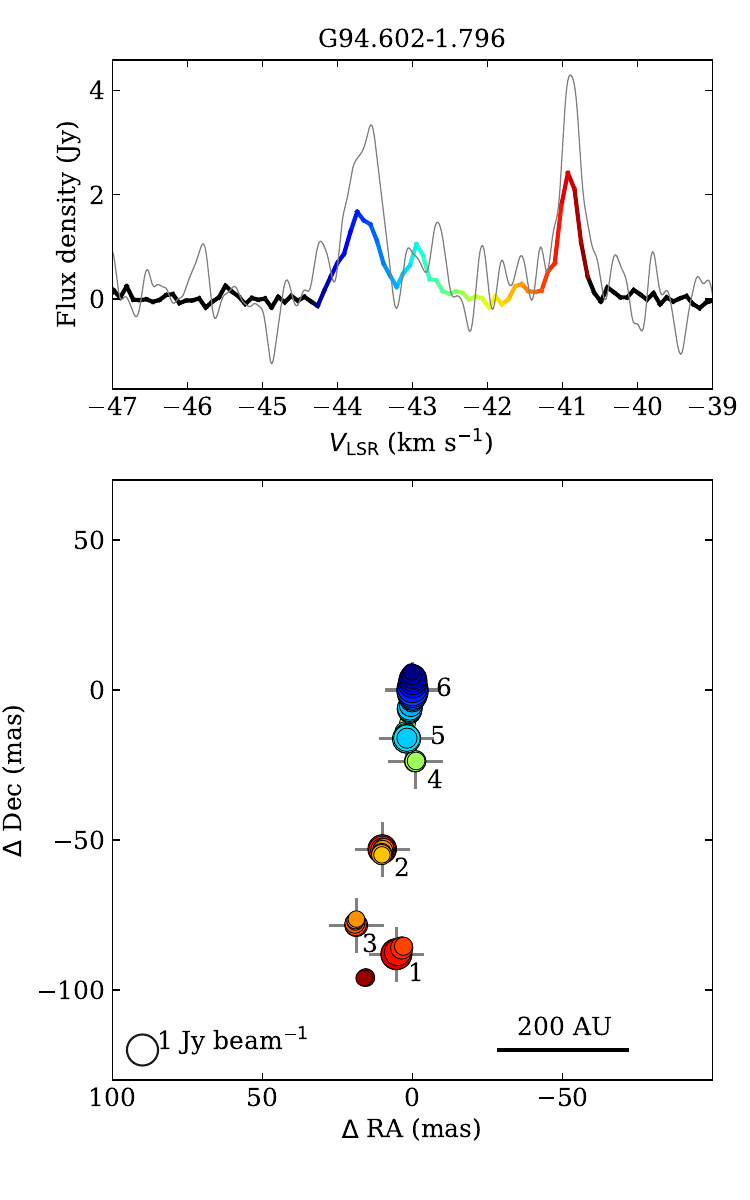} 
\end{minipage}\hfill
\caption{Top: Maser spectra of G78, G90 and G94 as obtained by EVN (colour) and the Irbene 16m radio telescope (grey line) on 31 October 2019. Bottom: Distribution of methanol maser spots. The colours are related to the LSR velocity as in the spectrum. The spot size is proportional to the logarithm of its brightness. 
The cloudlets (crosses) are numbered as listed in Tables\,\ref{g78_table}, \ref{g90_table} and \ref{g94_table}. 
}
\label{spots-maps}
\end{figure*}

The total intensity (0-th moment) maps of cloudlets for G78 are shown in Fig.\,\ref{fig:zerothmomnt_G78}, while those of the G90 and G94 sources are displayed in Figs.\,\ref{fig:g90momnt} and \ref{fig:g94momnt}, respectively. 
The positions of the brightest spot of each target are listed in Table\,\ref{details}. The maps of spot distribution and the spectra are presented in Figure\,\ref{spots-maps}. The parameters of maser cloudlets are given in Tables\,\ref{g78_table}, \ref{g90_table} and \ref{g94_table}. 
In this paper, a maser {\it cloudlet} is defined as a group of maser spots, with a signal-to-noise ratio higher than 10, in at least three adjacent spectral channels which coincide in position within half of the synthesized beam \citep{Bartkiewicz_2020}. 
By fitting the Gaussian function to the spectral shape of the emission of individual cloudlet we obtained the amplitude ($S_\mathrm{fit}$), the full width at half maximum (FWHM) and the peak velocity ($V_\mathrm{fit}$). The projected linear size of cloudlet ($L_\mathrm{proj}$) is estimated as the distance between two of the furthest spots. The velocity gradient ($V_\mathrm{grad}$) is estimated when a regular increase or decrease in velocity occurs, and it is defined as a maximum difference in the velocity between spots divided by their distance. Similar to \citet{moscadelli2011}, we neglect the gradient sign for analysis. The directions of $V_\mathrm{grad}$ are given by the position angle (PA) of the cloudlet major axis measured from the blue- to red-shifted wing taken as positive if increasing to the east. Note we used flux-weighted fits for PA estimation.

{\bf G78}.
98 maser spots with a velocity ranging from $-$8.37 to $-$4.77\kms are detected in 14 cloudlets, forming three clusters, which fall within a region of $\sim$290~mas$\times$150\,mas (480~au$\times$250\,au) (Fig.\,\ref{spots-maps}, Table\,\ref{g78_table}). 
The overall maser distribution is very similar to that observed 8-15\,yr ago (\citealt{moscadelli2011,surcis_2014}).

\begin{figure}
\centering
\includegraphics[scale=1.2]{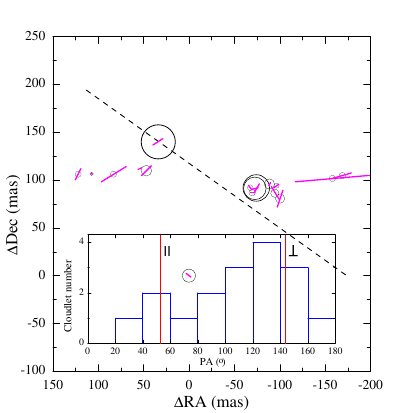}
\caption{Map of velocity gradient directions (magenta bars) for the  maser cloudlets in G78 from the 2019 EVN data. The length of the bars is proportional to the square of brightness, and the circle size is proportional to the velocity gradient. The diagonal black dashed line marks the direction of the disc plane \citep{cesaroni2005}. The inset shows the position angle (PA) histogram of the velocity gradient; the two vertical red lines mark directions parallel and orthogonal to the plane of the disc.}
\label{fig:g78-PA}  
\end{figure}

Most cloudlets usually show the linear or arched distribution of spots with an internal gradient of velocity with the position. Cloudlets 5 and 6 are complex with double Gaussian profiles and curved velocity gradients (see Figs.\,\ref{Cloudlet_5} and \ref{Cloudlet_6}). In further analysis, their components are considered separately. The angular extent of cloudlets ranges from 0.5 to 7\,mas, and the average value corresponds to $L_\mathrm{proj}$ = 4.2$\pm$0.3\,au. The velocity gradient ranges from 0.05 to 0.45\,kms\,au$^{-1}$ and the average value is 0.14$\pm$0.03\,kms\,au$^{-1}$. The mean FWHM is 0.31$\pm$0.02\kms. We were searching for relationships between cloudlet parameters but found a moderate correlation only between $L_\mathrm{proj}$ and $S_\mathrm{fit}$ (r = 0.596, p$<$0.015). Comparison of the EVN and Irbene spectra indicates that $\sim$57\,per\,cent of the emission is resolved out. 

Figure\,\ref{fig:g78-PA} shows the map of cloudlets with the velocity gradient directions. Here, the negative PA values of velocity gradient (Table\,\ref{g78_table}) are folded into the range 0$-$180\degr. The cloudlets with a gradient amplitude greater than 0.14\kms\,au$^{-1}$ tend to lay in the central part of the maser distribution. They have directions approximately orthogonal to the major disc axis, but the overall distribution of PA does not follow this trend (Fig.\,\ref{fig:g78-PA}), especially the cloudlets with low $V_\mathrm{grad}$ have a wide range of PA and might reflect complex kinematics of gas on a scale of $5-20$\,au.
For the most western emission (Cloudlet 5) the direction of the velocity gradient is consistent with the proper motion vectors reported in \citet{moscadelli2011mas} and may mark an outflowing gas. 

{\bf G90}. The map of 47 maser spots detected in the velocity range from $-$71.3 to $-$68.3\,\kms is shown in Figure\,\ref{spots-maps} and the cloudlet parameters are listed in Table\,\ref{g90_table}. The dominant emission is located in two clusters about 300\,au apart, the first being a blend of Cloudlets 1 and 2 and the second with Cloudlet 4. The other four cloudlets are weak and spread over a larger area; the overall size of maser distribution is $\sim$160~mas$\times$90\,mas  corresponding to $\sim$950~au$\times$530\,au. Most cloudlets show linear morphology (Fig.\,\ref{G90_cloudlets}), except for Cloudlet 1, which is slightly twisted. The double-peaked profiles are for Cloudlets 1 and 2; two faint cloudlets (3 and 7) appear as incomplete Gaussians. The projected sizes of cloudlets range from 6 to 100\,au, and the mean velocity gradient is 0.035$\pm$0.006\kms\,au$^{-1}$, the mean FWHM is 0.30$\pm$0.02\kms. The direction of the velocity gradient of the core cloudlets at PA of 62$-$100\degr\, generally follows the SW-NE elongation of maser spot distribution. The top panel of Fig.\,\ref{spots-maps} shows that the emission peaked at $-$70.3 and $-$69.2\kms is resolved out by 19 and 63\,per\,cent, respectively, while on average 57\,per\,cent of the flux density is missed. 
 
{\bf G94}.
We found 59 maser spots that formed 6 cloudlets in the LSR velocity range from $-$40.5 to $-$44.4\kms (Table~\ref{g94_table}). The emission is distributed over $\sim$33~mas$\times$115\,mas corresponding to $\sim$150~au$\times$520\,au (Fig.~\ref{spots-maps}). There is a clear major axis of the emission structure at PA = $-$10\degr\, with a velocity gradient of 0.035\kms\,mas$^{-1}$ corresponding to 0.008\,kms\,au$^{-1}$. Interestingly, the velocity gradient in the NS direction is seen in all cloudlets (Fig.\,\ref{G94_cloudlets}). The cloudlet parameters are basically similar to those in the two above sources. The ratio of EVN to Irbene velocity-integrated flux density is 0.42.

\section{Analysis and discussion}\label{discussion}
\subsection{Variability of G78}\label{sec:G78var}
The present study increases the number of 6.7\,GHz maser maps to five acquired over 15\,yr, allowing for a more detailed analysis of cloudlets variability. 
Previous EVN observations carried out in 2004, 2007, 2009 and 2011 were reported in \citet{Thorstensson_thesis2011}, \citet{moscadelli2011}, and \citet{surcis_2014}. We retrieved the archival data for the EL032, EM064C, EM064D, and ES066E projects\footnote{http://archive.jive.nl} and processed them according to the method described in Section\,\ref{sec:obser}. All the projects have the same observational setup, with the exception of a spectral resolution of 0.044~km~s$^{-1}$ in the 2011 experiment. 

The overall maser spot distributions in the five epochs are presented in Fig.~\ref{total}. Seven cloudlets persisted over 15\,yr and their parameters, obtained following the procedure applied by \citet{moscadelli2011mas}, are given in Table~\ref{5_epoch_cloudlet_table} and labelled in the last panel of Fig.~\ref{total}. In addition to the parameters used in Table\,\ref{g78_table}, we calculated the correlation coefficients of the linear fit to the spot locations on the sky plane, $r_\mathrm{s}$, and the velocity changes with location, $r_\mathrm{v}$, measured along the major axis of the spot distribution. Figs.\,\ref{Cloudlet_1}-\ref{Cloudlet_7} show seven persistent cloudlets' structures and spectra representing a wide range of morphology and intensity variations. Cloudlet 1 (Fig.~\ref{Cloudlet_1}) is the most compact ($\sim$1\,mas) in all five epochs, and consequently, some of its parameters ($r_\mathrm{s}$, $r_\mathrm{v}$, $V_\mathrm{grad}$ and PA) are poorly determined. In Cloudlet 2 the brightest spots form a linear structure with high values of $r_\mathrm{s}$, $r_\mathrm{v}$, and steady ($\pm$10\degr) PA (Fig.~\ref{Cloudlet_2}). Cloudlet 3 (Fig.~\ref{Cloudlet_3}) experienced significant changes in morphology and intensity from compactly distributed 3 spots in 2004 to the complex structure and double profiles since 2009. Cloudlet 4 (Fig.~\ref{Cloudlet_4}) is compact, and its parameters are uncertain similar to Cloudlet 1. The high-velocity spots of Cloudlet 5 (Fig.~\ref{Cloudlet_5}) form an uniquely stable curved distribution over 15\,yr. The low-velocity emission of Cloudlet 5 that appeared in 2009 preserved a stable linear morphology in the two consecutive observations. This cloudlet is the nearest one to the outflow traced by the 22\,GHz water masers \citep{moscadelli2011}, and the PA of the velocity gradient direction is nearly the same as that of the outflow direction. We note that variations of intensity of Cloudlet 5 significantly deviate from a general trend observed for all the rest cloudlets for which a maximum occurred in the 2009 epoch. This suggests that the behaviour of this cloudlet is less dependent on the activity of powering HMYSO. Cloudlet 6 has a complex and twisted morphology on $\sim$15\,mas scale with double-peaked profile (Fig.~\ref{Cloudlet_6}) and shows strong variability. Comparing maps from 2011 to 2019 may suggest a fragmentation of this cloudlet as new closely located cloudlets (9; 11; 13 and 14) appeared. Cloudlet 7 has had an arched structure since 2007 of $\sim$3\,mas in size, evolving towards less curvature. We conclude the morphology and intensity of seven persistent cloudlets show significant variability on $\sim$2-15\,yr timescales.

\begin{figure}
\centering
\includegraphics[scale=1.2]{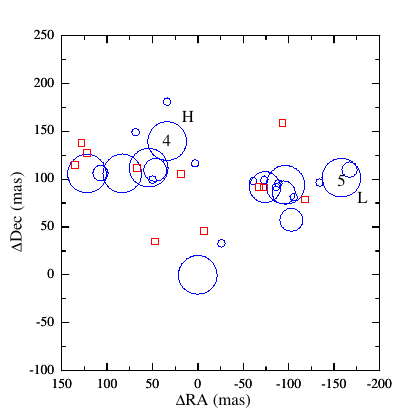}
\caption{Map of maser cloudlets detected with five epoch EVN observations in 2004-2019 interval in G78. The symbol size is proportional to the number of epochs in which the cloudlet appeared. The blue circles and red squares mark the cloudlets with and without Gaussian profiles, respectively. Cloudlets 4 and 5 (Table\,\ref{g78_table})
identified with the spectral features showing the highest and lowest variability evaluated from $\chi^2_\mathrm{r}$ (Table\,\ref{tab:monitoring}) are labelled with H and L, respectively.}
\label{fig:g78-5epoch-var}  
\end{figure}

In addition to the seven persistent cloudlets, there were 28 transient cloudlets in this, as many as 22 only in one epoch. Eleven of these short-lived cloudlets had no Gaussian profile, which may be because the VLBI telescope only detects the most compact parts of low-brightness masing gas. Figure\,\ref{fig:g78-5epoch-var} presents the distribution of the cloudlets where the symbol size denotes the number of epochs in which the emission was detected. It is striking that short-lived cloudlets lie to the north of or in the three main regions of persistent emission but not to the south. Cloudlets 4 and 5 have dominant contributions to the most and least variable spectral features, respectively, and are marked in Fig.\,\ref{fig:g78-5epoch-var} with the letters H and L. The 6.7\,GHz maser emission regions partly overlap with the CH$_3$CN (J = 12$-$11, K=3) emission zones for which the excitation temperature is 133\,K, while south of these zones, higher excitation CH$_3$CN lines occur (\citealt{cesaroni2014}, their Fig.\,5).  We, therefore, conclude that the transient emission avoidance zone is a simple consequence of the increase in the kinetic temperature, which is consistent with the standard methanol maser model (\citealt{cragg2002})

Comparison of the auto and cross-correlation spectra shows that 50-60\,per~cent of flux density is missed with the EVN beam, which is similar to that reported for a large sample of masers (\citealt{bartkiewicz2016}). The variability indices of $S_\mathrm{fit}$ for the persistent cloudlets on 15\,yr timescale (Table\,\ref{5_epoch_cloudlet_table}), nevertheless, follow a general trend observed for the spectral features (Table\,\ref{tab:monitoring}) on 5.5\,yr timescale. Cloudlet 5 largely contributes to the emission of the $-$6.1\,\kms feature and is much less variable than Cloudlet 4, mainly forming the $-$7.7\,\kms feature. This essential difference in variability of intensity and structure can be related to the location of Cloudlets 4 and 5 in a Keplerian disc and a disc-jet interface, respectively, as suggested by \citet{moscadelli2011}. 

The results of our 5\,yr monitoring are consistent with those obtained in the 2009-2013 period (\citealt{Szymczak2018}) where the $-$7.7\,\kms feature (Cloudlet 4) is strongly variable while that at $-$6.1\,\kms (Cloudlet 5) is little variable if any. The 6.7\,GHz spectra presented in earlier reports (\citealt{slysh1999, goedhart2004}) indicate that the object is highly variable on a time scale of up to 27\,yr.

\subsection{Relative proper motion in G78}\label{sec:g78_PM}

\begin{figure*}
\centering
\includegraphics[width=\textwidth]{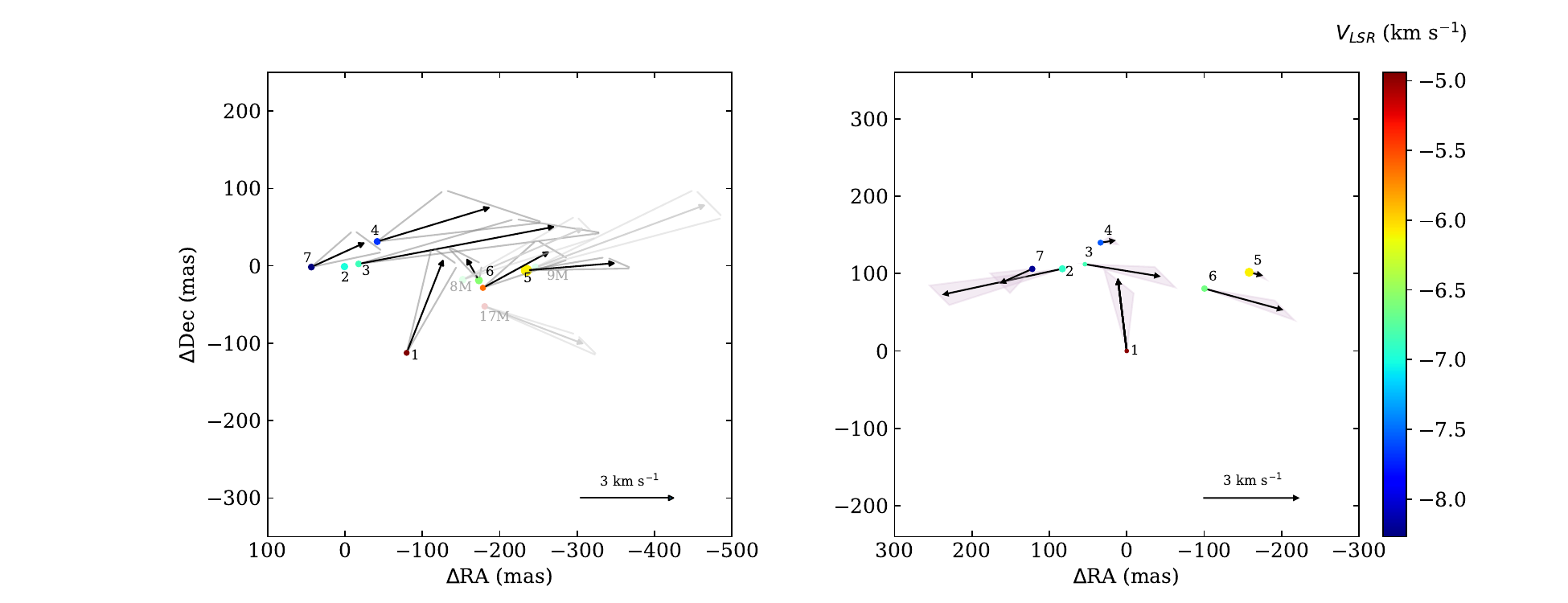} 
\caption{Relative proper motions of 6.7~GHz maser cloudlets in G78. {\bf Left:} Results for the model of the disk-jet interface scenario by \protect\cite{moscadelli2011} over a period of 2004-2019. The numbers correspond to the names of persistent cloudlets as given in Table~\ref{5_epoch_cloudlet_table}. The light-coloured circles and light grey vectors correspond to features 8, 9, and 17 from Table~1 in \protect\cite{moscadelli2011}; they were not detected in 2011 and 2019. The grey triangles correspond to the uncertainties of proper motion vectors. {\bf Right:} Relative proper motions derived from distance changes between cloudlet pairs as described in Section\,\ref{sec:g78_PM}. }
\label{g78_proper_motion}
\end{figure*}

We adopted the same approach as \cite{moscadelli2011} and used Cloudlet 2 (Table~\ref{5_epoch_cloudlet_table}), with the LSR velocity of $-$7.1~km\,s$^{-1}$ as a reference, to study the relative proper motions of methanol maser emission. The procedure was as follows: (1) we identified the cloudlets that were visible in all five epochs, (2) we checked if they showed reliable linear motions relative to Cloudlets 2, (3) finally, we derived the cloudlets' displacements via linear fits over five epochs. The relative proper motions and their uncertainties are presented in Fig.~\ref{g78_proper_motion} left. Our result is quite consistent with Moscadelli et al.'s result obtained from three epoch data spanning 5\,yr. For both measurements, the velocity is in a similar range of 1.5 to 6\kms and PA differences in the direction of proper motion vectors are less than 30\degr. In the eastern cluster, we noticed one more cloudlet showing proper motion, while in the western cluster, three cloudlets reported in \citet{moscadelli2011} disappeared. Our estimate of internal proper motion supports a view that the maser emission from the western part tracks
the gas lifted from the disc near the base of outflow/jet evidenced by the 22\,GHz water masers \citep{moscadelli2011}.

We also derived proper motions using the second method: determining the separation between cloudlet pairs. The procedure was as follows: (1) we measured relative distances between all seven cloudlets (i.e. 21 pairs) in all five epochs assuming that Cloudlet 1 is at (0,0) point in each epoch, (2) using the least squares method, we fitted linear motions for each pair to obtain the relative motions of each cloudlet within this pair (which means six vectors for each cloudlet), (3) we calculated the sum of all vectors for each cloudlet to obtain the proper motion. These results are presented in Fig.~\ref{g78_proper_motion} right. Cloudlet 1 moves towards the remaining cloudlets, similar to the results using the above-mentioned method, while Cloudlets 2 and 7 move away from  Cloudlets 3 and 4 and the western two cloudlets. The cloudlets located at the northern edge of distribution (Cloudlets 4, 5 and 7) have velocities smaller than 1\kms, i.e. significantly lower than the rest of cloudlets lying closer to the deriving star, which is placed $\sim$400\,au to the south \citep{cesaroni2013}. This picture is thus not consistent with the postulated dichotomy in the disc and disc-jet maser cloudlets (\citealt{moscadelli2011mas}); it may reflect the perturbed spatial-kinematic structure of disc clearly seen in the CH$_3$CN line maps \citep{cesaroni2014}.

\subsection{Structure and variability of G90 and G94}
The overall structure of the maser in G90 is similar to that obtained in 2012 during the BeSSeL\footnote{http://bessel.vlbi-astrometry.org/} survey using VLBA. The velocity and position of our Cloudlets 1, 2, 4 and 6 coincide with those given in the BeSSeL dataset. The rest cloudlets were not detected in 2012, likely due to a two times higher angular resolution than ours. The JVLA observation in 2012 \citep{Hu2016} revealed the emission over $\sim$300$\times$600\,mas ($\sim$1700$\times$3500\,au) region indicating the presence of extended emission resolved by the VLBA and EVN. The emission at $-$70.3\kms (mainly Cloudlet 4) preserved its morphology over 7\,yr. For the other cloudlets, we note a difference in PA of velocity gradient direction of less than 40\degr. We conclude the morphology of the source has not changed significantly on 7\,yr timespan.

\begin{figure}
\includegraphics[scale=1.]{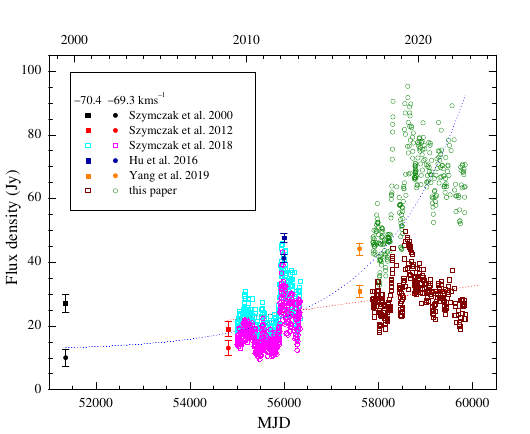}
\caption{Long term variations of the peak flux density of the $-$70.4 and $-$69.3\kms features in G90 marked as squares and circles, respectively. The flux calibration accuracy is $\sim$10\,per\,cent for the MJD 55005-56340 data  and $\sim$20\,per\,cent after MJD 57885. The red and blue dotted lines mark the best fits of linear (r=0.46, p$<$0.001) and exponential (r=0.80, p$<$0.001) curves to the $-$70.3 and $-$69.3\kms time series, respectively.}
\label{fig:g90-lc-long}  
\end{figure}

VLBI data for G90 implies that the emission of variable feature peaked at $-$69.2\kms comes from Cloudlet 1 (the origin of the map in Fig.\,\ref{spots-maps}), whereas the less variable feature at $-$70.4\kms is identified with Cloudlet 4. Our monitoring data indicate that for both features, low amplitude variations (20$-$50\,per\,cent) on timescales of 2$-$3\,months are tightly correlated, and the same is seen in variability on a timescale of 4\,yr. Comparison with the monitoring data in the period 2009-2013 \citep{Szymczak2018} implies the same trends in the variability of the two features. Inspection of the maser spectra available in the literature (\citealt{Szymczak_2000, Szymczak2012, Szymczak2018}; \citealt{Hu2016};\citealt{Yang2019}) and these obtained with the Irbene telescopes reveals that the ratio of peak flux density of features $-$70.4 and $-$69.2\kms monotonically decreases from 2.5 to 0.4 between 1999 and 2022 (Fig.\,\ref{fig:g90-lc-long}). As the 6.7\,GHz maser line is radiatively pumped \citep{cragg2002}, the occurrence of synchronised variations by a factor of up to 3.8 in timescales from $\sim$3\,months to a few years suggests global changes in the infrared photon flux driven by a central HMSYO. Figure\,\ref{fig:g90-lc-long} depicts that the $-$70.4\kms feature rises very slowly (0.75\,Jy\,yr$^{-1}$) in 2008-2022, whereas the $-$69.2\kms emission exponentially rise $\sim$7 times over $\sim$23\,yr. We hypothesise this long-term variability can be caused by changes in the amplification path length due to large-scale motions of the gas (\citealt{caswell1995}) in spiral arms within the accretion disc (\citealt{Bayandina_2019}).

The maser spectrum of G94 in the studied period changed significantly compared to that observed in 2009$-$2013 (\citealt{Szymczak2018}; \citealt{Hu2016}); the emission from $-$43.2 to $-$42.2\kms decreased by a factor of two. The shape of the spectrum is also significantly different from those observed $\sim$20$-$24\,yr ago (\citealt{slysh1999}; \citealt{Szymczak_2000}). We conclude the methanol maser shows appreciable variability on a two-decade timescale; it also displays low amplitude variability on shorter ($>$0.5-2\,yr) timescales (\citealt{Szymczak2018}) that was not seen in the present study likely due to low sensitivity.

G94 was observed with the VLBA (BeSSeL) in December 2012 and 12 spots imaged correspond to our Cloudlets 1, 2, 3, 5 and 6 given in Table 6. As the spectral resolution of VLBA data was 0.36\kms we cannot uncover the cloudlet's morphology but note the same overall linear structure of 430\,au with a pronounced velocity gradient in the N-S direction (PA=170\degr). We have attempted to identify the brightest spots in the 2012 and 2019 VLBI data at the same velocity in order to determine the relative motion. At both epochs, the brightest blue- and red-shifted spots have velocities of $-$42.94 and $-$40.78\kms, respectively. There is no change in the distance between the spots higher than 0.85\,mas. 
Observations with the VLA-C in March 2012 revealed a very similar morphology of $\sim$500\,au in size (\citealt{Hu2016}). The maser structure uncovered in 1998, and 2000 with five EVN telescopes (\citealt{Slysh2002}) of 0.4\,Jy\,beam$^{-1}$ sensitivity was composed of five spots, which distribution is consistent with ours. We conclude the overall kinematic-spatial structure of the maser emission in G94 is preserved over $\sim$21\,yr. The brightness of individual cloudlets likely varies over shorter timescales as suggested by time series of maser features (\citealt{Szymczak2018}). 

\begin{figure}
\centering
\includegraphics[scale=0.5]{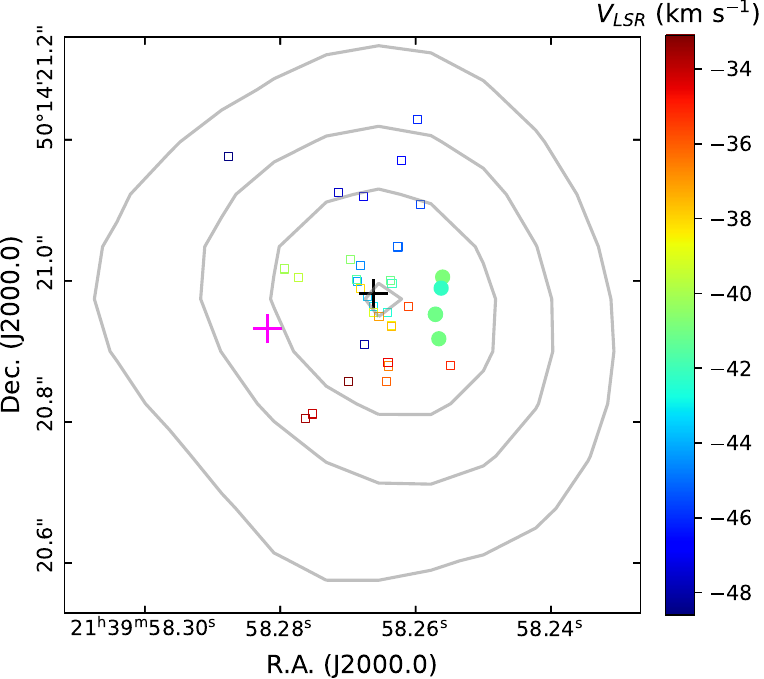}
\caption{Map of the 1.37\,mm continuum in G94 \citep{beuther2018}. The first contour is at 4.0\,mJy\,beam$^{-1}$ with steps increasing by 2.0\,mJy\,beam$^{-1}$. 
It is overlayed with the 6.7\,GHz maser cloudlets (filled circles) and the peaks of the CH$_3$CN 220747.261\,GHz line emission (open squares) retrieved from the NOEOMA data \citep{gieser2021}. The colour indicates the LSR velocity according to the scale shown in the wedge. The black and magenta crosses mark the peak of continuum emission at 5.8 and 44\,GHz, respectively \citep{purser2021}.
}
\label{fig:g94-disc-maser}  
\end{figure}

Near-infrared spectroscopy of G94 implied the fast jet/wind oriented  nearly towards the observer \citep{murakawa2013}. However, detection of a compact 0.23 and 0.52\,mJy continuum emission at 5.8 and 44\,GHz, respectively, with the position angle of the major axis of C-band emission of 91$\pm$71\degr\, can indicate this object as a jet candidate \citep{purser2021}. We note that the 5.8\,GHz emission coincides within 100\,mas with the maser position. The site was mapped
in the 1.37\,mm continuum emission with $\sim$400\,mas beam, and four objects were found \citep{beuther2018}. The strongest source coincides within $\sim$90\,mas with the 6.7\,GHz maser region (Fig.\,\ref{fig:g94-disc-maser}). The NOEMA spectral line data in the 1.3\,mm band \citep{gieser2021} allows us to retrieve the map of the CH$_3$CN line emission at 220747.261\,GHz which is a good tracer of molecular discs (e.g. \citealt{cesaroni2014}). The distribution of centroids of this thermal line emission (Fig.\,\ref{fig:g94-disc-maser}) is aligned in the N-S direction and shows a clear velocity gradient. These are exactly the same characteristics as we observed in the 6.7\,GHz methanol maser line. We conclude the maser emission probes the central parts of the western edge of the disc.

\section{Conclusions}
We have provided analysis of single-dish 4.5\,yr monitoring observations and single epoch EVN data of 6.7\,GHz methanol maser emission in three HMSFRs. For G78 source, we used the archival EVN data to study the maser structure for five epochs spanned $\sim$15\,yr. We conclude the following:  \\
\indent
(i) In G78 the persistent seven maser cloudlets and spectra represent a wide range of variations in intensity and morphology; the majority of them are variable over a few years. 
The cloudlets associated with the region tracing disc are more variable on short ($<$ 2 months) and 2-15\,yr time scales than the cloudlets likely tracing the outer regions of the jet.
The directions of the velocity gradient of cloudlets are largely consistent with the vectors of internal proper motion derived from 5 epoch observations spanned $\sim$15\,yr.
There were 27 transient cloudlets lying to the north or in the persistent cloudlet regions, which may indicate unsaturated amplification. \\
\indent
(ii) The overall structure of maser emission in G90 is stable on 7\,yr interval. Two main features exhibited remarkably different patterns of variability over $\sim$23\,yr; the one rises linearly at the rate of 0.75\,Jy\,yr$^{-1}$, while the second increase exponentially by a factor of seven. This long-term variability can be due to changes in the amplification path length.\\
\indent
(iii) Maser morphology in G94 is basically unchanged over 7\,yr, while the light curve of the main two features is a blend of $\sim$23\,yr lasting variations of low and high relative amplitudes of 0.52 and 5.82, respectively and $\sim$3\,months to a few years lasting synchronous changes of the relative amplitude 1.0$-$1.5. The latter is likely caused by changes in the pumping radiation flux and the former by large-scale gas motions.

\begin{table*}
\centering
\caption{Parameters of the 6.7~GHz methanol maser cloudlets in G78. $\Delta$RA and $\Delta$Dec correspond to the relative coordinates to the maser spot at RA=20$^\mathrm{h}$14$^\mathrm{m}$26\fs05839, Dec=+41\degr13\arcmin32\farcs5278 (J2000). $V_\mathrm{p}$ is the peak velocity,  $V_\mathrm{fit}$ is the fitted velocity, FWHM is the full-width at half maximum of the Gaussian profile, $S_\mathrm{p}$ is the peak brightness,  $S_\mathrm{fit}$ is the fitted brightness, $L_\mathrm{proj}$ is the projected linear size of cloudlet, $V_\mathrm{grad}$ is the velocity gradient and PA is the position angle of the major axis of cloudlet.
Entries in italic are uncertain. }
\label{g78_table}
\begin{tabular}{!c ^c ^c ^c ^c ^c ^c ^c ^c} 
 \hline
Cloudlet & $\Delta$RA & $\Delta$Dec  & $V_\mathrm{fit}$ & FWHM  & $S_\mathrm{fit}$ & $L_\mathrm{proj}$ & $V_\mathrm{grad}$ & PA\\ 
  & (mas) & (mas)  & (\kms) & (\kms) & (Jy beam$^{-1}$) & (mas(au)) & (\kms mas$^{-1}$(\kms au$^{-1}$)) & (\degr)\\
 \hline
1  & 0.0    & 0.0    & -4.94 & 0.26  & 0.12 & 1.3(2.1) & 0.29(0.173) & {\it 51} \\
2  & 83.2   & 106.2  & -6.93 & 0.33  & 1.38 & 5.5(8.8) & 0.13(0.079) & 122\\
3  & 54.1   & 112.1  & -6.85 & {\it 0.76} & 0.11 & 6.8(10.9) & 0.10(0.064) & -65\\
4  & 33.9   & 139.9  & -7.60 & 0.27  & 0.42 & 0.6(1.0) & 0.72(0.451) & 119\\
5  & -168.9 & 104.6  & -6.72 & 0.31  & 0.80 & 1.1(1.7) & 0.14(0.085) & 108\\
   & -158.1 & 101.7  & -6.08 & 0.41  & 8.09 & 6.9(11.1) & 0.14(0.085) & 92\\
6  & -96.5  & 94.1   & -6.60 & 0.24  & 0.22 & 1.6(2.6) & 0.11(0.067) & 139\\
   & -100.4 & 80.6   & -5.69 & 0.40  & 0.68 & 2.7(4.3) & 0.20(0.123) & -114\\
7  & 122.1  & 106.0  & -8.27 & 0.29  & 0.45 & 4.2(6.7) & 0.13(0.079) & -137\\
8  & 107.2  & 106.5  & -7.75 & 0.31  &  0.04 & 2.4(3.8) & 0.07(0.046) & -107\\
9  & -74.2  & 91.7   & -6.71 & 0.36  & 0.37 & 0.6(1.0) & 0.56(0.352) & {\it -30}\\
10 & 47.0   & 109.8  & -7.10 & 0.25  & 0.51 & 1.2(1.9) & 0.23(0.142) & -44\\
11 & -94.6  & 86.2   & -6.41 & 0.37  & 0.58 & 2.5(4.0) & 0.17(0.109) & -133\\ 
12 & -67.5  & 92.0   & -6.70 & 0.33  & 0.17 & 1.4(2.2) & 0.12(0.076) & 45\\
13 & -88.7  & 95.8   & -     & -     & -     & 0.8(1.3) & 0.22(0.136) & {\it -31}\\
14 & -94.9 & -80.9 & -5.86 &  0.46 & 0.41 &  3.3(5.3) & 0.11(0.066) & -67 \\
\hline
\end{tabular}
\end{table*}

\begin{table*}
\centering
\caption{The same as Table~\ref{g78_table} but for G90.925$+$1.486. Coordinates of the (0,0) point are RA=21$^\mathrm{h}$09$^\mathrm{m}$12\fs97472, Dec=$+$50\degr01\arcmin03\farcs6578 (J2000) and corresponds to the brightest maser spot.}
\begin{tabular}{c c c c c c c c c} 
 \hline
Cloudlet & $\Delta$RA & $\Delta$Dec  & $V_\mathrm{fit}$ & FWHM  & $S_\mathrm{fit}$ & $L_\mathrm{proj}$ & $V_\mathrm{grad}$ & PA\\ 
  & (mas) & (mas)  & (\kms) & (\kms) & (Jy beam$^{-1}$) & (mas(au)) & (\kms mas$^{-1}$(\kms au$^{-1}$)) & (\degr)\\
\hline
1       & 0.0   & 0.0    & -69.21 & 0.30   & 21.71 & 2.8(12.4) & 0.18(0.031) & 93 \\
        & 5.1   & -1.4   & -68.79 & 0.24   & 3.10  & 4.3(19.5) & 0.14(0.024) & 62 \\ 
2       & -9.5  & -13.3  & -69.41 & 0.28   & 1.64  & 0.6(3.3) & 0.35(0.060) & 68\\
        & -14.1 & -7.7   & -69.82 & 0.24   & 0.38  & 1.9(11.5) & 0.10(0.018) & 62\\
3       & 42.5  & 9.9    & -69.59 & 0.36   & 0.19  & 1.1(6.4)  & 0.18(0.031) & {\it -96} \\
4       & -38.6 & -22.6  & -70.34 & 0.34   & 9.37  & 17.5(103.2) & 0.05(0.008) & 100\\
5       & -92.0 & 27.0   & -70.49 & 0.32   & 0.65  & 0.7(4.0) & 0.29(0.050) & 6 \\
6       & 47.9  & 68.2   & -71.14 & 0.19   & 0.40  & 0.9(5.4) & 0.32(0.055) & {\it 143} \\
7       & 67.6  & 17.6   & -69.60 & 0.34   & 0.10  & 1.0(5.7) & 0.20(0.035) & {\it 146} \\
\hline
\end{tabular}
\label{g90_table}
\end{table*}

\begin{table*}
\centering
\caption{The same as Table~\ref{g78_table} but for G94.602$-$1.796. Coordinates of the (0,0) point are: RA=21$^\mathrm{h}$39$^\mathrm{m}$58\fs25561, Dec=$+$50\degr14\arcmin20\farcs9108 (J2000) and corresponds to the brightest maser spot.}
\begin{tabular}{c c c c c c c c c } 
 \hline
Cloudlet & $\Delta$RA & $\Delta$Dec  & $V_\mathrm{fit}$ & FWHM  & $S_\mathrm{fit}$ & $L_\mathrm{proj}$ & $V_\mathrm{grad}$ & PA\\ 
  & (mas) & (mas)  & (\kms) & (\kms) & (Jy beam$^{-1}$) & (mas(au)) & (\kms mas$^{-1}$(\kms au$^{-1}$)) & (\degr)\\
 \hline
1       & 5.3  & -88.1   & -40.85 & {\it 0.53} & 1.05 & 5.1(23.0) & 0.11(0.024) & 137\\
2       & 10.1 & -53.1   & -40.98 & 0.30  & 0.59 & 2.3(10.4) & 0.36(0.081) & -30 \\
        & 10.1 & -53.7   & -41.43 & {\it 0.58}  & 0.10 & 2.6(12.0) & 0.42(0.091) & -21 \\        
3       & 18.6 & -78.4   & -41.07 & {\it 0.56}  & 0.15 & 2.1(9.5) & 0.23(0.051) & 179 \\
4       & -0.9 & -23.7   & -42.20 & 0.19  & 0.12 & 0.5(2.3) & 0.43(0.96) & {\it  85} \\
5       & 2.0  & -16.2   & -42.94 & 0.36  & 0.54 & 5.7(25.7) & 0.14(0.031) & 3\\
6       & 0.0  & 0.0     & -43.65 & {\it0.54}   & 0.96 & 14.0(63.0) & 0.11(0.024) & 7\\
\hline
\end{tabular}
\label{g94_table}
\end{table*}

\section*{Acknowledgements}
We thank a referee for their comments and suggestions, which improved the manuscript.
We thank Dr. Luca Moscadelli from INAF-Osservatorio Astrofisico di Arcetri, Italy, for the valuable discussion about proper motion analysis.
This publication has received funding from the European Union’s Horizon 2020 research and innovation program under RadioNet grant agreement No 730562. 
I.S. acknowledges support from the ERDF project “Physical and chemical processes in the interstellar medium”, No.1.1.1.1/16/A/213. 
A.B. and her Torun collaborators acknowledge support from the National Science Centre, Poland, through grant 2021/43/B/ST9/02008.

%%%%%%%%%%%%%%%%%%%%%%%%%%%%%%%%%%%%%%%%%%%%%%%%%%
\section*{Data Availability}

EVN data are available at http://archive.jive.nl. The single-dish data presented in the paper will be shared on reasonable request to the corresponding author.



\bibliographystyle{mnras}
\bibliography{librarian} % if your bibtex file is called example.bib

\begin{thebibliography}{}
\makeatletter
\relax
\def\mn@urlcharsother{\let\do\@makeother \do\$\do\&\do\#\do\^\do\_\do\%\do\~}
\def\mn@doi{\begingroup\mn@urlcharsother \@ifnextchar [ {\mn@doi@}
  {\mn@doi@[]}}
\def\mn@doi@[#1]#2{\def\@tempa{#1}\ifx\@tempa\@empty \href
  {http://dx.doi.org/#2} {doi:#2}\else \href {http://dx.doi.org/#2} {#1}\fi
  \endgroup}
\def\mn@eprint#1#2{\mn@eprint@#1:#2::\@nil}
\def\mn@eprint@arXiv#1{\href {http://arxiv.org/abs/#1} {{\tt arXiv:#1}}}
\def\mn@eprint@dblp#1{\href {http://dblp.uni-trier.de/rec/bibtex/#1.xml}
  {dblp:#1}}
\def\mn@eprint@#1:#2:#3:#4\@nil{\def\@tempa {#1}\def\@tempb {#2}\def\@tempc
  {#3}\ifx \@tempc \@empty \let \@tempc \@tempb \let \@tempb \@tempa \fi \ifx
  \@tempb \@empty \def\@tempb {arXiv}\fi \@ifundefined
  {mn@eprint@\@tempb}{\@tempb:\@tempc}{\expandafter \expandafter \csname
  mn@eprint@\@tempb\endcsname \expandafter{\@tempc}}}

\bibitem[\protect\citeauthoryear{{Aberfelds}, {Shmeld}  \&
  {Berzins}}{{Aberfelds} et~al.}{2018}]{aberfelds2018}
{Aberfelds} A.,  {Shmeld} I.,   {Berzins} K.,  2018, in {Tarchi} A.,  {Reid}
  M.~J.,   {Castangia} P.,  eds,  IAU Symposium Vol. 336, Astrophysical Masers:
  Unlocking the Mysteries of the Universe. pp 277--278,
  \mn@doi{10.1017/S1743921317009437}

\bibitem[\protect\citeauthoryear{{Aberfelds}, {Steinbergs}, {Shmeld}  \&
  {Bartkiewicz}}{{Aberfelds} et~al.}{2021}]{aberfelds2021}
{Aberfelds} A.,  {Steinbergs} J.,  {Shmeld} I.,   {Bartkiewicz} A.,  2021,
  Astronomical and Astrophysical Transactions, \href
  {https://ui.adsabs.harvard.edu/abs/2021A&AT...32..383A} {32, 383}

\bibitem[\protect\citeauthoryear{Aller, Aller  \& Hughes}{Aller
  et~al.}{2003}]{Aller_2003}
Aller M.~F.,  Aller H.~D.,   Hughes P.~A.,  2003, \mn@doi [\apj]
  {10.1086/367538}, 586, 33–51

\bibitem[\protect\citeauthoryear{{Bartkiewicz}, {Szymczak}  \& {van
  Langevelde}}{{Bartkiewicz} et~al.}{2005}]{bartkiewicz2005}
{Bartkiewicz} A.,  {Szymczak} M.,   {van Langevelde} H.~J.,  2005, \mn@doi
  [\aap] {10.1051/0004-6361:200500190}, \href
  {https://ui.adsabs.harvard.edu/abs/2005A&A...442L..61B} {442, L61}

\bibitem[\protect\citeauthoryear{{Bartkiewicz}, {Szymczak}, {van Langevelde},
  {Richards}  \& {Pihlstr{\"o}m}}{{Bartkiewicz} et~al.}{2009}]{bartkiewicz2009}
{Bartkiewicz} A.,  {Szymczak} M.,  {van Langevelde} H.~J.,  {Richards}
  A.~M.~S.,   {Pihlstr{\"o}m} Y.~M.,  2009, \mn@doi [\aap]
  {10.1051/0004-6361/200912250}, \href
  {https://ui.adsabs.harvard.edu/abs/2009A&A...502..155B} {502, 155}

\bibitem[\protect\citeauthoryear{{Bartkiewicz}, {Szymczak}  \& {van
  Langevelde}}{{Bartkiewicz} et~al.}{2014}]{bartkiewicz2014}
{Bartkiewicz} A.,  {Szymczak} M.,   {van Langevelde} H.~J.,  2014, \mn@doi
  [\aap] {10.1051/0004-6361/201322629}, \href
  {https://ui.adsabs.harvard.edu/abs/2014A&A...564A.110B} {564, A110}

\bibitem[\protect\citeauthoryear{{Bartkiewicz}, {Szymczak}  \& {van
  Langevelde}}{{Bartkiewicz} et~al.}{2016}]{bartkiewicz2016}
{Bartkiewicz} A.,  {Szymczak} M.,   {van Langevelde} H.~J.,  2016, \mn@doi
  [\aap] {10.1051/0004-6361/201527541}, \href
  {https://ui.adsabs.harvard.edu/abs/2016A&A...587A.104B} {587, A104}

\bibitem[\protect\citeauthoryear{{Bartkiewicz}, {Sanna}, {Szymczak},
  {Moscadelli}, {van Langevelde}  \& {Wolak}}{{Bartkiewicz}
  et~al.}{2020}]{Bartkiewicz_2020}
{Bartkiewicz} A.,  {Sanna} A.,  {Szymczak} M.,  {Moscadelli} L.,  {van
  Langevelde} H.~J.,   {Wolak} P.,  2020, \mn@doi [\aap]
  {10.1051/0004-6361/202037562}, \href
  {https://ui.adsabs.harvard.edu/abs/2020A&A...637A..15B} {637, A15}

\bibitem[\protect\citeauthoryear{Bayandina, Burns, Kurtz, Shakhvorostova  \&
  Val’tts}{Bayandina et~al.}{2019}]{Bayandina_2019}
Bayandina O.~S.,  Burns R.~A.,  Kurtz S.~E.,  Shakhvorostova N.~N.,   Val’tts
  I.~E.,  2019, \mn@doi [\aj] {10.3847/1538-4357/ab3fa4}, 884, 140

\bibitem[\protect\citeauthoryear{{Beuther} et~al.,}{{Beuther}
  et~al.}{2018}]{beuther2018}
{Beuther} H.,  et~al., 2018, \mn@doi [\aap] {10.1051/0004-6361/201833021},
  \href {https://ui.adsabs.harvard.edu/abs/2018A&A...617A.100B} {617, A100}

\bibitem[\protect\citeauthoryear{{Burns} et~al.,}{{Burns}
  et~al.}{2020}]{burns2020}
{Burns} R.~A.,  et~al., 2020, \mn@doi [Nature Astronomy]
  {10.1038/s41550-019-0989-3}, \href
  {https://ui.adsabs.harvard.edu/abs/2020NatAs...4..506B} {4, 506}

\bibitem[\protect\citeauthoryear{{Caratti o Garatti} et~al.,}{{Caratti o
  Garatti} et~al.}{2017}]{caratti2017}
{Caratti o Garatti} A.,  et~al., 2017, \mn@doi [Nature Physics]
  {10.1038/nphys3942}, \href
  {https://ui.adsabs.harvard.edu/abs/2017NatPh..13..276C} {13, 276}

\bibitem[\protect\citeauthoryear{{Caswell}, {Vaile}  \& {Ellingsen}}{{Caswell}
  et~al.}{1995}]{caswell1995}
{Caswell} J.~L.,  {Vaile} R.~A.,   {Ellingsen} S.~P.,  1995, \mn@doi [\pasa]
  {10.1017/S1323358000020026}, \href
  {https://ui.adsabs.harvard.edu/abs/1995PASA...12...37C} {12, 37}

\bibitem[\protect\citeauthoryear{{Cesaroni}, {Felli}, {Testi}, {Walmsley}  \&
  {Olmi}}{{Cesaroni} et~al.}{1997}]{cesaroni1997}
{Cesaroni} R.,  {Felli} M.,  {Testi} L.,  {Walmsley} C.~M.,   {Olmi} L.,  1997,
  \aap, \href {https://ui.adsabs.harvard.edu/abs/1997A&A...325..725C} {325,
  725}

\bibitem[\protect\citeauthoryear{{Cesaroni}, {Neri}, {Olmi}, {Testi},
  {Walmsley}  \& {Hofner}}{{Cesaroni} et~al.}{2005}]{cesaroni2005}
{Cesaroni} R.,  {Neri} R.,  {Olmi} L.,  {Testi} L.,  {Walmsley} C.~M.,
  {Hofner} P.,  2005, \mn@doi [\aap] {10.1051/0004-6361:20041639}, \href
  {https://ui.adsabs.harvard.edu/abs/2005A&A...434.1039C} {434, 1039}

\bibitem[\protect\citeauthoryear{{Cesaroni} et~al.,}{{Cesaroni}
  et~al.}{2013}]{cesaroni2013}
{Cesaroni} R.,  et~al., 2013, \mn@doi [\aap] {10.1051/0004-6361/201220609},
  \href {https://ui.adsabs.harvard.edu/abs/2013A&A...549A.146C} {549, A146}

\bibitem[\protect\citeauthoryear{{Cesaroni}, {Galli}, {Neri}  \&
  {Walmsley}}{{Cesaroni} et~al.}{2014}]{cesaroni2014}
{Cesaroni} R.,  {Galli} D.,  {Neri} R.,   {Walmsley} C.~M.,  2014, \mn@doi
  [\aap] {10.1051/0004-6361/201323065}, \href
  {https://ui.adsabs.harvard.edu/abs/2014A&A...566A..73C} {566, A73}

\bibitem[\protect\citeauthoryear{Choi, Hachisuka, Reid, Xu, Brunthaler, Menten
  \& Dame}{Choi et~al.}{2014}]{Choi_2014}
Choi Y.~K.,  Hachisuka K.,  Reid M.~J.,  Xu Y.,  Brunthaler A.,  Menten K.~M.,
   Dame T.~M.,  2014, \mn@doi [\apj] {10.1088/0004-637x/790/2/99}, 790, 99

\bibitem[\protect\citeauthoryear{Clarke, Lumsden, Oudmaijer, Busfield, Hoare,
  Moore, Sheret  \& Urquhart}{Clarke et~al.}{2006}]{Clarke_2006}
Clarke A.~J.,  Lumsden S.~L.,  Oudmaijer R.~D.,  Busfield A.~L.,  Hoare M.~G.,
  Moore T. J.~T.,  Sheret T.~L.,   Urquhart J.~S.,  2006, \mn@doi [\aap]
  {10.1051/0004-6361:20064839}, 457, 183–188

\bibitem[\protect\citeauthoryear{{Cohen}}{{Cohen}}{1977}]{cohen1977}
{Cohen} M.,  1977, \mn@doi [\apj] {10.1086/155386}, \href
  {https://ui.adsabs.harvard.edu/abs/1977ApJ...215..533C} {215, 533}

\bibitem[\protect\citeauthoryear{{Cragg}, {Sobolev}  \& {Godfrey}}{{Cragg}
  et~al.}{2002}]{cragg2002}
{Cragg} D.~M.,  {Sobolev} A.~M.,   {Godfrey} P.~D.,  2002, \mn@doi [\mnras]
  {10.1046/j.1365-8711.2002.05226.x}, \href
  {https://ui.adsabs.harvard.edu/abs/2002MNRAS.331..521C} {331, 521}

\bibitem[\protect\citeauthoryear{{Dodson}, {Ojha}  \& {Ellingsen}}{{Dodson}
  et~al.}{2004}]{dodson2004}
{Dodson} R.,  {Ojha} R.,   {Ellingsen} S.~P.,  2004, \mn@doi [\mnras]
  {10.1111/j.1365-2966.2004.07844.x}, \href
  {https://ui.adsabs.harvard.edu/abs/2004MNRAS.351..779D} {351, 779}

\bibitem[\protect\citeauthoryear{Durjasz, Szymczak  \& Olech}{Durjasz
  et~al.}{2019}]{durjasz2019}
Durjasz M.,  Szymczak M.,   Olech M.,  2019, \mn@doi [\mnras]
  {10.1093/mnras/stz472}, 485, 777

\bibitem[\protect\citeauthoryear{{Fujisawa} et~al.,}{{Fujisawa}
  et~al.}{2014a}]{fujisawa2014a}
{Fujisawa} K.,  et~al., 2014a, \mn@doi [\pasj] {10.1093/pasj/psu015}, \href
  {https://ui.adsabs.harvard.edu/abs/2014PASJ...66...31F} {66, 31}

\bibitem[\protect\citeauthoryear{{Fujisawa} et~al.,}{{Fujisawa}
  et~al.}{2014b}]{fujisawa2014b}
{Fujisawa} K.,  et~al., 2014b, \mn@doi [\pasj] {10.1093/pasj/psu053}, \href
  {https://ui.adsabs.harvard.edu/abs/2014PASJ...66...78F} {66, 78}

\bibitem[\protect\citeauthoryear{{Fujisawa}, {Yonekura}, {Sugiyama},
  {Horiuchi}, {Hayashi}, {Hachisuka}, {Matsumoto}  \& {Niinuma}}{{Fujisawa}
  et~al.}{2015}]{fujisawa2015}
{Fujisawa} K.,  {Yonekura} Y.,  {Sugiyama} K.,  {Horiuchi} H.,  {Hayashi} T.,
  {Hachisuka} K.,  {Matsumoto} N.,   {Niinuma} K.,  2015, The Astronomer's
  Telegram, \href {https://ui.adsabs.harvard.edu/abs/2015ATel.8286....1F}
  {8286, 1}

\bibitem[\protect\citeauthoryear{{Gieser} et~al.,}{{Gieser}
  et~al.}{2021}]{gieser2021}
{Gieser} C.,  et~al., 2021, \mn@doi [\aap] {10.1051/0004-6361/202039670}, \href
  {https://ui.adsabs.harvard.edu/abs/2021A&A...648A..66G} {648, A66}

\bibitem[\protect\citeauthoryear{{Goedhart}, {Gaylard}  \& {van der
  Walt}}{{Goedhart} et~al.}{2004}]{goedhart2004}
{Goedhart} S.,  {Gaylard} M.~J.,   {van der Walt} D.~J.,  2004, \mn@doi
  [\mnras] {10.1111/j.1365-2966.2004.08340.x}, \href
  {https://ui.adsabs.harvard.edu/abs/2004MNRAS.355..553G} {355, 553}

\bibitem[\protect\citeauthoryear{{Goedhart}, {Maswanganye}, {Gaylard}  \& {van
  der Walt}}{{Goedhart} et~al.}{2014}]{goedhart2014b}
{Goedhart} S.,  {Maswanganye} J.~P.,  {Gaylard} M.~J.,   {van der Walt} D.~J.,
  2014, \mn@doi [\mnras] {10.1093/mnras/stt2009}, \href
  {https://ui.adsabs.harvard.edu/abs/2014MNRAS.437.1808G} {437, 1808}

\bibitem[\protect\citeauthoryear{Hu, Menten, Wu, Bartkiewicz, Rygl, Reid,
  Urquhart  \& Zheng}{Hu et~al.}{2016}]{Hu2016}
Hu B.,  Menten K.~M.,  Wu Y.,  Bartkiewicz A.,  Rygl K.,  Reid M.~J.,  Urquhart
  J.~S.,   Zheng X.,  2016, \mn@doi [\apj] {10.3847/0004-637x/833/1/18}, 833,
  18

\bibitem[\protect\citeauthoryear{Keimpema et~al.,}{Keimpema
  et~al.}{2015}]{sfxc_2016}
Keimpema A.,  et~al., 2015, \mn@doi [Experimental Astronomy]
  {10.1007/s10686-015-9446-1}, 39, 259

\bibitem[\protect\citeauthoryear{{Kobak} et~al.,}{{Kobak}
  et~al.}{2023}]{kobak2023}
{Kobak} A.,  et~al., 2023, \mn@doi [\aap] {10.1051/0004-6361/202244772}, \href
  {https://ui.adsabs.harvard.edu/abs/2023A&A...671A.135K} {671, A135}

\bibitem[\protect\citeauthoryear{{MacLeod} et~al.,}{{MacLeod}
  et~al.}{2018}]{macleod2018}
{MacLeod} G.~C.,  et~al., 2018, \mn@doi [\mnras] {10.1093/mnras/sty996}, \href
  {https://ui.adsabs.harvard.edu/abs/2018MNRAS.478.1077M} {478, 1077}

\bibitem[\protect\citeauthoryear{{Menten}}{{Menten}}{1991}]{menten1991}
{Menten} K.~M.,  1991, \mn@doi [\apjl] {10.1086/186177}, \href
  {https://ui.adsabs.harvard.edu/abs/1991ApJ...380L..75M} {380, L75}

\bibitem[\protect\citeauthoryear{{Minier}, {Booth}  \& {Conway}}{{Minier}
  et~al.}{2000}]{minier2000}
{Minier} V.,  {Booth} R.~S.,   {Conway} J.~E.,  2000, \aap, \href
  {https://ui.adsabs.harvard.edu/abs/2000A&A...362.1093M} {362, 1093}

\bibitem[\protect\citeauthoryear{{Moscadelli}, {Cesaroni}, {Rioja}, {Dodson}
  \& {Reid}}{{Moscadelli} et~al.}{2011a}]{moscadelli2011}
{Moscadelli} L.,  {Cesaroni} R.,  {Rioja} M.~J.,  {Dodson} R.,   {Reid} M.~J.,
  2011a, \mn@doi [\aap] {10.1051/0004-6361/201015641}, \href
  {https://ui.adsabs.harvard.edu/abs/2011A&A...526A..66M} {526, A66}

\bibitem[\protect\citeauthoryear{{Moscadelli}, {Sanna}  \&
  {Goddi}}{{Moscadelli} et~al.}{2011b}]{moscadelli2011mas}
{Moscadelli} L.,  {Sanna} A.,   {Goddi} C.,  2011b, \mn@doi [\aap]
  {10.1051/0004-6361/201117791}, \href
  {https://ui.adsabs.harvard.edu/abs/2011A&A...536A..38M} {536, A38}

\bibitem[\protect\citeauthoryear{{Murakawa}, {Lumsden}, {Oudmaijer}, {Davies},
  {Wheelwright}, {Hoare}  \& {Ilee}}{{Murakawa} et~al.}{2013}]{murakawa2013}
{Murakawa} K.,  {Lumsden} S.~L.,  {Oudmaijer} R.~D.,  {Davies} B.,
  {Wheelwright} H.~E.,  {Hoare} M.~G.,   {Ilee} J.~D.,  2013, \mn@doi [\mnras]
  {10.1093/mnras/stt1592}, \href
  {https://ui.adsabs.harvard.edu/abs/2013MNRAS.436..511M} {436, 511}

\bibitem[\protect\citeauthoryear{{Nagayama} et~al.,}{{Nagayama}
  et~al.}{2015}]{Nagayama2015}
{Nagayama} T.,  et~al., 2015, \mn@doi [\pasj] {10.1093/pasj/psu133}, \href
  {https://ui.adsabs.harvard.edu/abs/2015PASJ...67...66N} {67, 66}

\bibitem[\protect\citeauthoryear{{Norris}, {Whiteoak}, {Caswell}, {Wieringa}
  \& {Gough}}{{Norris} et~al.}{1993}]{norris1993}
{Norris} R.~P.,  {Whiteoak} J.~B.,  {Caswell} J.~L.,  {Wieringa} M.~H.,
  {Gough} R.~G.,  1993, \mn@doi [\apj] {10.1086/172914}, \href
  {https://ui.adsabs.harvard.edu/abs/1993ApJ...412..222N} {412, 222}

\bibitem[\protect\citeauthoryear{{Oh}, {Kobayashi}, {Honma}, {Hirota}, {Sato}
  \& {Ueno}}{{Oh} et~al.}{2010}]{Oh2010}
{Oh} C.~S.,  {Kobayashi} H.,  {Honma} M.,  {Hirota} T.,  {Sato} K.,   {Ueno}
  Y.,  2010, \mn@doi [\pasj] {10.1093/pasj/62.1.101}, \href
  {https://ui.adsabs.harvard.edu/abs/2010PASJ...62..101O} {62, 101}

\bibitem[\protect\citeauthoryear{{Olech}, {Szymczak}, {Wolak}, {Sarniak}  \&
  {Bartkiewicz}}{{Olech} et~al.}{2019}]{olech2019}
{Olech} M.,  {Szymczak} M.,  {Wolak} P.,  {Sarniak} R.,   {Bartkiewicz} A.,
  2019, \mn@doi [\mnras] {10.1093/mnras/stz926}, \href
  {https://ui.adsabs.harvard.edu/abs/2019MNRAS.486.1236O} {486, 1236}

\bibitem[\protect\citeauthoryear{{Olech}, {Szymczak}, {Wolak}, {G{\'e}rard}  \&
  {Bartkiewicz}}{{Olech} et~al.}{2020}]{olech2020}
{Olech} M.,  {Szymczak} M.,  {Wolak} P.,  {G{\'e}rard} E.,   {Bartkiewicz} A.,
  2020, \mn@doi [\aap] {10.1051/0004-6361/201936943}, \href
  {https://ui.adsabs.harvard.edu/abs/2020A&A...634A..41O} {634, A41}

\bibitem[\protect\citeauthoryear{{Olech}, {Durjasz}, {Szymczak}  \&
  {Bartkiewicz}}{{Olech} et~al.}{2022}]{olech2022}
{Olech} M.,  {Durjasz} M.,  {Szymczak} M.,   {Bartkiewicz} A.,  2022, \mn@doi
  [\aap] {10.1051/0004-6361/202243108}, \href
  {https://ui.adsabs.harvard.edu/abs/2022A&A...661A.114O} {661, A114}

\bibitem[\protect\citeauthoryear{{Purser}, {Lumsden}, {Hoare}  \&
  {Kurtz}}{{Purser} et~al.}{2021}]{purser2021}
{Purser} S.~J.~D.,  {Lumsden} S.~L.,  {Hoare} M.~G.,   {Kurtz} S.,  2021,
  \mn@doi [\mnras] {10.1093/mnras/stab747}, \href
  {https://ui.adsabs.harvard.edu/abs/2021MNRAS.504..338P} {504, 338}

\bibitem[\protect\citeauthoryear{Reid et~al.,}{Reid et~al.}{2019}]{reid2019}
Reid M.~J.,  et~al., 2019, \mn@doi [\apj] {10.3847/1538-4357/ab4a11}, 885, 131

\bibitem[\protect\citeauthoryear{{Rygl}, {Brunthaler}, {Reid}, {Menten}, {van
  Langevelde}  \& {Xu}}{{Rygl} et~al.}{2010}]{rygl2010}
{Rygl} K.~L.~J.,  {Brunthaler} A.,  {Reid} M.~J.,  {Menten} K.~M.,  {van
  Langevelde} H.~J.,   {Xu} Y.,  2010, \mn@doi [\aap]
  {10.1051/0004-6361/200913135}, \href
  {https://ui.adsabs.harvard.edu/abs/2010A&A...511A...2R} {511, A2}

\bibitem[\protect\citeauthoryear{{Sakai}, {Reid}, {Menten}, {Brunthaler}  \&
  {Dame}}{{Sakai} et~al.}{2019}]{Sakai2019}
{Sakai} N.,  {Reid} M.~J.,  {Menten} K.~M.,  {Brunthaler} A.,   {Dame} T.~M.,
  2019, \mn@doi [\apj] {10.3847/1538-4357/ab12e0}, \href
  {https://ui.adsabs.harvard.edu/abs/2019ApJ...876...30S} {876, 30}

\bibitem[\protect\citeauthoryear{{Sanna}, {Moscadelli}, {Cesaroni}, {Tarchi},
  {Furuya}  \& {Goddi}}{{Sanna} et~al.}{2010a}]{sanna2010a}
{Sanna} A.,  {Moscadelli} L.,  {Cesaroni} R.,  {Tarchi} A.,  {Furuya} R.~S.,
  {Goddi} C.,  2010a, \mn@doi [\aap] {10.1051/0004-6361/201014233}, \href
  {https://ui.adsabs.harvard.edu/abs/2010A&A...517A..71S} {517, A71}

\bibitem[\protect\citeauthoryear{{Sanna}, {Moscadelli}, {Cesaroni}, {Tarchi},
  {Furuya}  \& {Goddi}}{{Sanna} et~al.}{2010b}]{sanna2010b}
{Sanna} A.,  {Moscadelli} L.,  {Cesaroni} R.,  {Tarchi} A.,  {Furuya} R.~S.,
  {Goddi} C.,  2010b, \mn@doi [\aap] {10.1051/0004-6361/201014234}, \href
  {https://ui.adsabs.harvard.edu/abs/2010A&A...517A..78S} {517, A78}

\bibitem[\protect\citeauthoryear{{Sanna}, {Moscadelli}, {Surcis}, {van
  Langevelde}, {Torstensson}  \& {Sobolev}}{{Sanna} et~al.}{2017}]{sanna2017}
{Sanna} A.,  {Moscadelli} L.,  {Surcis} G.,  {van Langevelde} H.~J.,
  {Torstensson} K.~J.~E.,   {Sobolev} A.~M.,  2017, \mn@doi [\aap]
  {10.1051/0004-6361/201730773}, \href
  {https://ui.adsabs.harvard.edu/abs/2017A&A...603A..94S} {603, A94}

\bibitem[\protect\citeauthoryear{{Scargle}}{{Scargle}}{1982}]{scargle1982}
{Scargle} J.~D.,  1982, \mn@doi [\apj] {10.1086/160554}, \href
  {http://adsabs.harvard.edu/abs/1982ApJ...263..835S} {263, 835}

\bibitem[\protect\citeauthoryear{{Slysh}, {Val'tts}, {Kalenskii}, {Voronkov},
  {Palagi}, {Tofani}  \& {Catarzi}}{{Slysh} et~al.}{1999}]{slysh1999}
{Slysh} V.~I.,  {Val'tts} I.~E.,  {Kalenskii} S.~V.,  {Voronkov} M.~A.,
  {Palagi} F.,  {Tofani} G.,   {Catarzi} M.,  1999, \mn@doi [\aaps]
  {10.1051/aas:1999127}, \href
  {https://ui.adsabs.harvard.edu/abs/1999A&AS..134..115S} {134, 115}

\bibitem[\protect\citeauthoryear{{Slysh}, {Voronkov}, {Val'tts}  \&
  {Migenes}}{{Slysh} et~al.}{2002}]{Slysh2002}
{Slysh} V.~I.,  {Voronkov} M.~A.,  {Val'tts} I.~E.,   {Migenes} V.,  2002,
  \mn@doi [Astronomy Reports] {10.1134/1.1529255}, \href
  {https://ui.adsabs.harvard.edu/abs/2002ARep...46..969S} {46, 969}

\bibitem[\protect\citeauthoryear{{Stecklum} et~al.,}{{Stecklum}
  et~al.}{2021}]{stecklum2021}
{Stecklum} B.,  et~al., 2021, \mn@doi [\aap] {10.1051/0004-6361/202039645},
  \href {https://ui.adsabs.harvard.edu/abs/2021A&A...646A.161S} {646, A161}

\bibitem[\protect\citeauthoryear{{Sugiyama} et~al.,}{{Sugiyama}
  et~al.}{2014}]{sugiyama2014}
{Sugiyama} K.,  et~al., 2014, \mn@doi [\aap] {10.1051/0004-6361/201321278},
  \href {https://ui.adsabs.harvard.edu/abs/2014A&A...562A..82S} {562, A82}

\bibitem[\protect\citeauthoryear{{Surcis}, {Vlemmings, W. H. T.}, {van
  Langevelde, H. J.}, {Moscadelli, L.}  \& {Hutawarakorn Kramer, B.}}{{Surcis}
  et~al.}{2014}]{surcis_2014}
{Surcis} G.,  {Vlemmings, W. H. T.} {van Langevelde, H. J.} {Moscadelli, L.}
  {Hutawarakorn Kramer, B.} 2014, \mn@doi [A\&A] {10.1051/0004-6361/201322795},
  563, A30

\bibitem[\protect\citeauthoryear{{Szymczak}, {Hrynek, G.}  \& {Kus, A.
  J.}}{{Szymczak} et~al.}{2000}]{Szymczak_2000}
{Szymczak} M.,  {Hrynek, G.}  {Kus, A. J.} 2000, \mn@doi [A&AS]
  {10.1051/aas:2000334}, 143, 269

\bibitem[\protect\citeauthoryear{{Szymczak}, {Wolak}, {Bartkiewicz}  \&
  {Borkowski}}{{Szymczak} et~al.}{2012}]{Szymczak2012}
{Szymczak} M.,  {Wolak} P.,  {Bartkiewicz} A.,   {Borkowski} K.~M.,  2012,
  \mn@doi [Astronomische Nachrichten] {10.1002/asna.201211702}, \href
  {https://ui.adsabs.harvard.edu/abs/2012AN....333..634S} {333, 634}

\bibitem[\protect\citeauthoryear{{Szymczak}, {Wolak}  \&
  {Bartkiewicz}}{{Szymczak} et~al.}{2014}]{szymczak2014}
{Szymczak} M.,  {Wolak} P.,   {Bartkiewicz} A.,  2014, \mn@doi [\mnras]
  {10.1093/mnras/stu019}, \href
  {https://ui.adsabs.harvard.edu/abs/2014MNRAS.439..407S} {439, 407}

\bibitem[\protect\citeauthoryear{Szymczak, Olech, Wolak, Bartkiewicz  \&
  Gawroński}{Szymczak et~al.}{2016}]{g107_period}
Szymczak M.,  Olech M.,  Wolak P.,  Bartkiewicz A.,   Gawroński M.,  2016,
  \mn@doi [\mnras] {10.1093/mnrasl/slw044}, 459, L56–L60

\bibitem[\protect\citeauthoryear{{Szymczak}, {Olech}, {Sarniak}, {Wolak}  \&
  {Bartkiewicz}}{{Szymczak} et~al.}{2018}]{Szymczak2018}
{Szymczak} M.,  {Olech} M.,  {Sarniak} R.,  {Wolak} P.,   {Bartkiewicz} A.,
  2018, \mn@doi [\mnras] {10.1093/mnras/stx2693}, \href
  {https://ui.adsabs.harvard.edu/abs/2018MNRAS.474..219S} {474, 219}

\bibitem[\protect\citeauthoryear{{Torstensson}}{{Torstensson}}{2011}]{Thorstensson_thesis2011}
{Torstensson} K. J.~E.,  2011, PhD thesis, Leiden Observatory, The Netherlands

\bibitem[\protect\citeauthoryear{{Yang} et~al.,}{{Yang}
  et~al.}{2019}]{Yang2019}
{Yang} K.,  et~al., 2019, \mn@doi [\apjs] {10.3847/1538-4365/ab06fb}, \href
  {https://ui.adsabs.harvard.edu/abs/2019ApJS..241...18Y} {241, 18}

\bibitem[\protect\citeauthoryear{Zinnecker \& Yorke}{Zinnecker \&
  Yorke}{2007}]{Zinnecker2007}
Zinnecker H.,  Yorke H.~W.,  2007, \mn@doi [Ann. Rev. Astron. Astrophys.]
  {10.1146/annurev.astro.44.051905.092549}, 45, 481

\bibitem[\protect\citeauthoryear{{{\v{S}}teinbergs}, {Aberfelds}, {Bleiders}
  \& {Shmeld}}{{{\v{S}}teinbergs} et~al.}{2021}]{steinbergs2021}
{{\v{S}}teinbergs} J.,  {Aberfelds} A.,  {Bleiders} M.,   {Shmeld} I.,  2021,
  Astronomical and Astrophysical Transactions, \href
  {https://ui.adsabs.harvard.edu/abs/2021A&AT...32..227S} {32, 227}

\makeatother
\end{thebibliography}

%\begin{thebibliography}{99}
%\end{thebibliography}

%%%%%%%%%%%%%%%%%%%%%%%%%%%%%%%%%%%%%%%%%%%%%%%%%%

%%%%%%%%%%%%%%%%% APPENDICES %%%%%%%%%%%%%%%%%%%%%

\appendix
\section{Variability parameters}\label{appendix-A}
Statistical tools to classify significance of flux changes in light curves. 
Variability index ($VI$) \citep{Aller_2003}:
\begin{equation}
VI = \frac{(S_{max} -\sigma_{max})-(S_{min} +\sigma_{min})}{(S_{max} -\sigma_{max})+(S_{min} +\sigma_{min})}.
\end{equation}
Here $S_{max}$ and $S_{min}$ are highest and lowest flux density values but $\sigma_{max}$ and $\sigma_{min}$ are absolute uncertainties in these measurements. $VI$ values are in range from 0 for no variable to 1 for strongly variable lines.

Fluctuation index ($FI$) \citep{Aller_2003}:
\begin{equation}
FI = \left[ \frac{N}{\sum_{i=1}^{N}\sigma_{i}^2 }\left(\frac{\sum_{i=1}^{N}S_{i}^2\sigma_{i}^2 - \overline{S}\sum_{i=1}^{N}S_{i}\sigma_{i}^2}{N-1}-1\right) \right]^{0.5}/\overline{S}.
\end{equation}
Here $S_{i}$ is individual flux density measured in specific epoch $i$, $\overline{S}$ average flux density over the time interval under the evaluation, $\sigma_{i}$ individual measurement uncertainties and N the number of measurements.
In essence fluctuation index measures a spread around a mean value.

The reduced $\chi^2_{r}$ test \citep{Szymczak2018}:
\begin{equation}
\chi^2_{r}=\frac{1}{N-1}\sum_{i=1}^{N}\left(\frac{S_{i}-\overline{S}}{\sigma_{i}}\right)^2
\label{chi2red_formula}
\end{equation}
Provides wide range of values, for quiescent sources there will be around one and will increasing by significance of variability.
For more details of methanol maser variability we suggest \citet{goedhart2004} and \citet{Szymczak2018}.

\section{G78 EVN multi epoch}

\begin{figure*}
\centering
\includegraphics[width=\textwidth]{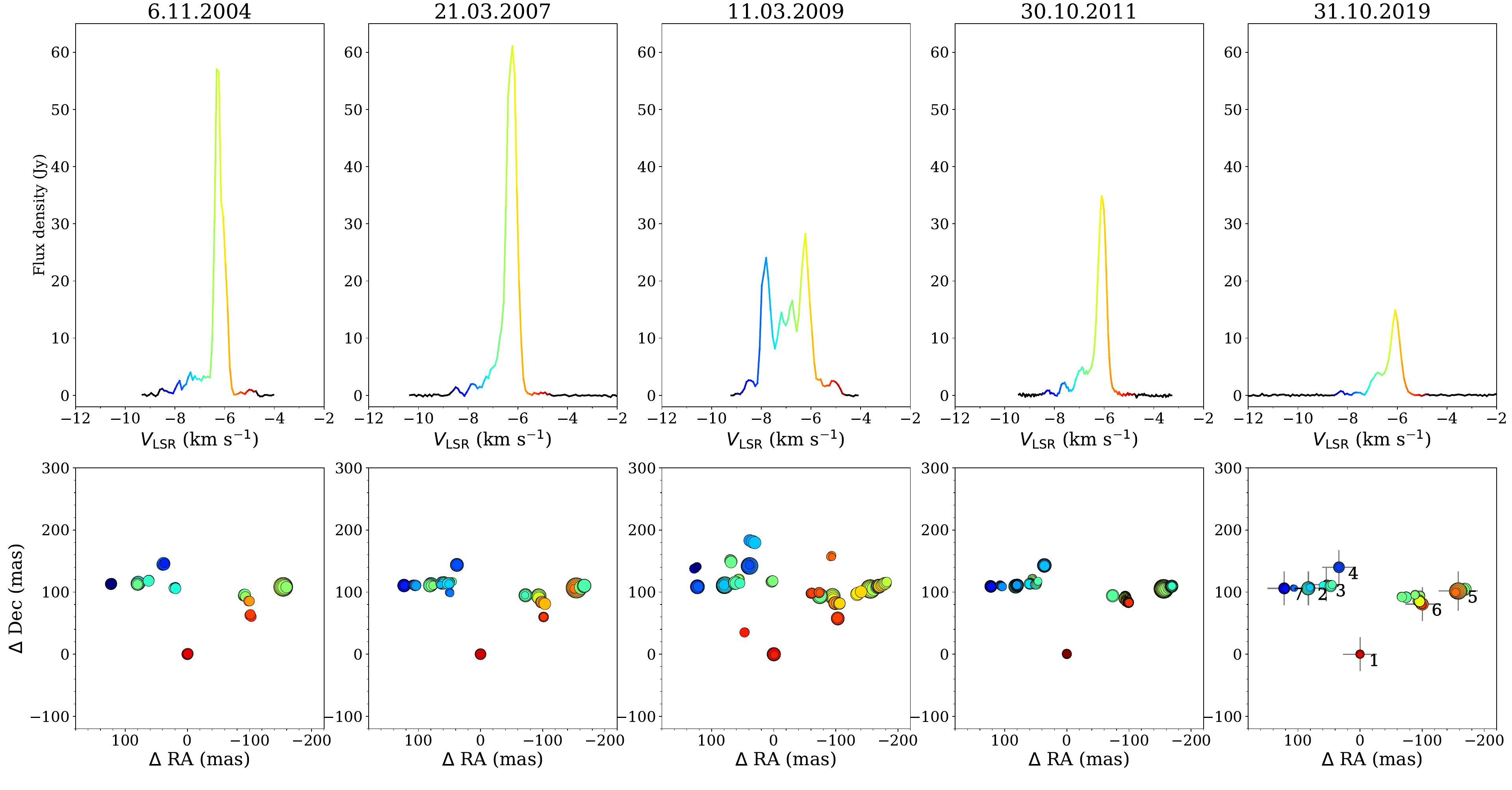}
\caption{{\it Top:} The 6.7~GHz methanol maser spectra of G78.122$+$3.633 at five EVN epochs that were retrieved from the EVN Archive (projects EL032, EM064C, EM064D and ES066E) and our own. {\it Bottom:} Distributions of the methanol maser spots. The circle size represents the logarithm of its flux density and the colour corresponds to the LSR velocity as in the top panel. Coordinate centre (0,0) is the position of the spot at the LSR velocity of $-$4.9~km~s$^{-1}$ that was compact and stable in all epochs. Its coordinates are given in Table~\ref{g78_table}. Grey crosses and numbers on the most right panel mark positions of seven cloudlets that persisted over 15~yr. }
\label{total}  
\end{figure*}

\begin{table*}
\centering
\caption{Parameters of the 6.7~GHz maser cloudlets in G78.122$+$3.633 that persisted 15~years. The parameters are the same as in Table~\ref{g78_table} but for all five epochs. We also list the correlation coefficients of maser spot distribution (r$_s$) and offset of major axes (r$_v$) and the position angle of the major axis of maser spots within each cloudlet (PA).
}
\begin{tabular}{c c c c c c c c c c} 
\hline\hline
 Cloud & $\Delta$RA & $\Delta$Dec & $V_\mathrm{fit}$ & FWHM  & $S_\mathrm{fit}$ & r$_s$ & r$_v$ & $V_\mathrm{grad}$ & PA\\ 
Epoch &  (mas) & (mas) & (km s$^{-1}$) & (km s$^{-1}$) & (Jy beam$^{-1}$) & & &(km s$^{-1}$ mas$^{-1}$(AU)$^{-1}$)& ($^{\circ}$)\\
\hline
{\bf Cloudlet 1} \\
2004  & 0.0    & 0.0  & -4.88  & 0.26  & 0.4  & -0.66 & 0.36  & 0.20(0.12)  & {\it 127}\\
2007  & 0.0    & 0.0  & -4.85  & 0.25  & 0.4  & 0.58  & 0.88  & 0.26(0.16)  & {\it 73} \\
2009  & -0.2   & -0.1 & -4.86  & 0.26  & 1.5  & 0.90  & 0.93  & 0.29(0.18)  & {\it 57} \\
2011  & 0.3    & 0.8  & -4.90  & 0.27  & 0.2  & -0.51 & -0.92 & 0.18(0.12)  & {\it -82} \\
2019  & 0.0    & 0.0  & -4.94  & 0.26  & 0.1  & 0.83  & 0.95  & 0.28(0.17)  & {\it 51} \\
{\bf Cloudlet 2 }\\
2004 & 79.3  & 114.4 & -7.15  & 0.17 & 1.9  & -0.90 & 0.94  & 0.28(0.18) & 142\\
2007 & 79.7  & 112.4 & -7.14  & {\it 0.52}& 1.7 & -0.89 & 0.77  & 0.13(0.08) & 134\\
2009 & 78.6  & 111.2 & -7.10  & 0.42 & 9.2  & -0.65 & 0.59  & 0.15(0.09) & 129 \\
2011 & 82.3  & 109.4 & -6.93  & 0.20 & 1.2  & -0.99 & 0.99  & 0.12(0.07) & 123\\
     & -     &   -    &  -7.07 & {\it 0.47} & 1.0 & - & - & - & - \\
2019 & 83.2  & 106.2 & -6.93  & 0.33 & 1.4  & -0.99 & 0.99  & 0.13(0.08) & 122\\
{\bf Cloudlet 3 } \\
2004 & 62.4 & 118.0 & -       & -    & -     & -0.99 & -0.90 & 0.26(0.16) & {\it -29} \\
2007 & 62.2 & 114.5 & -7.00   & 0.43 & 0.9   & -0.72 & -0.79 & 0.08(0.05) & {\it -100}\\
2009 & 61.6 & 114.9 & -6.94   & 0.26 &  1.4  & -0.54 & 0.24  & 0.04(0.03) & 88 \\
     & 56.3 & 120.0 & -6.56   & 0.23 & 0.5   & -0.94 & -0.99 & 0.07(0.05) & -31 \\
2011 & 60.1 & 113.3 & -7.00   & 0.41 & 0.4   & -0.90 & -0.90 & 0.19(0.12) & -53 \\
     & 55.8 & 121.5 & -6.49   & 0.26 & 0.1   & -0.91 & -0.97 & 0.04(0.03) & -26\\
2019 & 54.1 & 112.1 & -6.85   & {\it 0.76} &  0.1  & -0.94  & -0.98 & 0.10(0.06) & -65\\
{\bf Cloudlet 4} \\
2004 & 38.3 & 145.6 & -7.67 & 0.29 & 1.1  & -0.59 & 0.94  & 0.10(0.06) & {\it 109 } \\
2007 & 38.1 & 143.7 & -7.64 & 0.29 & 1.1  & -0.67 & 0.82  & 0.10(0.06) & {\it 99}  \\
2009 & 38.8 & 142.2 & -7.73 & {\it0.28} & 9.2  & -0.10 & -0.21 & 0.19(0.12) & {\it 101} \\
2011 & 36.3 & 142.9 & -7.62 & 0.27 & 1.8  & 0.91  & -0.98 & 0.48(0.30) & 153 \\
2019 & 33.9 & 139.9 & -7.60 & 0.27 & 0.4  & -0.72 & 0.95  & 0.72(0.45) & 119\\
{\bf Cloudlet 5 }\\
2004 & -154.2 & 108.4 & -6.10  & 0.33 & 27.8 & 0.80  & 0.97 & 0.11(0.07) & 79 \\
2007 & -154.3 & 106.6 & -6.09  & 0.36 & 60.7 & -*    & 0.97 & 0.12(0.07) & 76 \\
2009 & -168.8 & 109.3 & -6.74  & 0.32 & 2.2  & -0.99 & 0.95 & 0.10(0.06) & 118 \\
     & -155.6 & 105.0 & -6.08  & 0.39 & 23.7 & -*    & 0.92 & 0.11(0.07) & 77  \\    
2011 & -168.9 & 109.4 & -6.76  & 0.35 & 0.9  & -0.99 & 0.75 & 0.09(0.05) & 115\\
     & -155.9 & 105.1 & -6.08  & 0.38 & 26.4 & 0.46  & 0.78 & 0.11(0.07) & 75 \\
2019 & -168.9 & 104.6 & -6.71  & 0.31 & 0.8  & -0.56 & 0.65 & 0.14(0.09) & 108 \\
     & -158.1 & 101.7 & -6.08  & 0.41 & 8.3  & -0.81 & 0.99 & 0.14(0.09) & 92 \\
{\bf Cloudlet 6} \\
2004 & -91.7  & 95.3 & -6.64  & 0.32 & 0.6 & 0.38  & -0.73  & 0.37(0.23) & {\it 97} \\
     & -98.5  & 85.5 & -      &  -   &   - &  -    &  -     & -          &  - \\
2007 & -93.8  & 93.4 & -6.52  & 0.33 & 2.9 & 0.62  & 0.99   & 0.11(0.07) & -128 \\
     & -98.1  & 83.8 & -5.49  & 0.45 & 0.5 & 0.67  & 0.79   & 0.28(0.18) & 70 \\
2009 & -94.7  & 93.4 & -6.52  & 0.30 & 4.2 & 0.87 & -0.98  & 0.09(0.06) & -162 \\
     & -99.0  & 82.2 & -5.50  & 0.48 & 1.2 & -0.88  & -0.97  & 0.27(0.17) & -72 \\   
2011 & -93.4  & 89.8 & -6.39  & 0.44 & 0.3 & 0.71  & -0.99  & 0.09(0.06) & -138  \\
     & -99.0  & 83.1 & -5.74  & {\it 0.82}& 0.4 & 0.91  & -0.95  & 0.14(0.08) & -112\\
2019 & -96.5  & 94.1 & -6.60  & 0.24 & 0.2 & 0.79 & -0.94  & 0.11(0.07) & 139\\
     & -100.4 & 80.6 & -5.70  & 0.40 & 0.7 & 0.70  & -0.80  & 0.20(0.12) & -114\\
{\bf Cloudlet 7 }\\
2004 & 123.1 & 113.0 & -8.34  & 0.34 & 0.5 & -0.96 & 0.41  & 0.17(0.11) & {\it 127} \\
2007 & 123.2 & 110.5 & -8.30  & 0.32 & 1.0 & 0.52  & -0.95 & 0.30(0.19) & -147\\
2009 & 122.9 & 108.5 & -8.30  & 0.32 & 2.1 & 0.33  & -0.98 & 0.26(0.16) & -132\\
2011 & 122.5 & 109.0 & -8.27  & 0.30 & 0.6 & 0.85  & -0.99 & 0.15(0.09) & -140\\
2019 & 122.1 & 106.0 & -8.27  & 0.29 & 0.5 & 0.87  & -0.98 & 0.13(0.08) & -137\\            
\hline
\end{tabular}\\
$^*$The correlation coefficient is undefined because the EW spot distribution, i.e. variance of Y is zero.
\label{5_epoch_cloudlet_table}
\end{table*}

\section{Cloudlet structure}
% This is new!
\begin{figure*}
\centering 
\includegraphics[scale=0.23,trim=100 100 0 200]{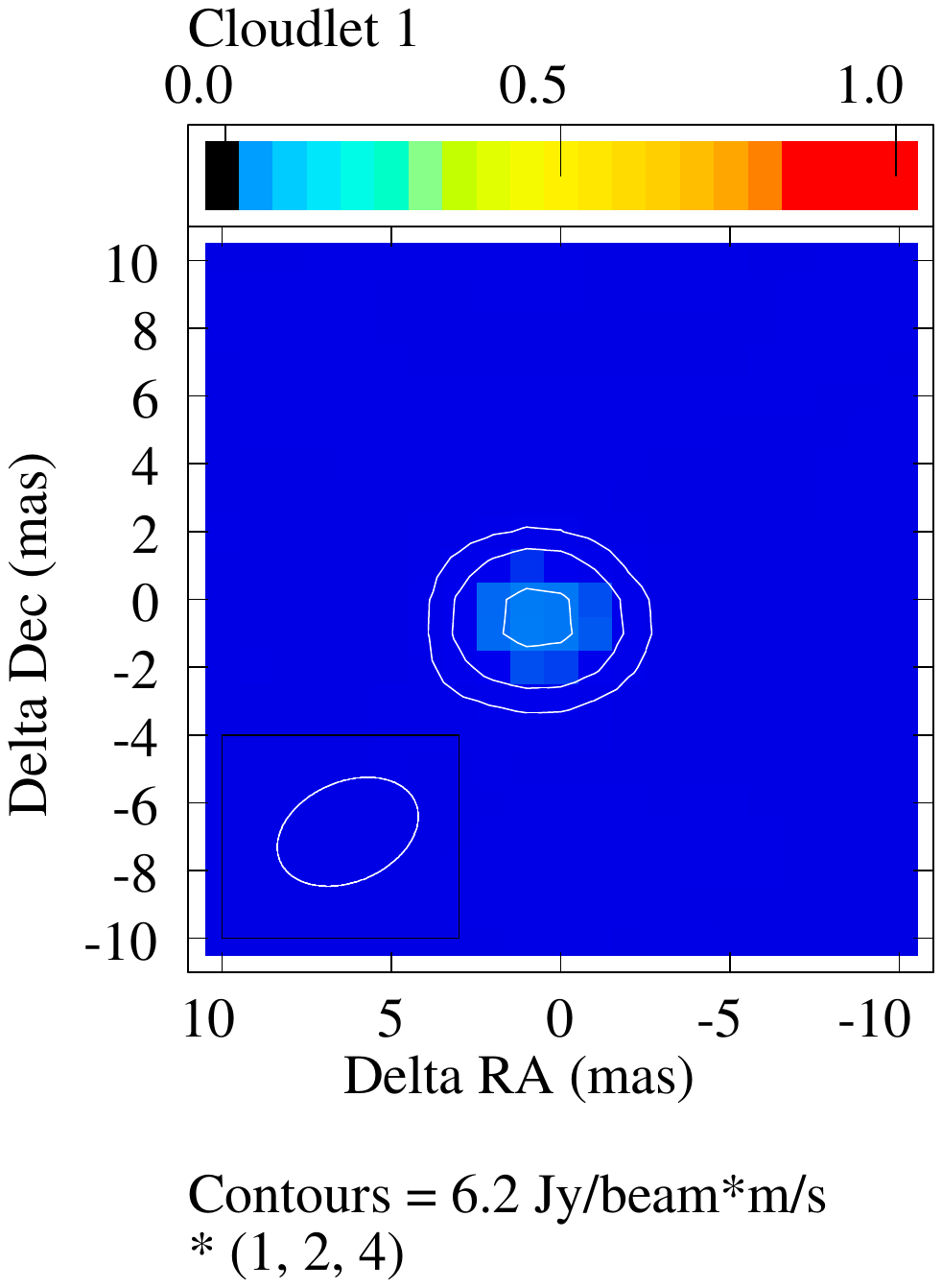}
\includegraphics[scale=0.23,trim=120 100 0 200]{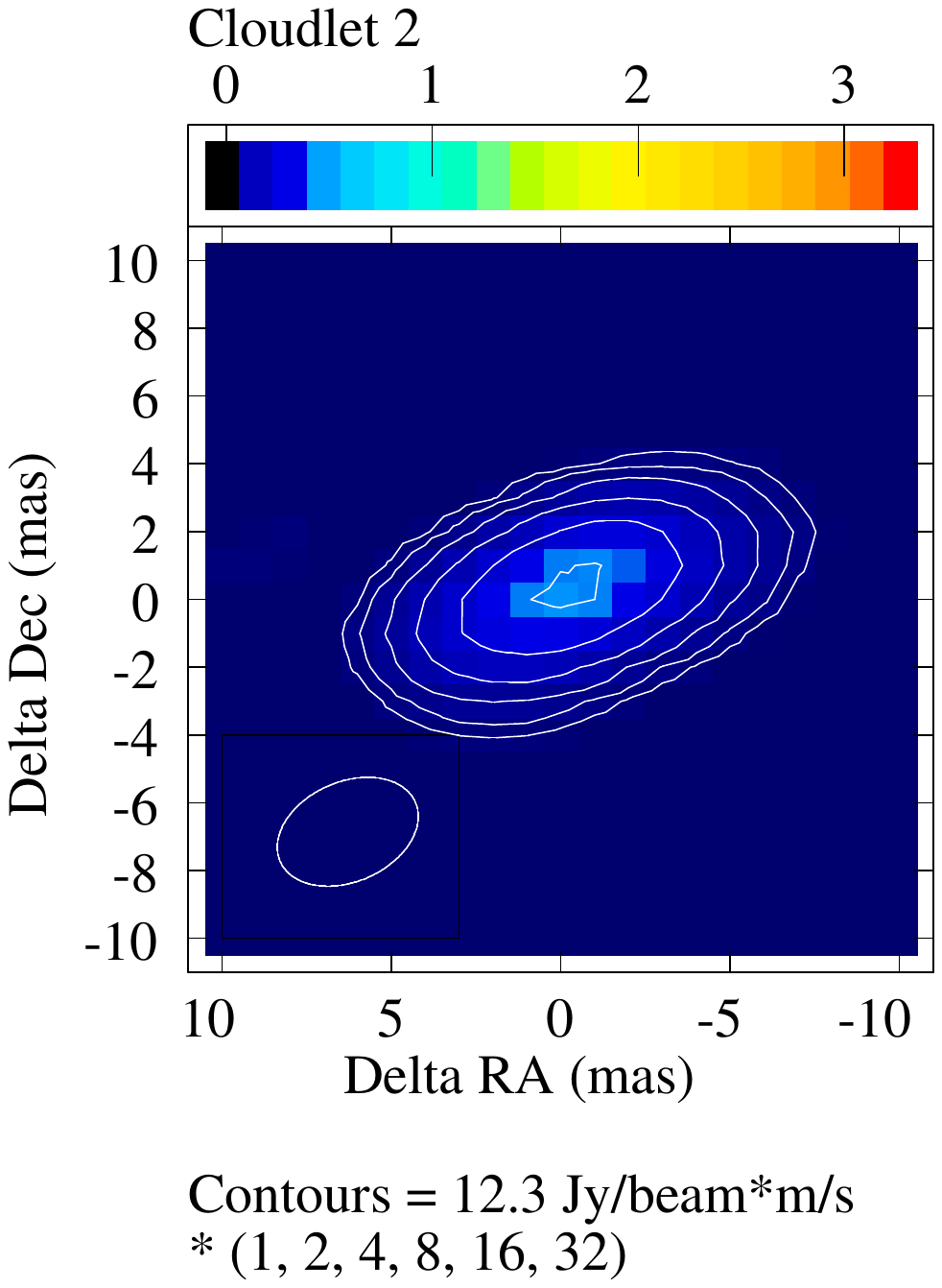}
\includegraphics[scale=0.23,trim=120 100 0 200]{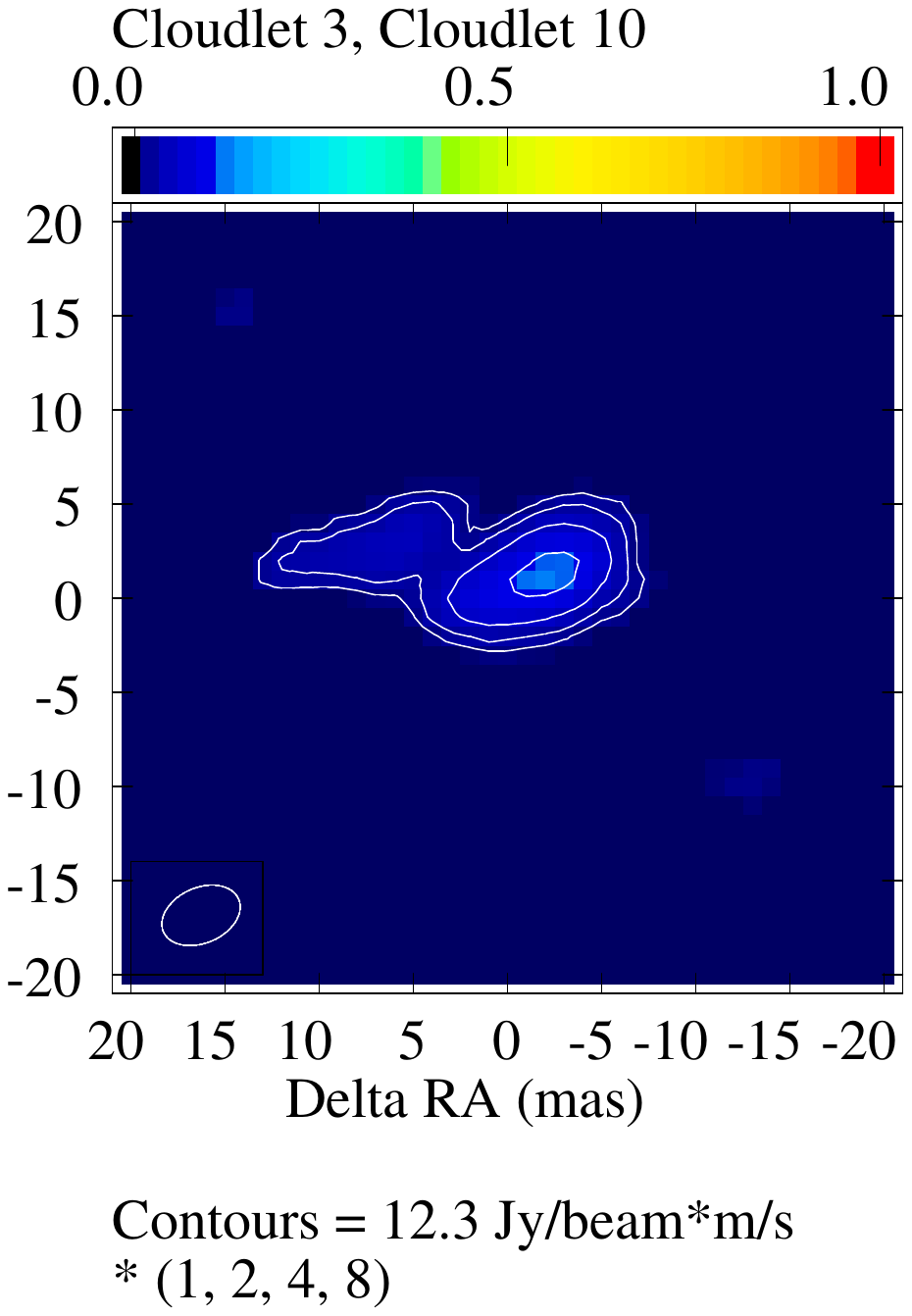}
\includegraphics[scale=0.23,trim=120 100 0 200]{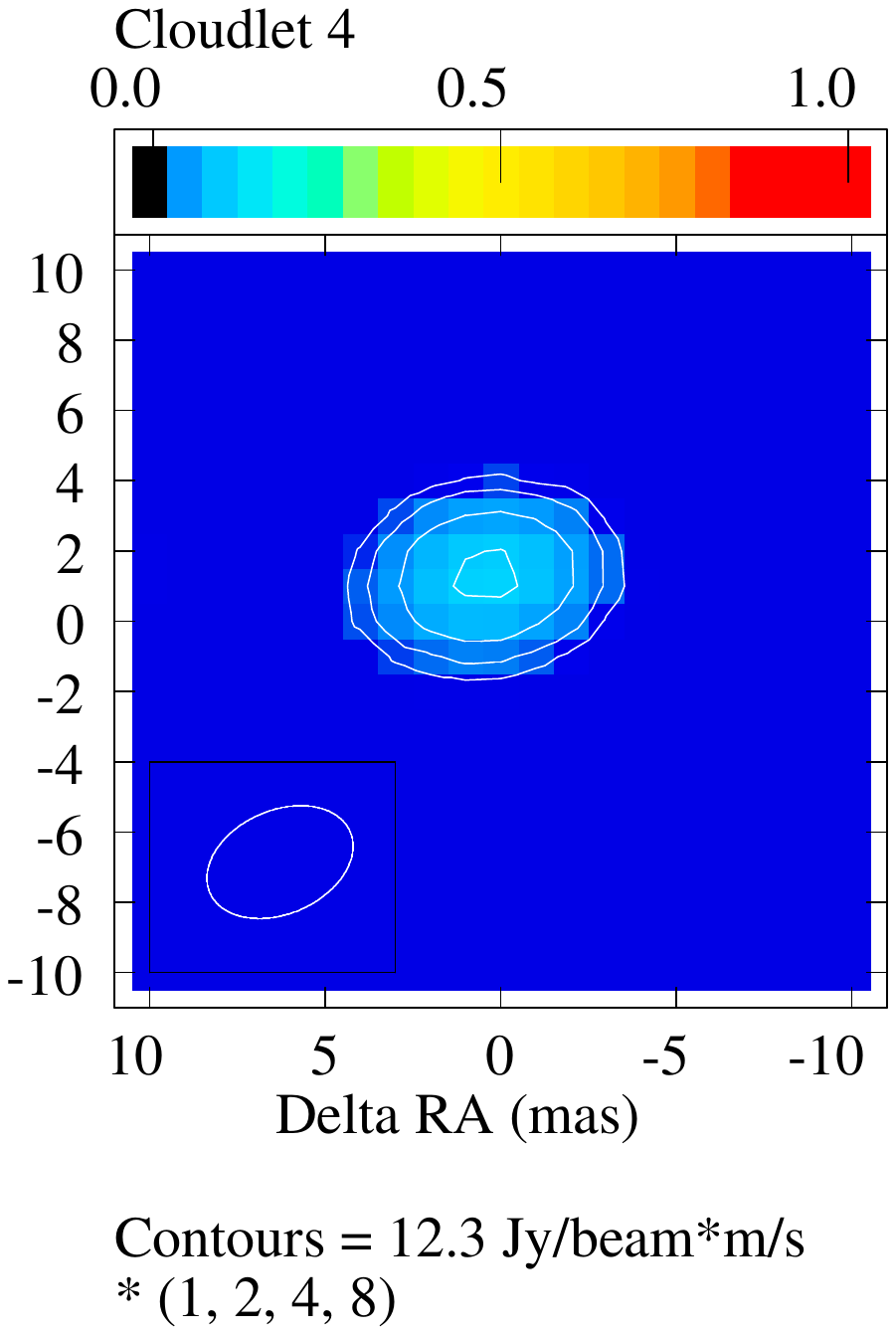}
\includegraphics[scale=0.23,trim=120 100 0 200]{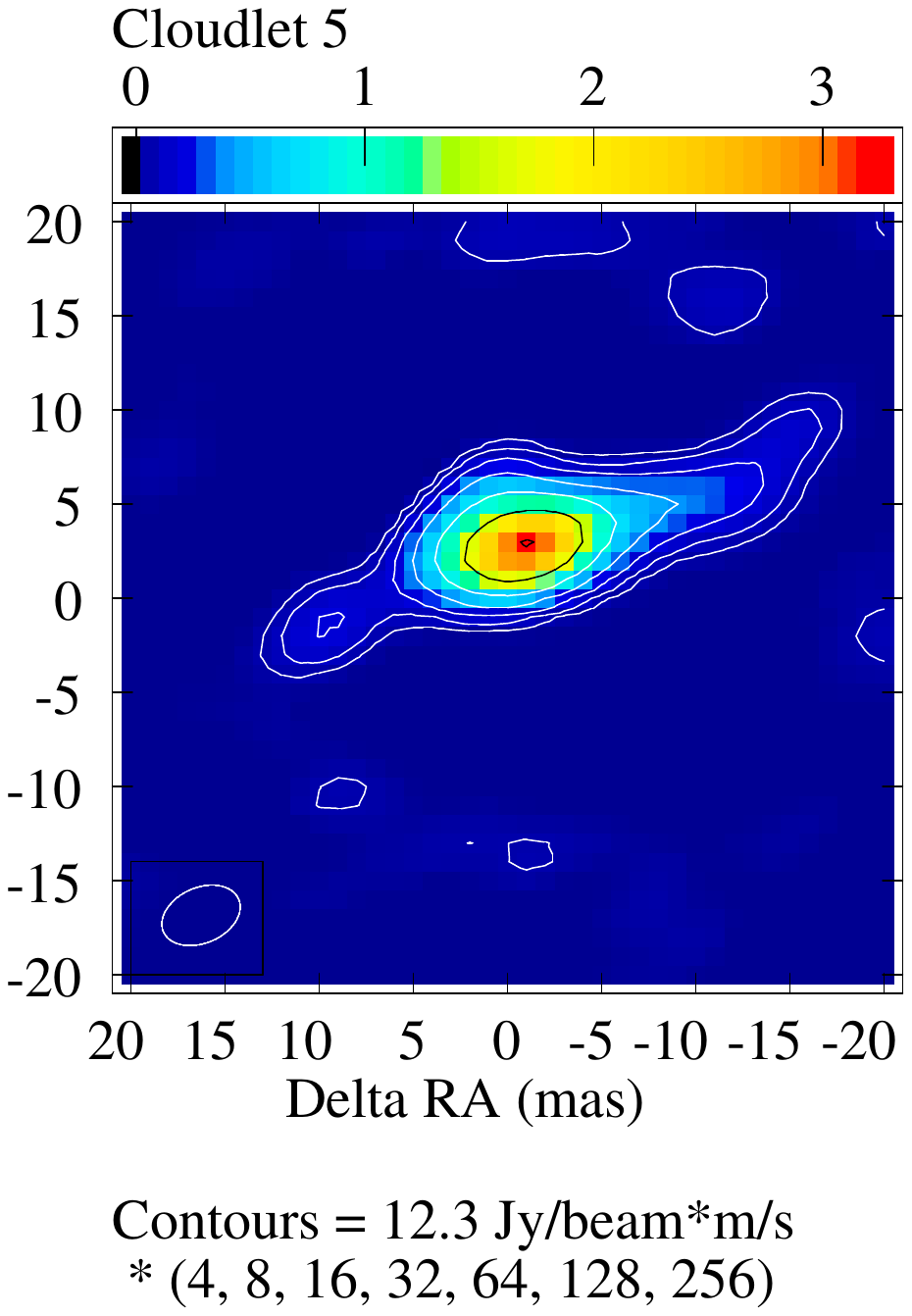}
\includegraphics[scale=0.23,trim=120 100 0 100]{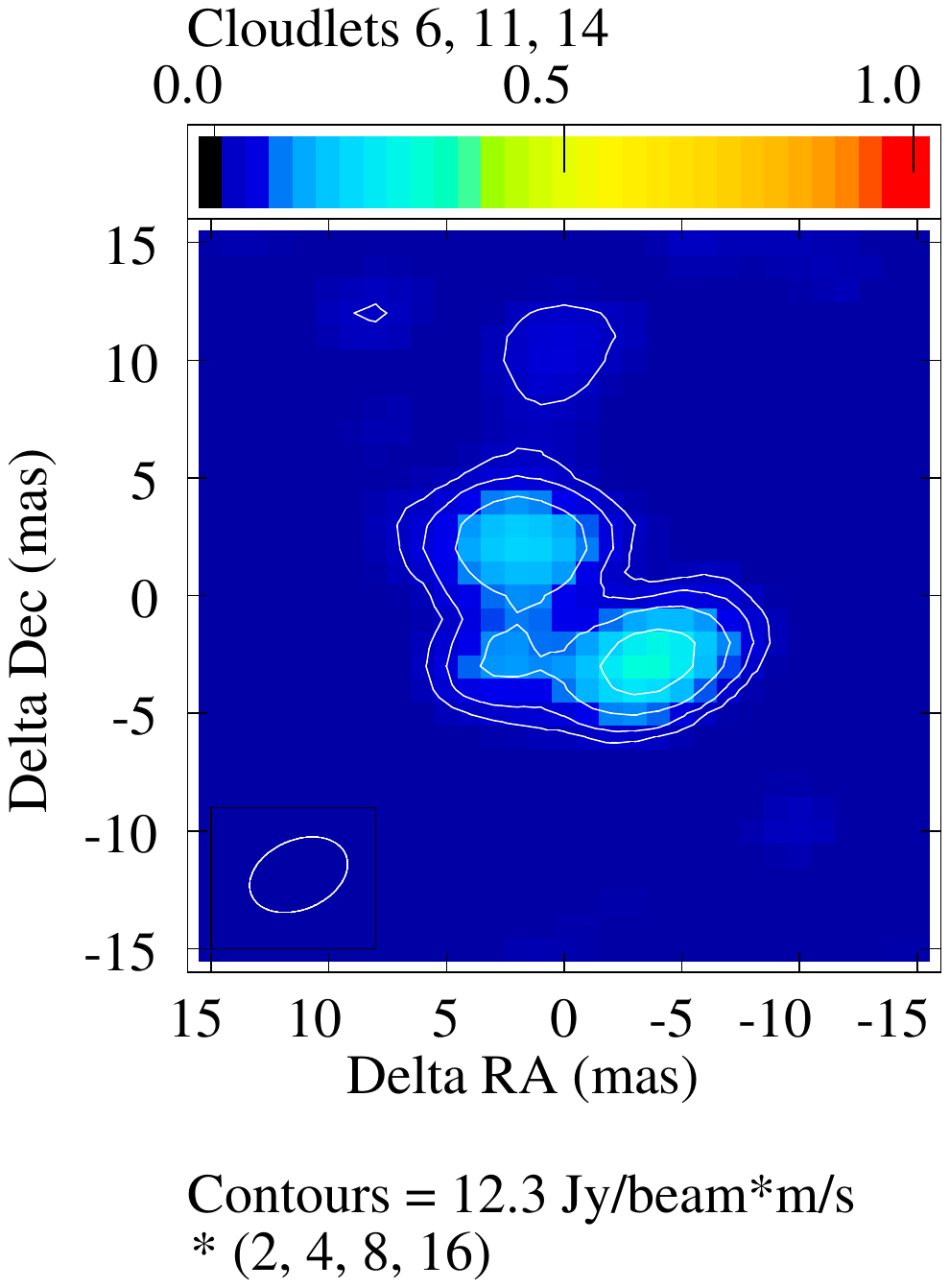}
\includegraphics[scale=0.23,trim=120 100 0 100]{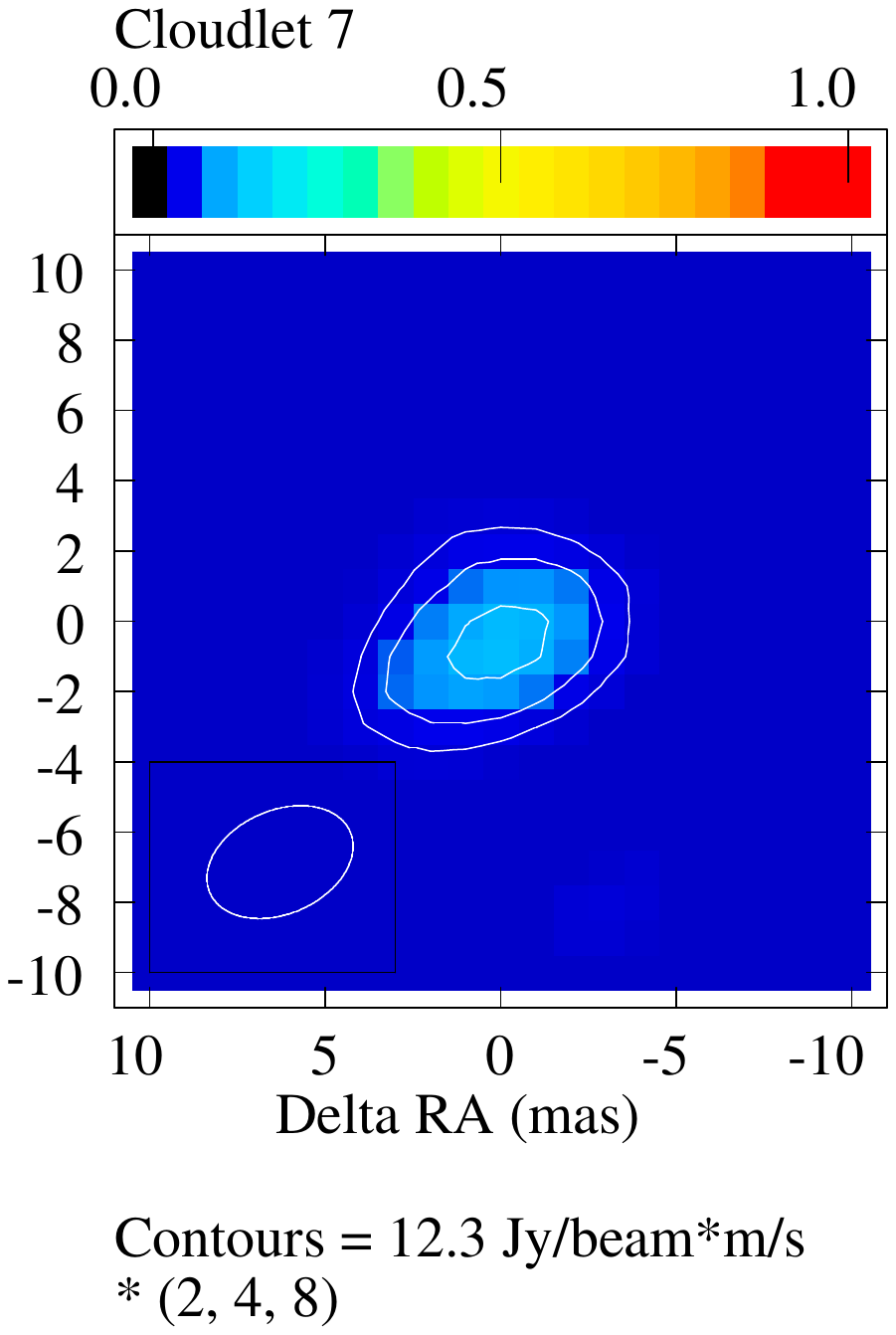}
\includegraphics[scale=0.23,trim=120 100 0 100]{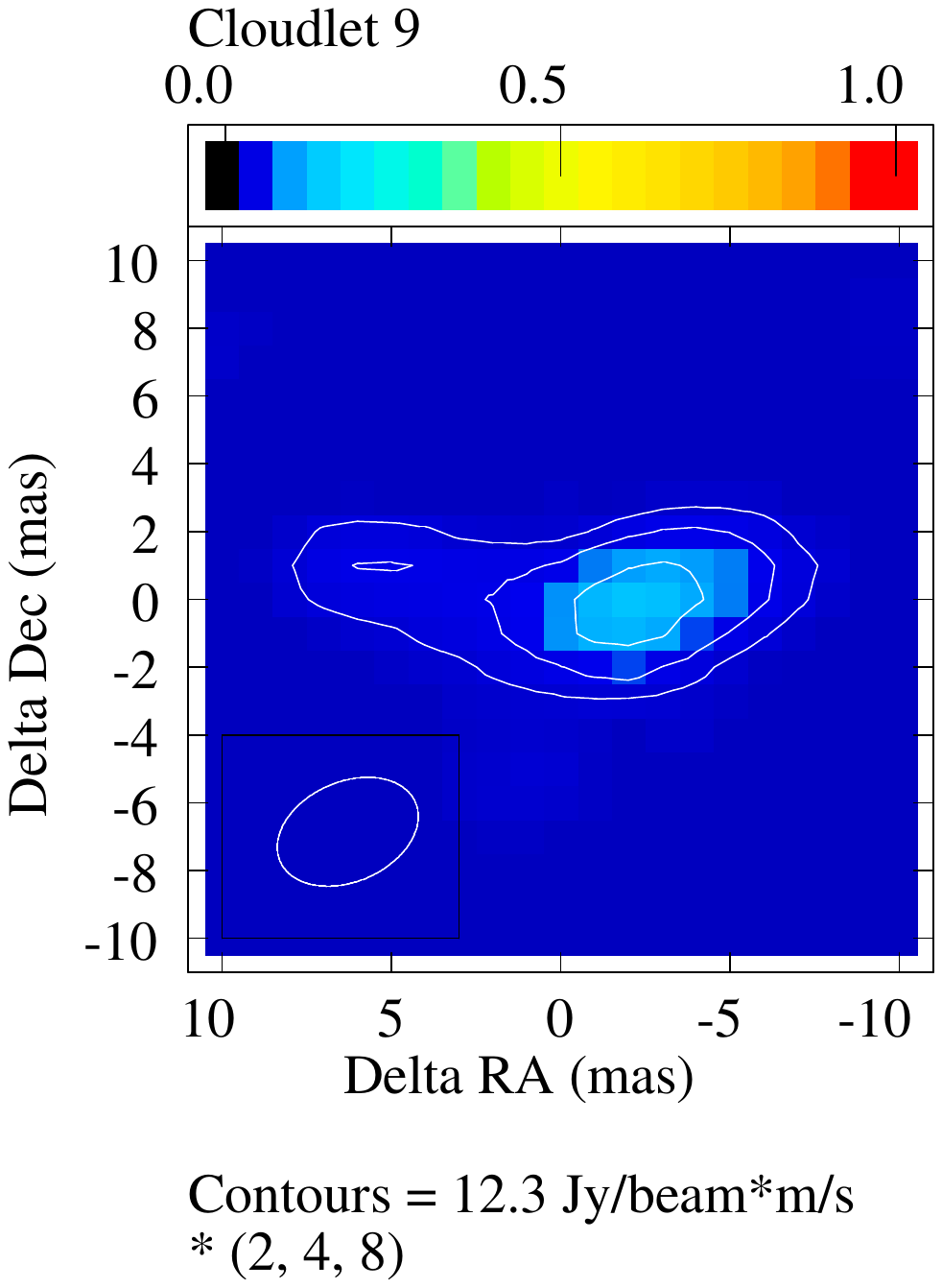}
\caption{Total intensity (zeroth moment) maps of cloudlets in G78.122$+$3.633 as observed in 2019. The (0,0) point corresponds to a centre of a given cloudlet. As indicated in the top wedges, the colour scale varies linearly in Jy~beam$^{-1}\times $~km~s$^{-1}$. The ellipses in the bottom left-hand corners of the images indicate the beam. Note, the panels are in different sizes to present the best resolution and contain the whole emission of each cloudlet or a group of cloudlets.
%these sources if possible since they will more clearly show the true structure of the maser emission and kinematics of the gas in the cloudlets.  
}
\label{fig:zerothmomnt_G78} 
\end{figure*}

\begin{figure*}
\centering
\includegraphics[width=\textwidth]{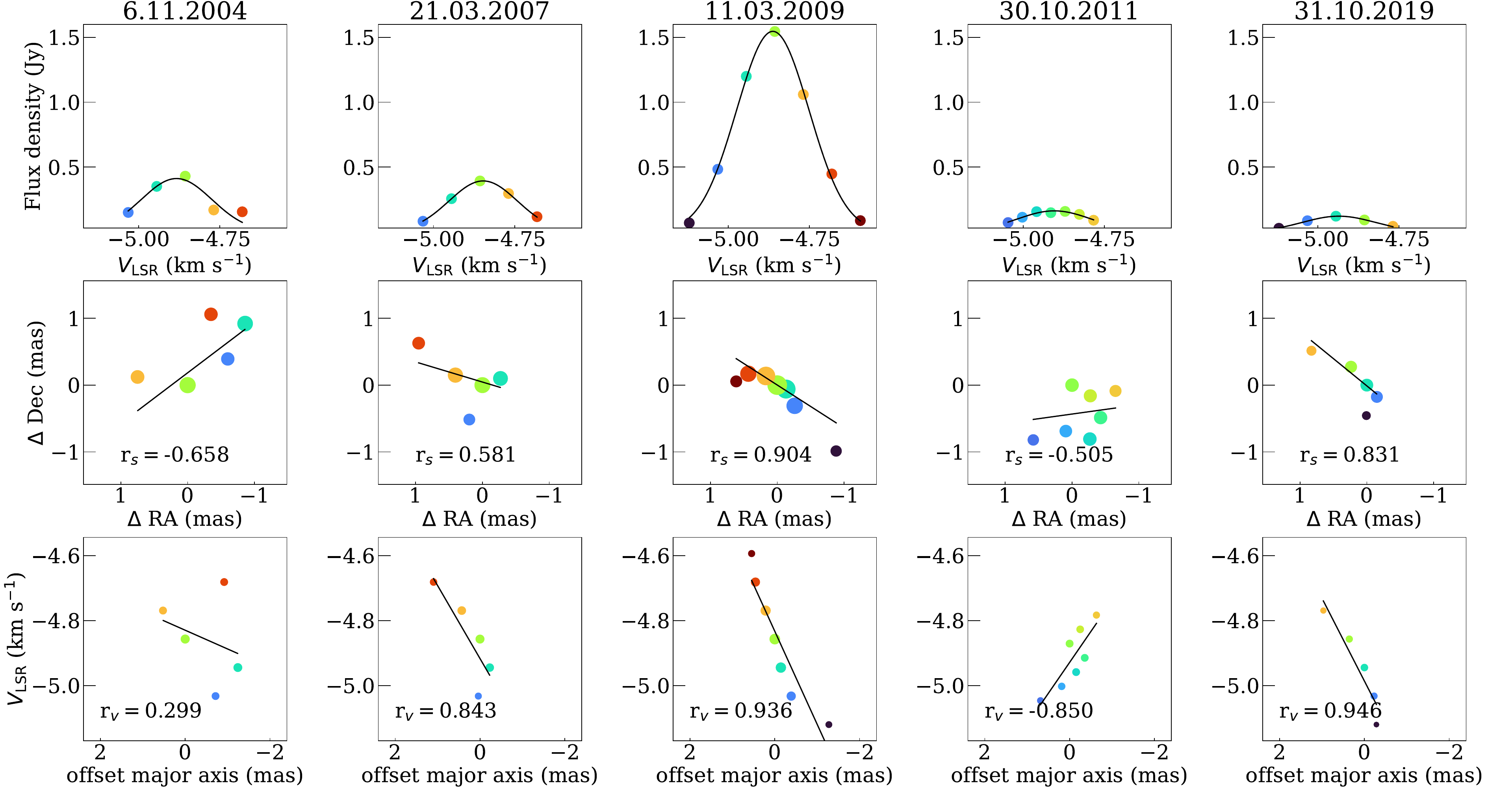}
\caption{{\it Top:} Spectra of Cloudlet~1 as in Table~\ref{total} with a Gaussian velocity profile for five epochs. {\it Middle:} The spot distribution with the major-axis fit. A circle size is proportional to the logarithm of spot's flux density. The (0,0) point corresponds to spot position with max flux density at given epoch. {\it Bottom:} The spot LSR velocity vs position offset along the major axes. The correlation coefficients r$_s$, r$_v$ are listed.}
\label{Cloudlet_1}  
\end{figure*}

\begin{figure*}
\centering
\includegraphics[width=\textwidth]{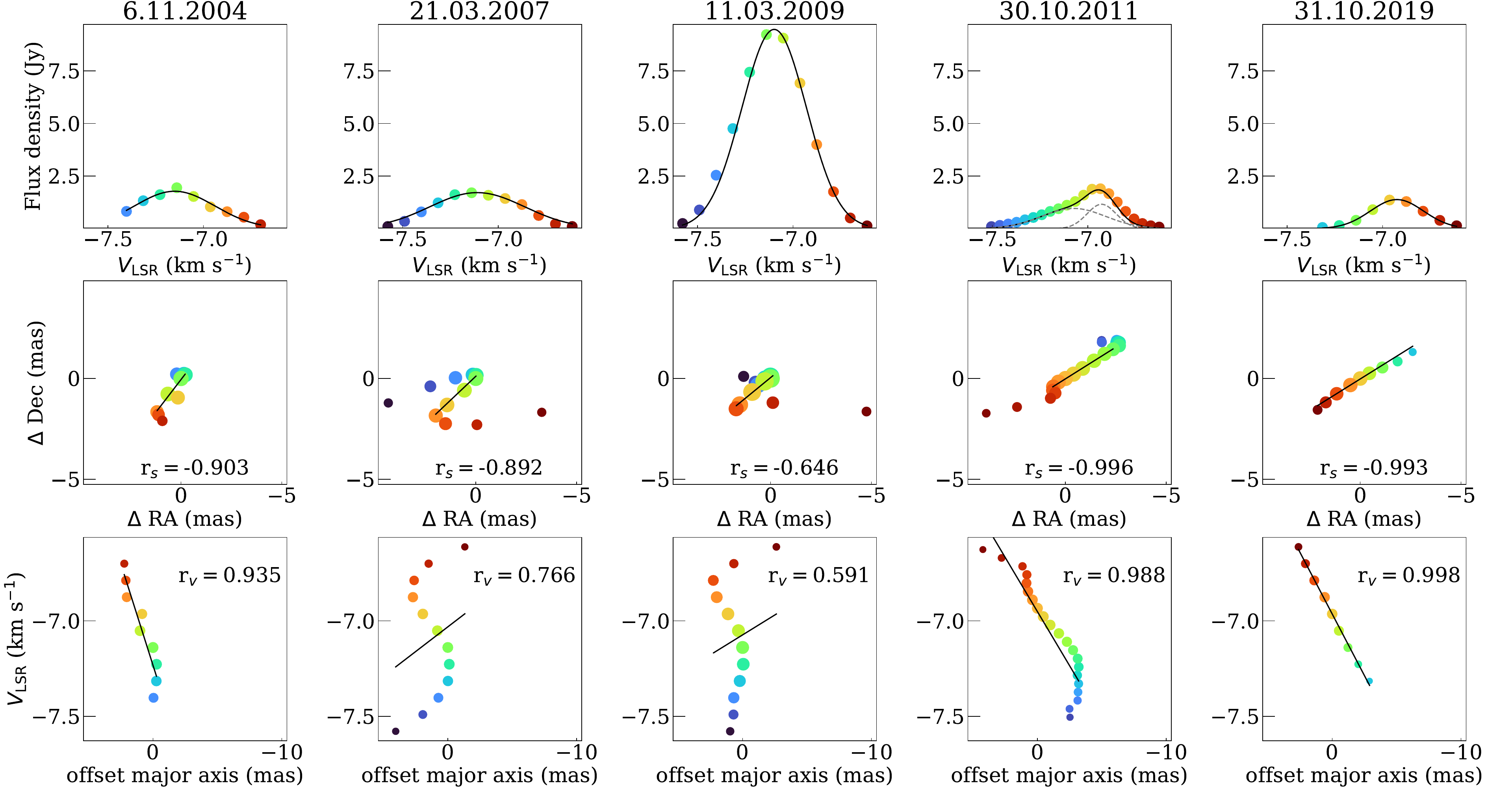}
\caption{The same as Fig.~\ref{Cloudlet_1} but for the Cloudlet~2.}
\label{Cloudlet_2}  
\end{figure*}

\begin{figure*}
\centering
\includegraphics[width=\textwidth]{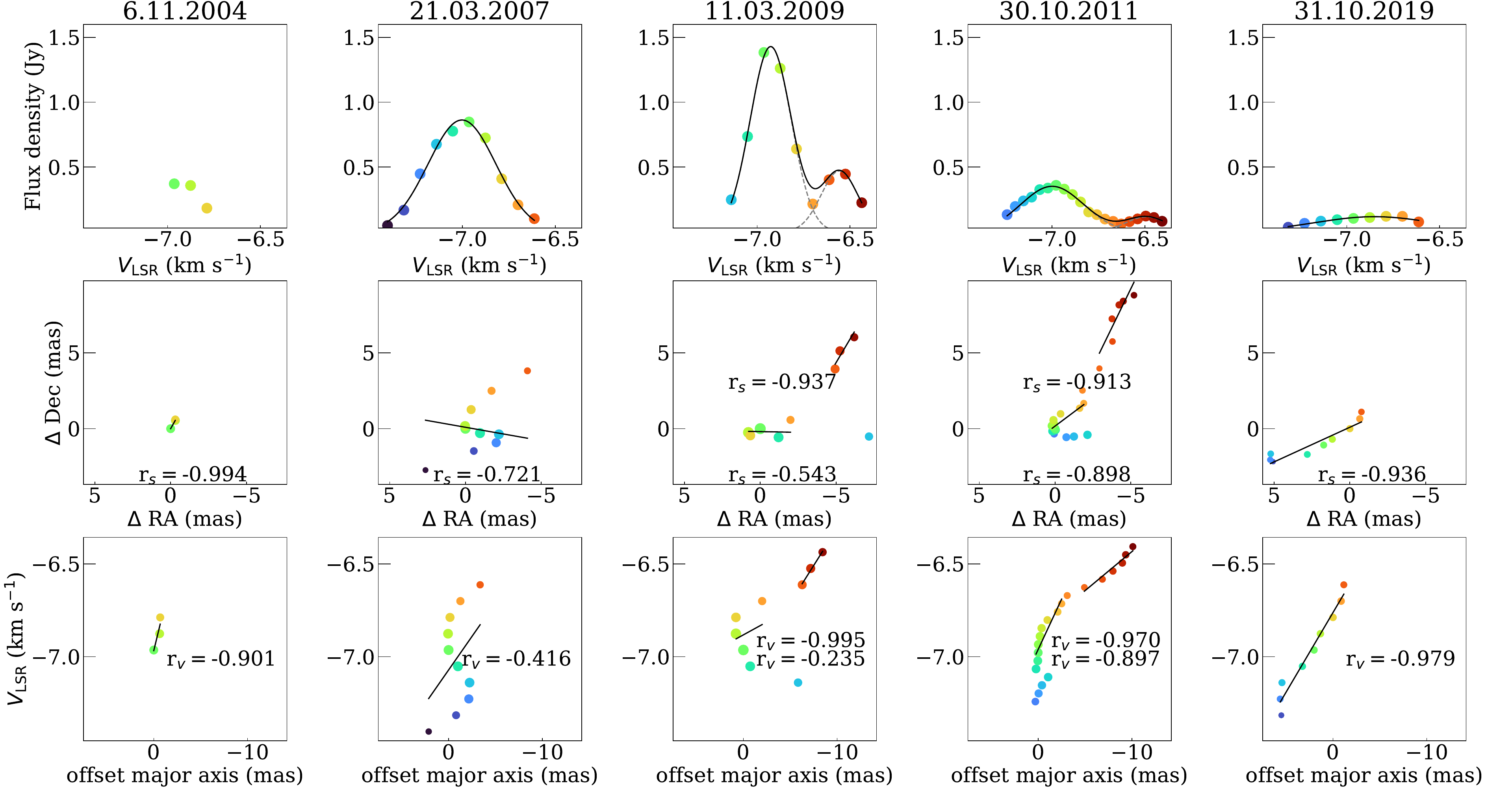}
\caption{The same as Fig.~\ref{Cloudlet_1} but for the Cloudlet~3.}
\label{Cloudlet_3}  
\end{figure*}

\begin{figure*}
\centering
\includegraphics[width=\textwidth]{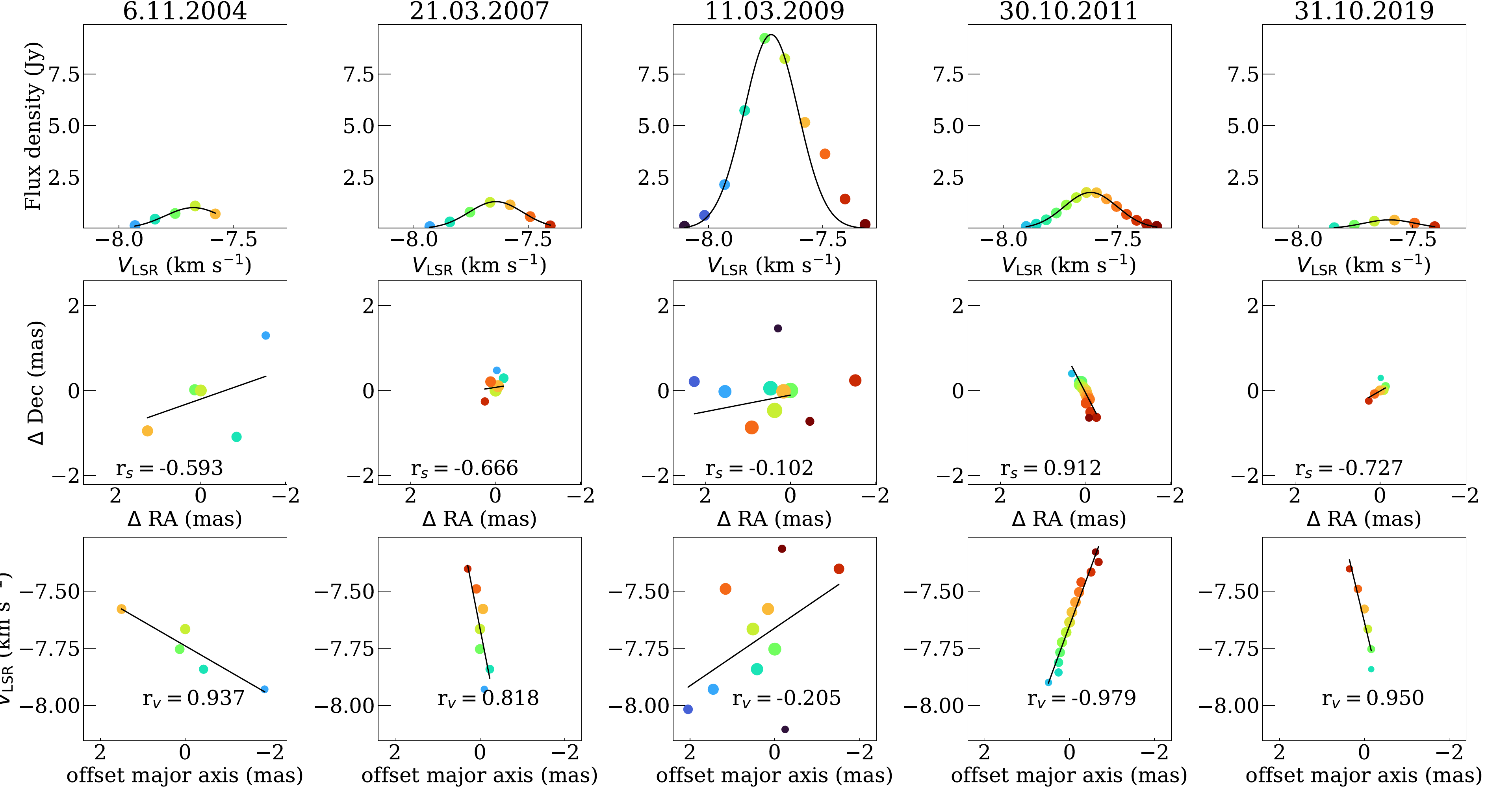}
\caption{The same as Fig.~\ref{Cloudlet_1} but for the Cloudlet~4.}
\label{Cloudlet_4}  
\end{figure*}

\begin{figure*}
\centering
\includegraphics[width=\textwidth]{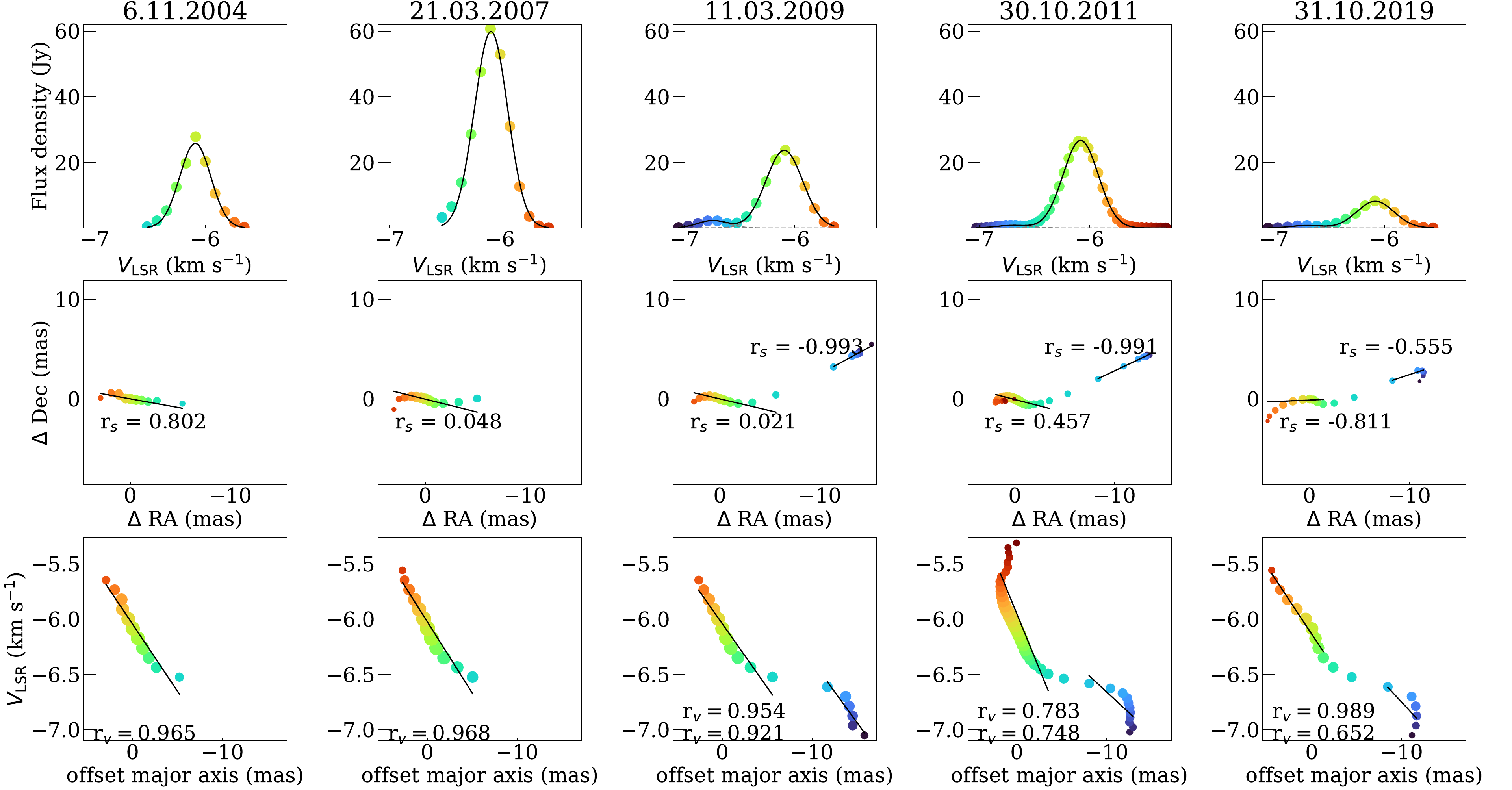}
\caption{The same as Fig.~\ref{Cloudlet_1} but for the Cloudlet~5.}
\label{Cloudlet_5} 
\end{figure*}

\begin{figure*}
\includegraphics[width=\textwidth]{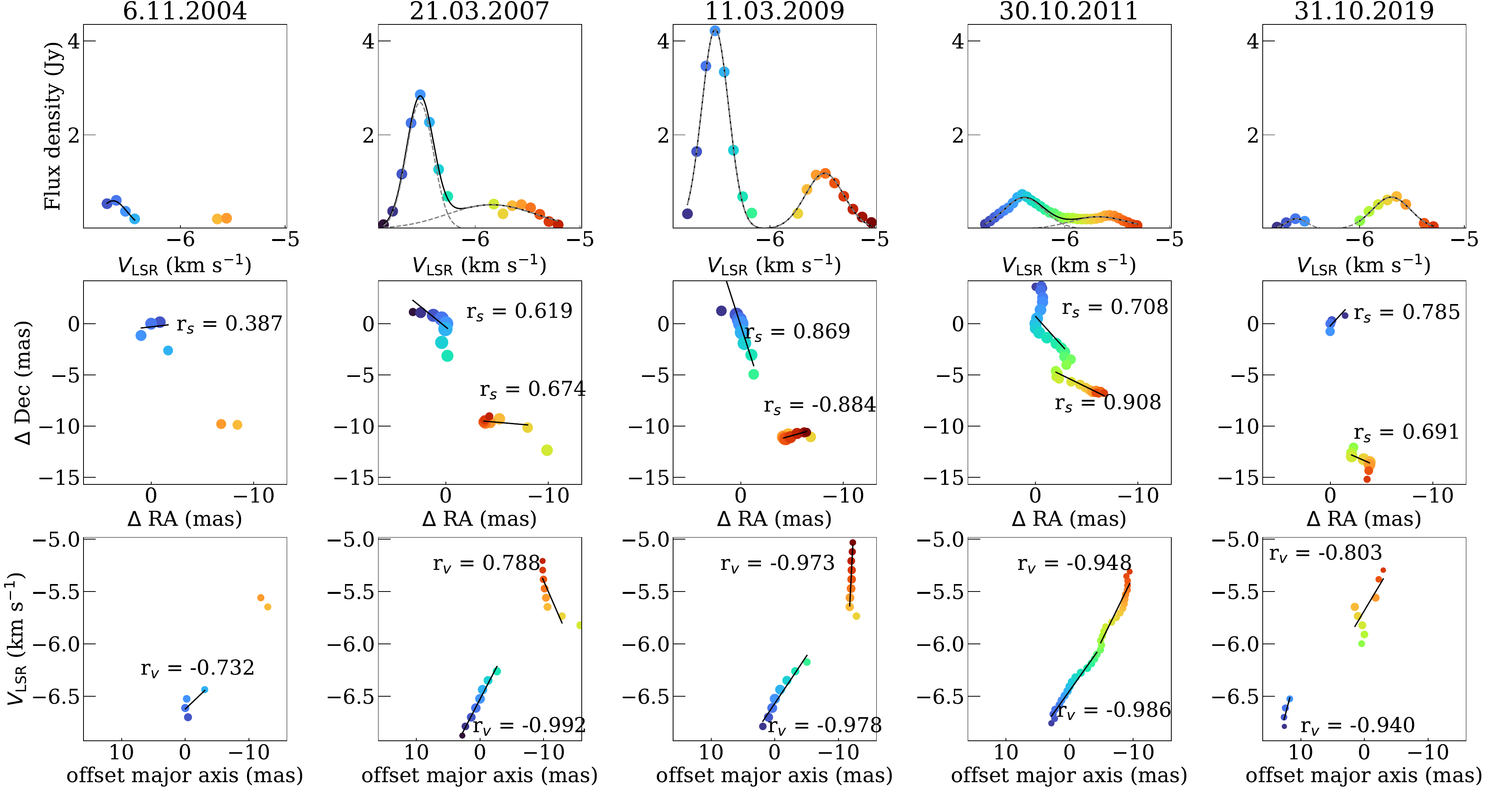}
\caption{The same as Fig.~\ref{Cloudlet_1} but for the Cloudlet~6. }
\label{Cloudlet_6}  
\end{figure*}

\begin{figure*}
\includegraphics[width=\textwidth]{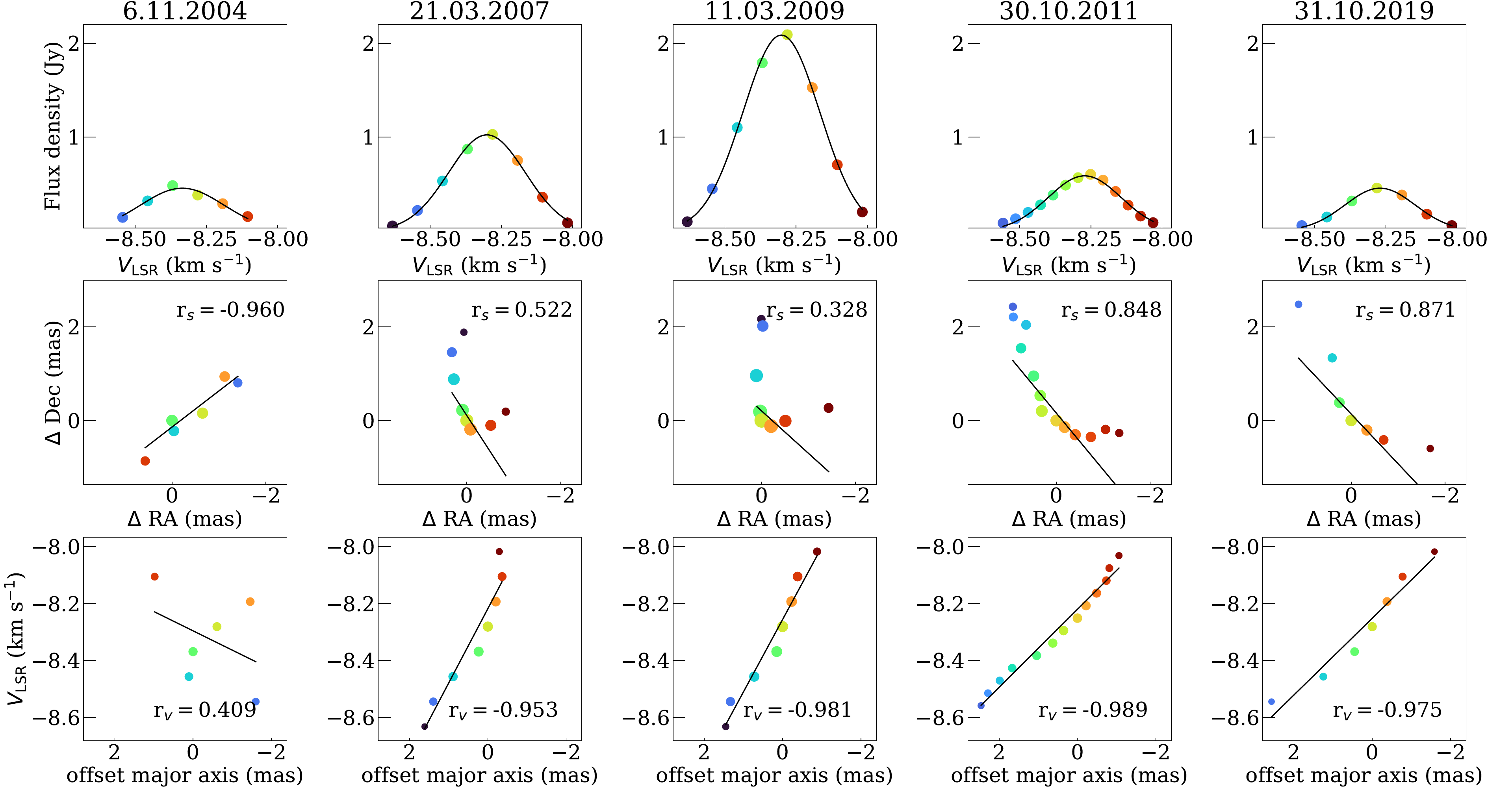}
\caption{The same as Fig.~\ref{Cloudlet_1} but for the Cloudlet~7.}
\label{Cloudlet_7}  
\end{figure*}

\begin{figure*}
\includegraphics[scale=0.75]{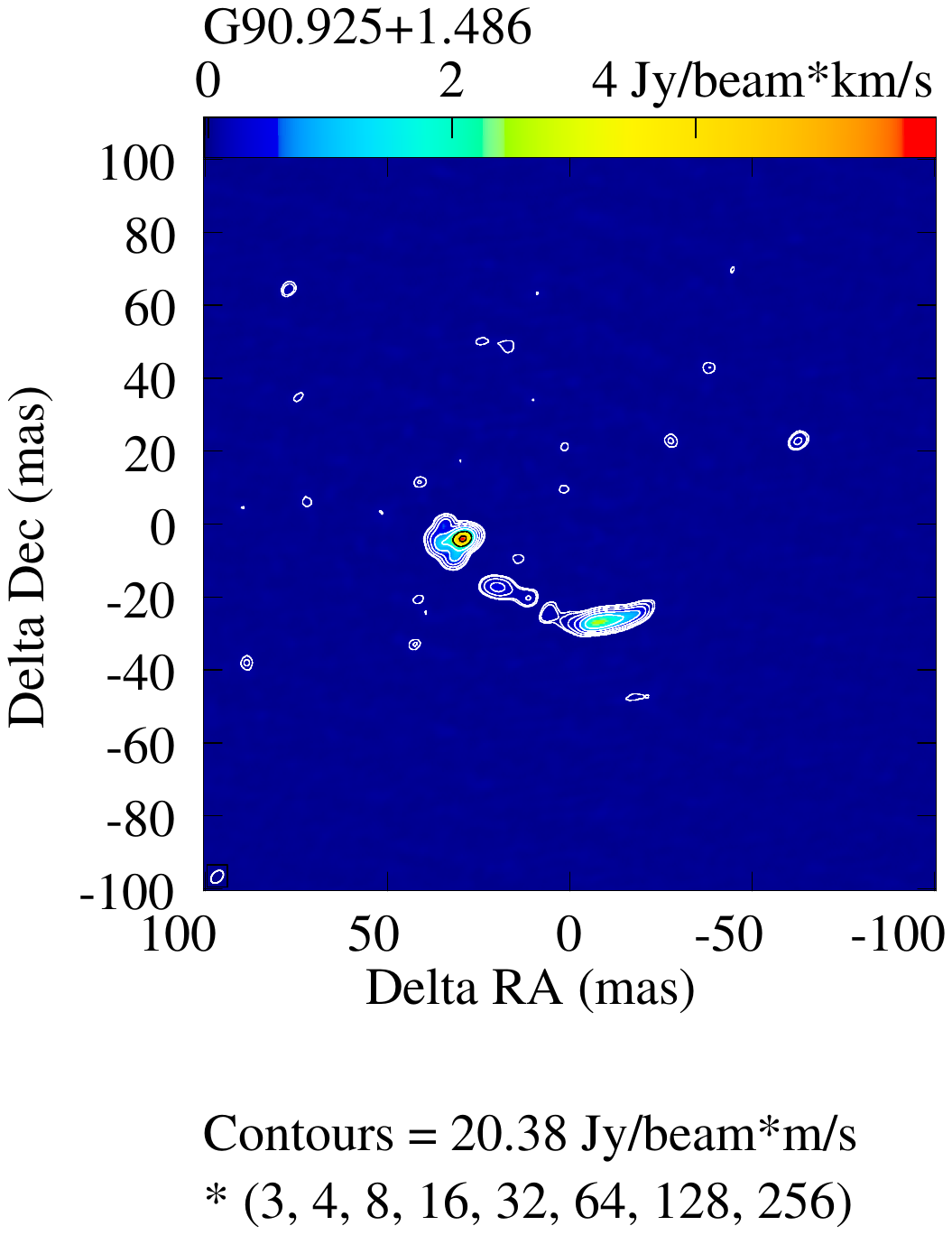}
\caption{ Total intensity (zeroth moment) map of G90. The ellipse in the bottom left-hand corner of the images indicates the beam.}
\label{fig:g90momnt}  
\end{figure*}

\begin{figure*}
\centering
\includegraphics[scale=0.13]{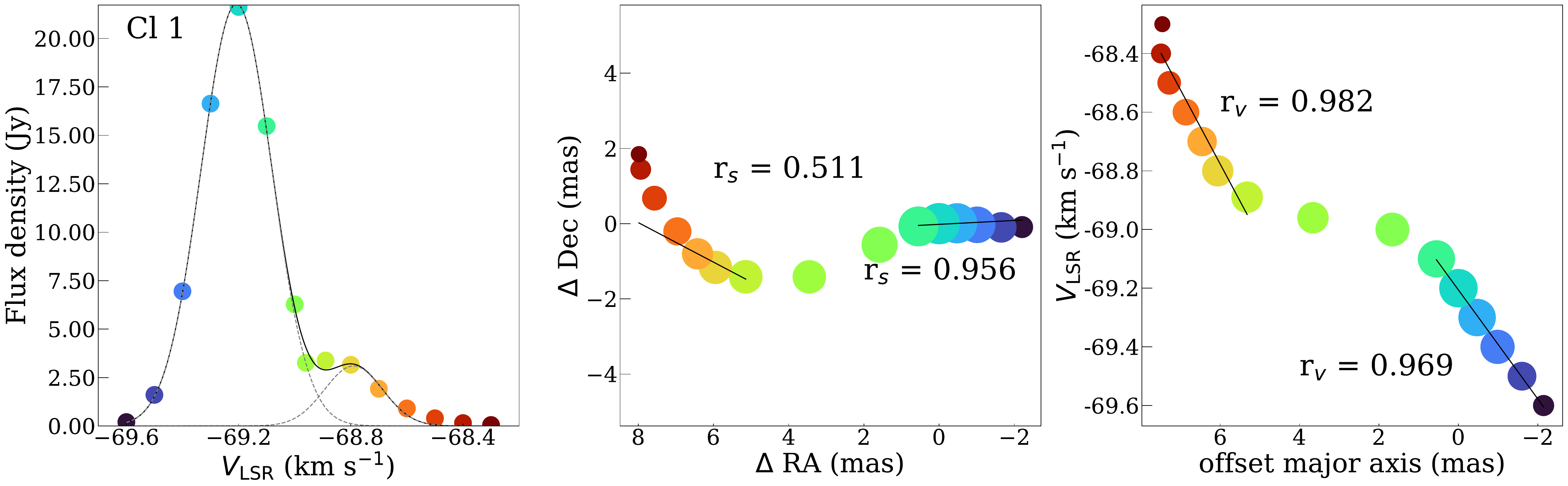}
\includegraphics[scale=0.13]{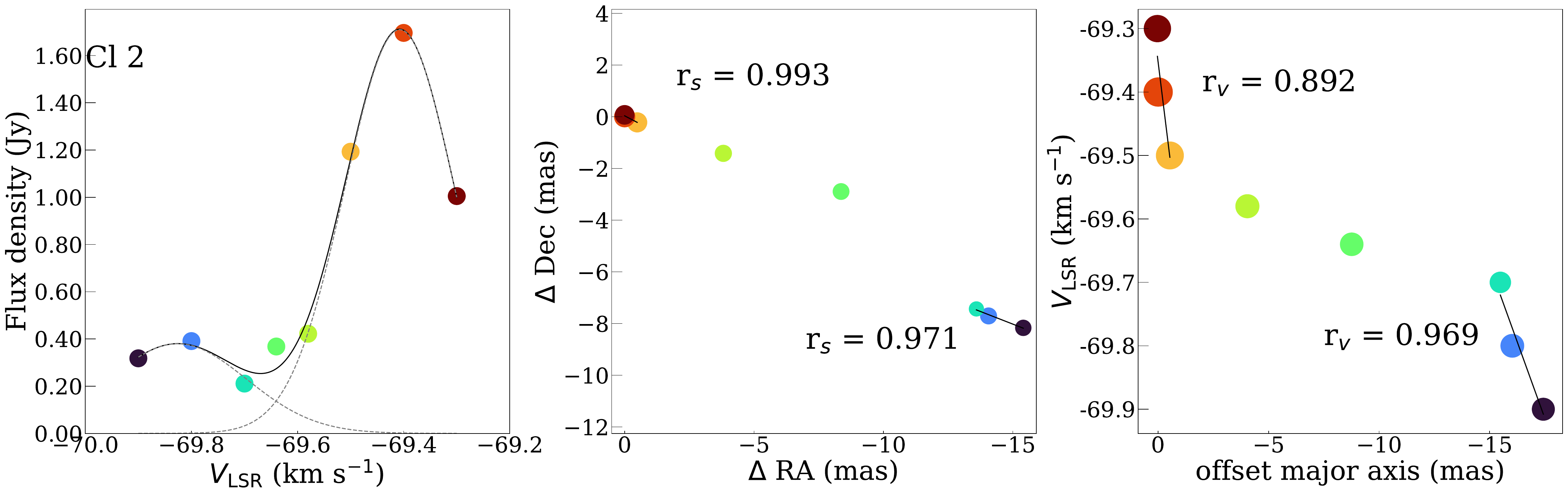}
\includegraphics[scale=0.13]{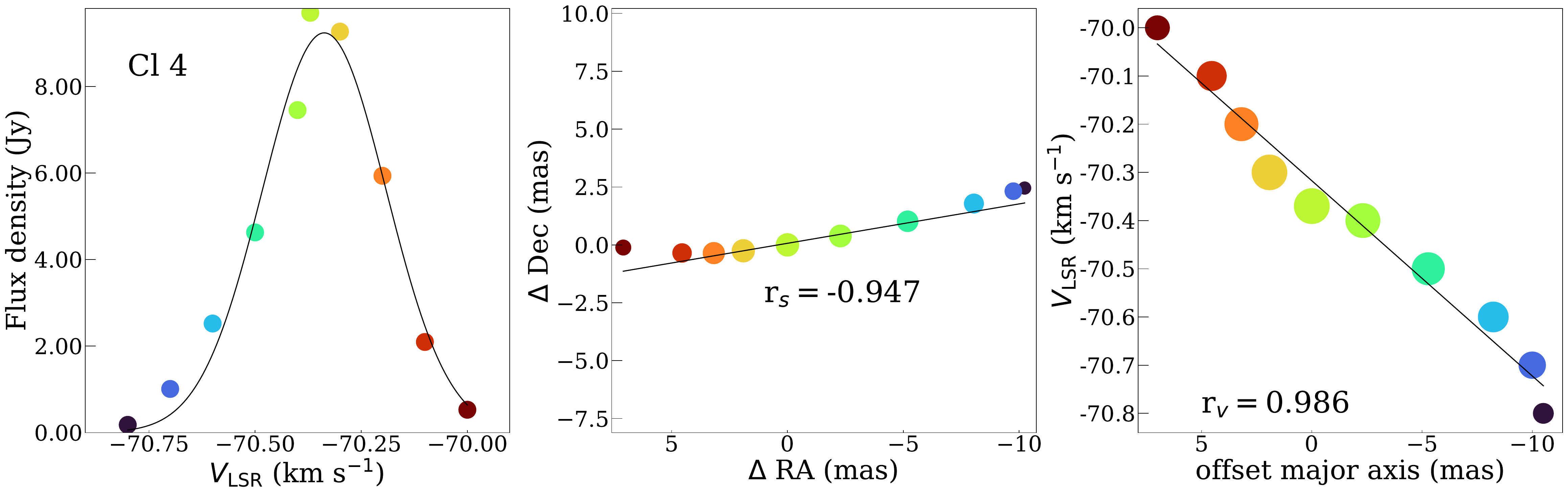}
\includegraphics[scale=0.13]{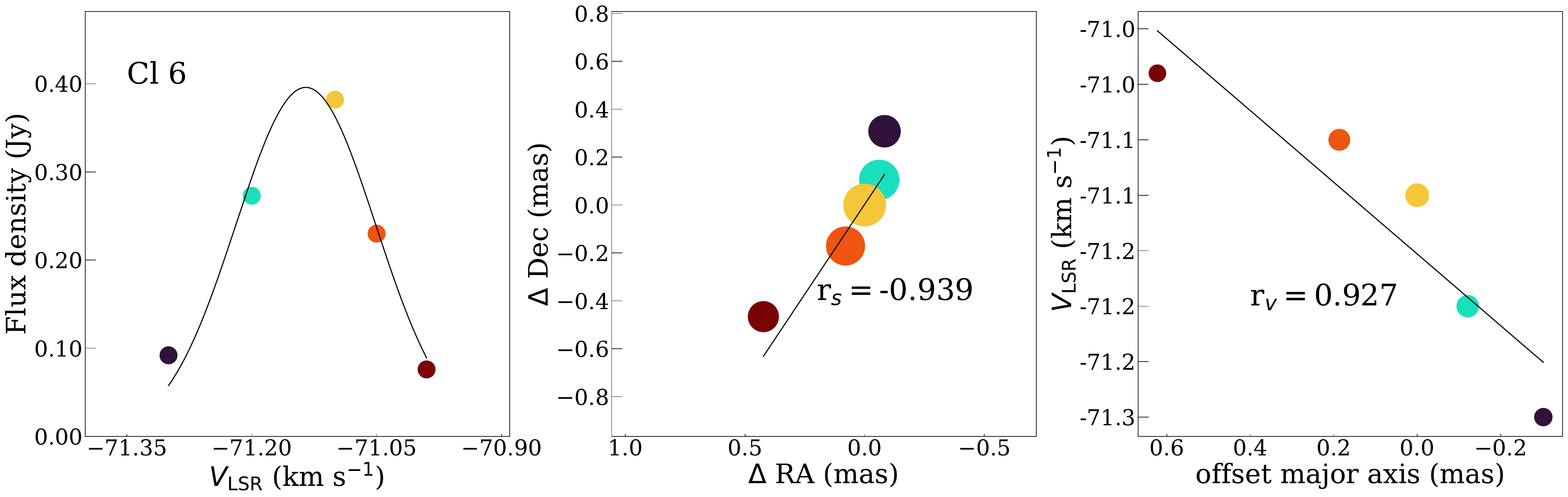}
\caption{Cloudlets of G90.925$+$1.486, as obtained in the EVN observation in 2019. The same information as in Fig.~\ref{Cloudlet_1} (but panels are orientated horizontally now). We show cloudlets 1, 2, 4, and 6, since  3, 5 and 7 structures are very simple.
}
\label{G90_cloudlets}  
\end{figure*}

\begin{figure*}
\centering
\includegraphics[scale=0.13]{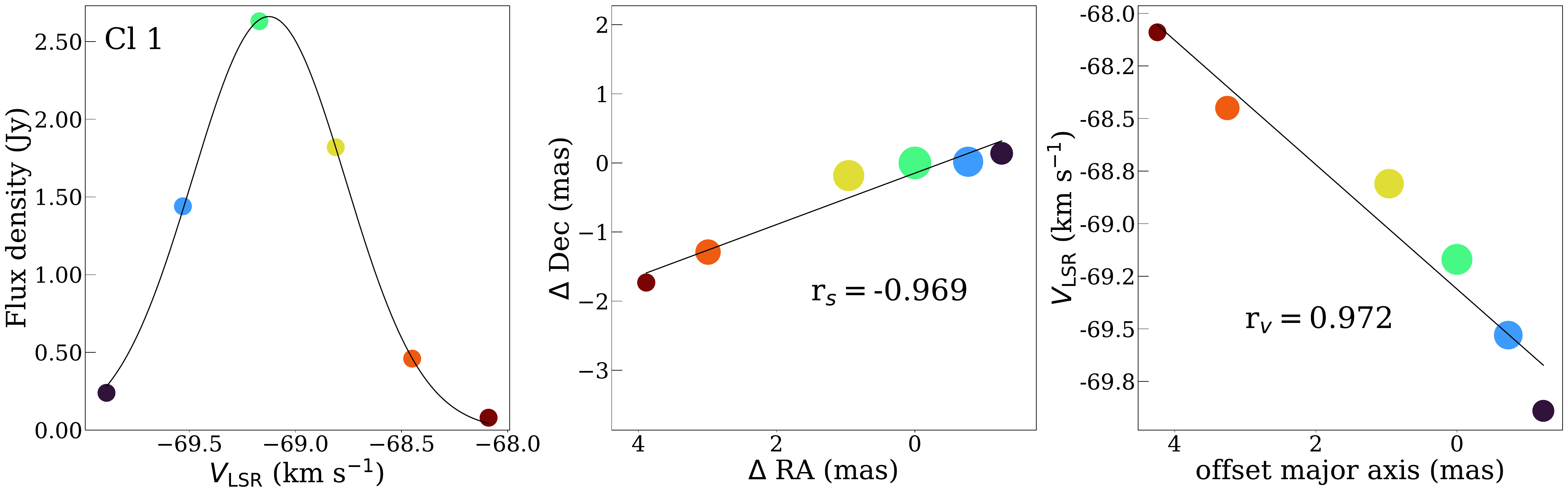}
\includegraphics[scale=0.13]{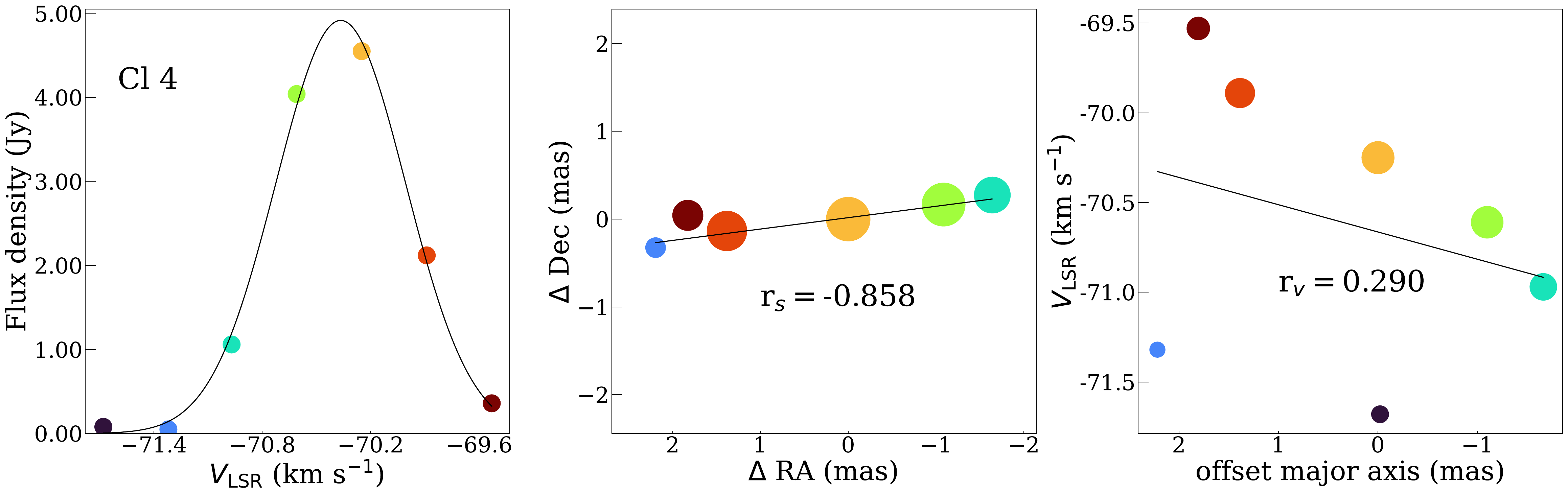}
\caption{Cloudlets of G90.925$+$1.486 as obtained by the BeSSeL team. The same information as in Fig.~\ref{G90_cloudlets}. } 
\label{G90_BESSEL}  
\end{figure*}

\begin{figure*}
\includegraphics[scale=0.75]{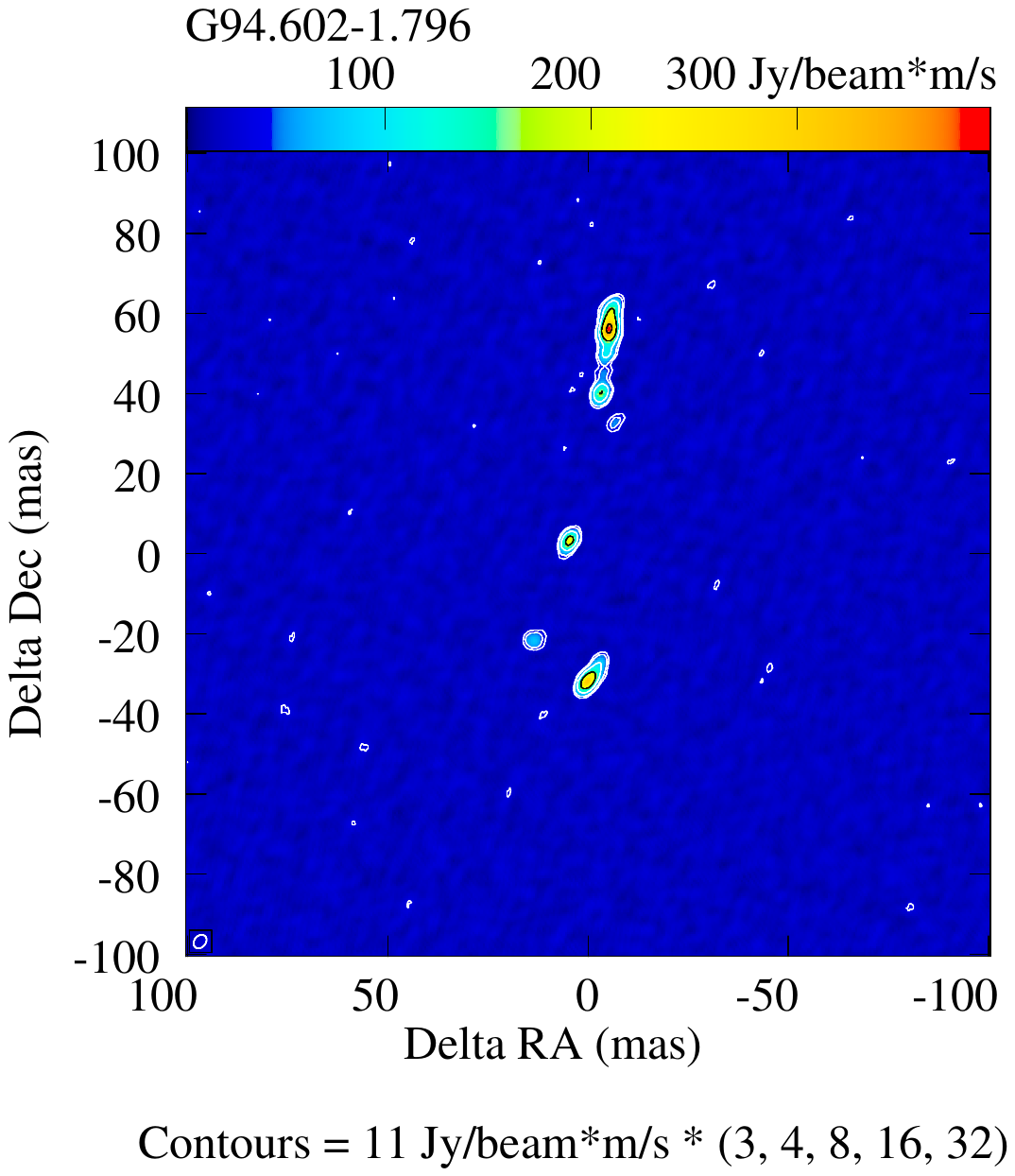}
\caption{ Total intensity (zeroth moment) map of G94. The ellipse in the bottom left-hand corner of the images indicates the beam.}
\label{fig:g94momnt}  
\end{figure*}

\begin{figure*}
\centering
\includegraphics[scale=0.15]{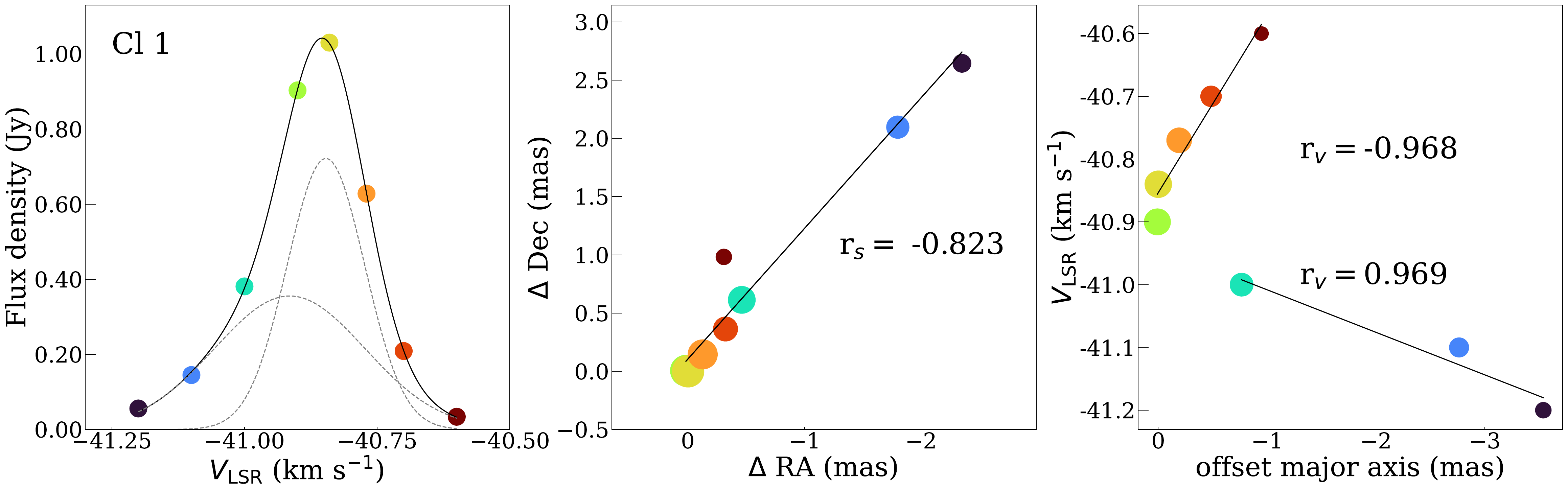}
\includegraphics[scale=0.15]{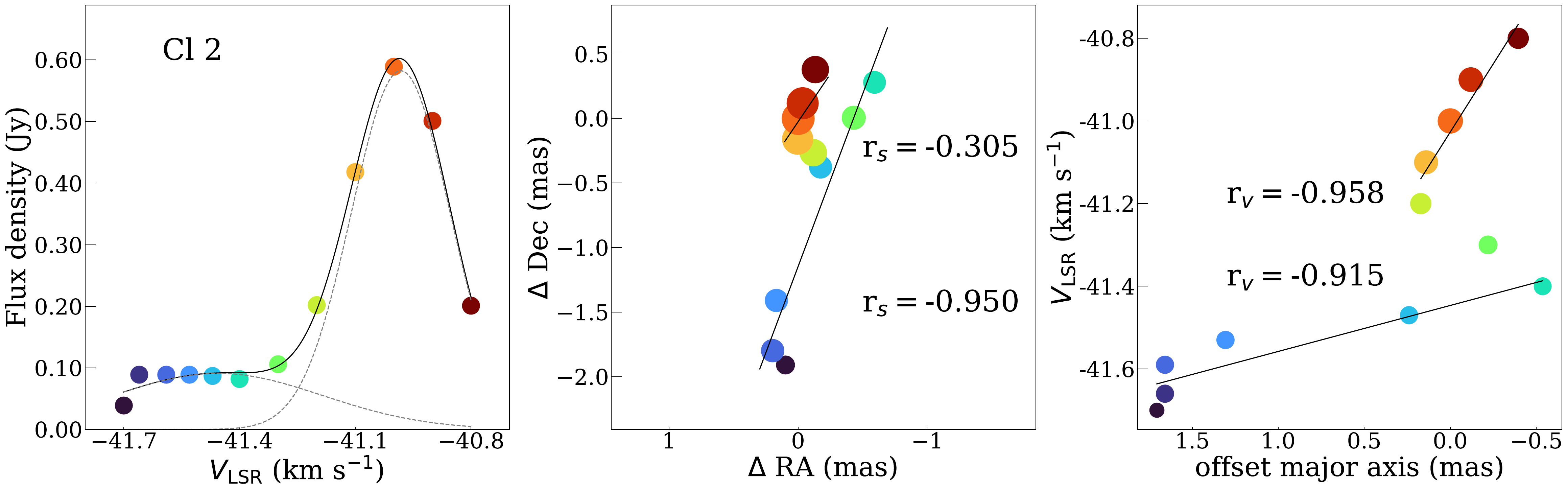}
\includegraphics[scale=0.15]{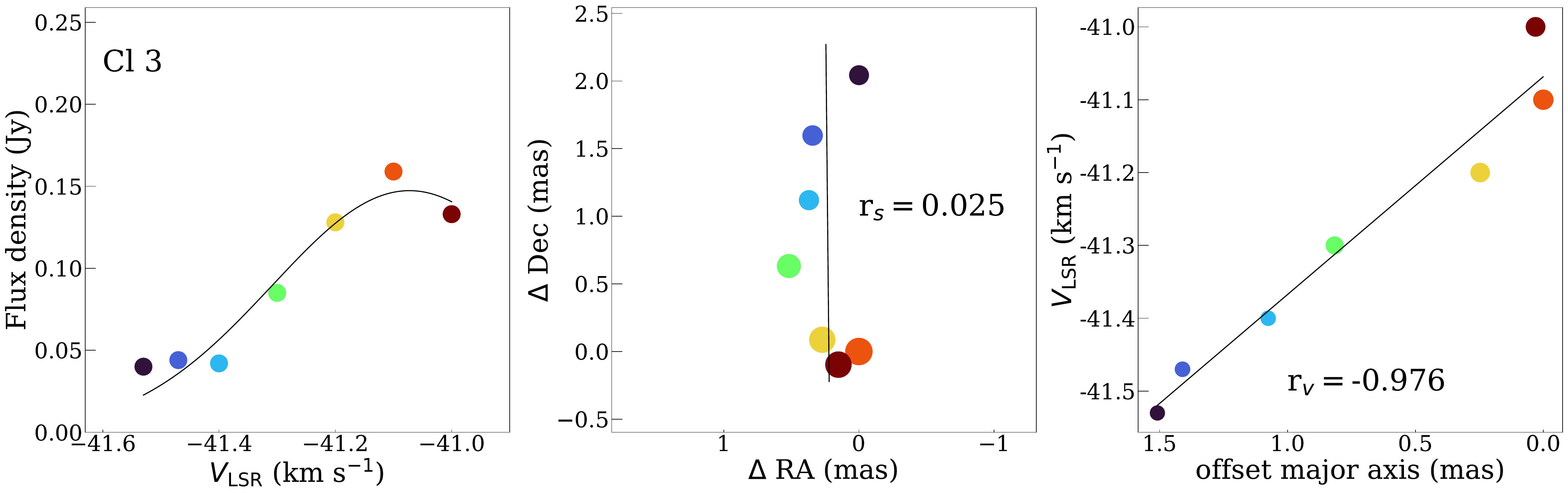}
\includegraphics[scale=0.15]{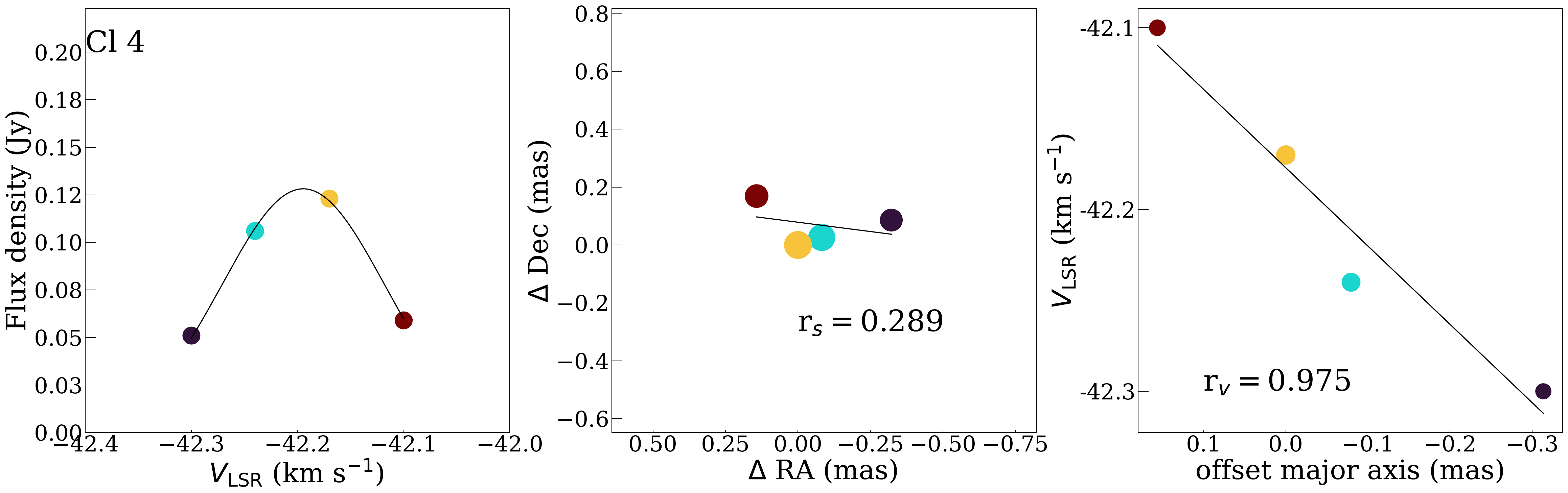}
\includegraphics[scale=0.15]{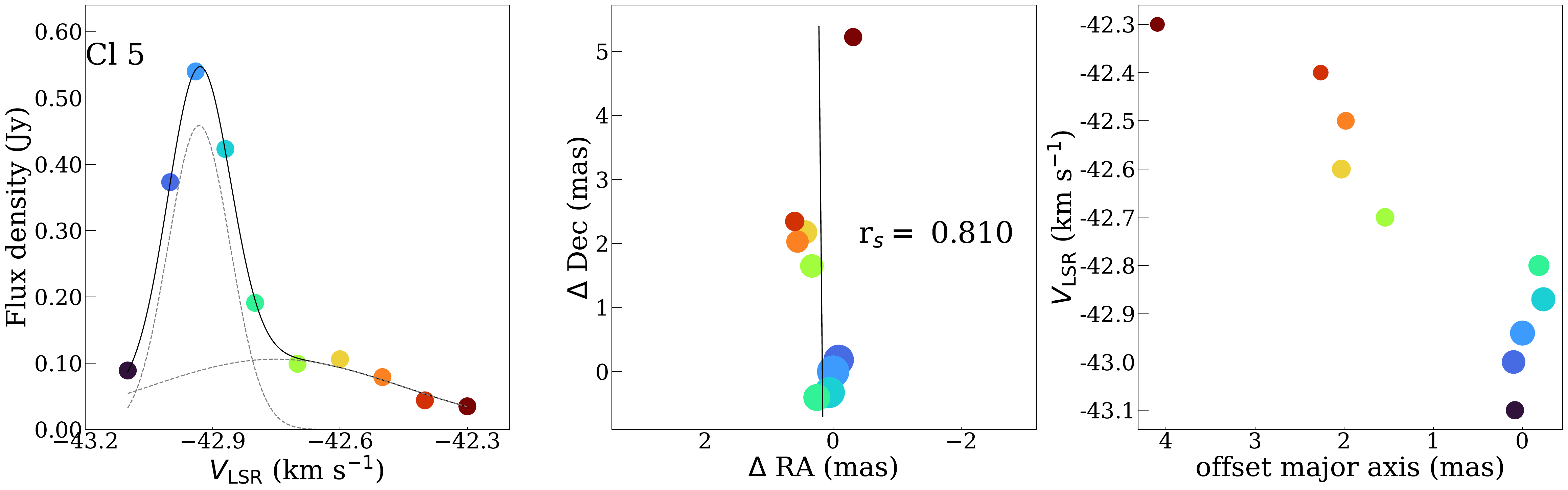}
\includegraphics[scale=0.15]{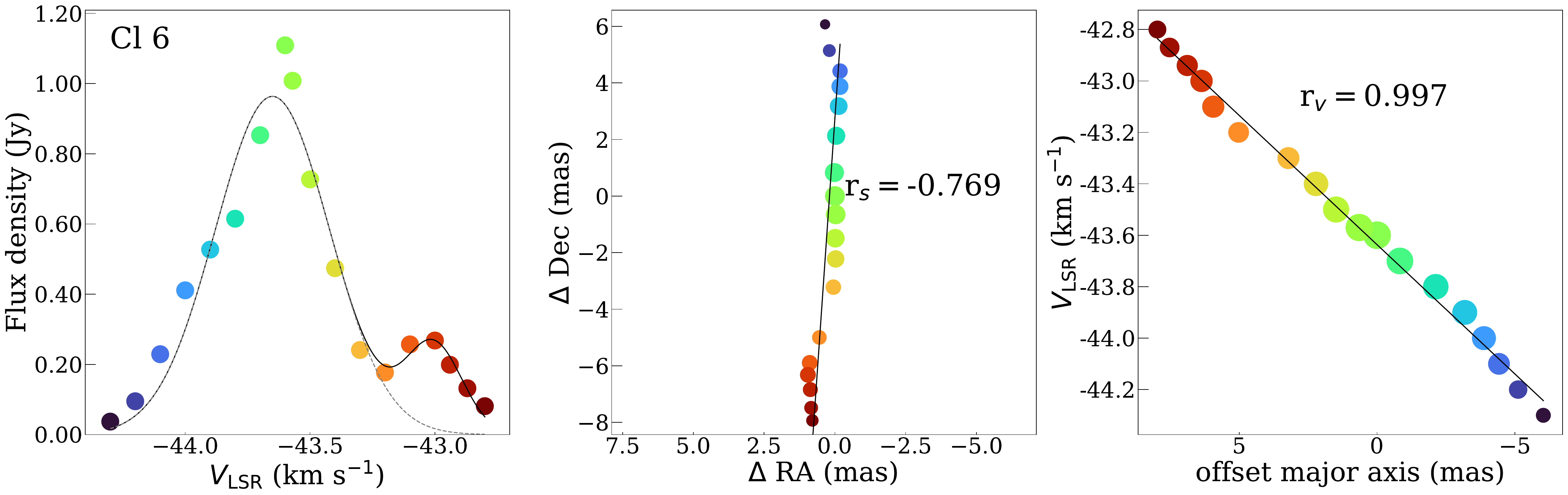}
\caption{Cloudlets of G94.602$-$1.796 as obtained in the EVN observation in 2019. The same information as in Fig.~\ref{G90_cloudlets}.}
\label{G94_cloudlets}  
\end{figure*}


%%%%%%%%%%%%%%%%%%%%%%%%%%%%%%%%%%%%%%%%%%%%%%%%%%

% Don't change these lines
\bsp	% typesetting comment
\label{lastpage}
\end{document}